\documentclass[10pt]{article}
\pdfoutput=1

\usepackage[utf8]{inputenc}
\usepackage[T2A]{fontenc}

\input{epsf}
\usepackage{epsfig}
\usepackage{amssymb}
\usepackage{amsfonts}
\usepackage{amsbsy}
\usepackage[cmtip,all]{xy}
\usepackage{amsmath}
\usepackage{esint}
\usepackage{collectbox}
\usepackage{amscd}
\usepackage{mathrsfs}
\usepackage{amsmath,amsthm}
\usepackage{enumitem}
\usepackage{dsfont}
\usepackage{soul}
\usepackage{comment}
\usepackage{cite} 
\usepackage[most]{tcolorbox}
\usepackage{stmaryrd}
\usepackage{xcolor}
\definecolor{burgundy}{rgb}{0.5, 0.0, 0.13}
\definecolor{olive}{rgb}{0.50, 0.50, 0.0}
\usepackage[linktocpage=true,colorlinks=true,linkcolor=burgundy,citecolor=black!20!blue,urlcolor=violet]{hyperref}
\usepackage{accents}
\usepackage{xfrac}
\usepackage{bm}

\usepackage{tikz}
\usepackage{tikz-cd}
\usetikzlibrary{arrows}
\usetikzlibrary{arrows.meta}
\usetikzlibrary{positioning}
\usetikzlibrary{shapes}
\usetikzlibrary{fit}
\usetikzlibrary{decorations.pathmorphing,decorations.pathreplacing,decorations.markings}
\usetikzlibrary{calc}

\usepackage[font=footnotesize,labelfont=bf]{caption}

\theoremstyle{definition}

\DeclareMathAlphabet{\mathpzc}{OT1}{pzc}{m}{it}

\def\exp{{\rm exp}}

\def\I{{\rm i}}

\def\log{{\rm log}}

\def\p{\partial}

\def\CA{{\cal A}}
\def\CB{{\cal B}}

\def\CB {{\cal B}}

\def\IC{\mathbb{C}}

\def\IR{{\mathbb{R}}}

\def\fe{\mathfrak{e}}
\def\ff{\mathfrak{f}}

\def\fg{\mathfrak{g}}

\def\fl{\mathfrak{l}}
\def\fm{\mathfrak{m}}

\def\fp{\mathfrak{p}}

\def\fs{\mathfrak{s}}

\def\fp{\mathfrak{p}}

\def\fs{\mathfrak{s}}

\def\fu{\mathfrak{u}}

\def\fB{\mathfrak{B}}

\def\lm{\limits}
\def\be{\begin{eqnarray}}
\def\ee{\end{eqnarray}}
\def\nn{\nonumber}

\hoffset= - 1.0in

\numberwithin{equation}{section}

\tcbset{boxrule=0.6pt,colback=white!97!orange}

\DeclareSymbolFont{bbsymbol}{U}{bbold}{m}{n}
\DeclareMathSymbol{\bbzero}{\mathbin}{bbsymbol}{"30}
\DeclareMathSymbol{\bbone}{\mathbin}{bbsymbol}{"31}
\DeclareMathSymbol{\bbtwo}{\mathbin}{bbsymbol}{"32}
\DeclareMathSymbol{\bbthree}{\mathbin}{bbsymbol}{"33}
\DeclareMathSymbol{\bbfour}{\mathbin}{bbsymbol}{"34}
\DeclareMathSymbol{\bbfive}{\mathbin}{bbsymbol}{"35}
\DeclareMathSymbol{\bbsix}{\mathbin}{bbsymbol}{"36}
\DeclareMathSymbol{\bbseven}{\mathbin}{bbsymbol}{"37}
\DeclareMathSymbol{\bbeight}{\mathbin}{bbsymbol}{"38}
\DeclareMathSymbol{\bbnine}{\mathbin}{bbsymbol}{"39}

\def\myblue{white!40!blue}
\def\mygreen{black!40!green}

\definecolor{palette1}{rgb}{0.603922, 0.466667, 0.811765}
\definecolor{palette2}{rgb}{0.329412, 0.219608, 0.517647}
\definecolor{palette3}{rgb}{0.0156863, 0.282353, 0.333333}
\definecolor{palette4}{rgb}{0.631373, 0.211765, 0.439216}
\definecolor{palette5}{rgb}{0.92549, 0.254902, 0.462745}
\definecolor{palette6}{rgb}{1., 0.643137, 0.368627}
\definecolor{palette7}{rgb}{0.313725, 0.45098, 0.85098}

\definecolor{Xmagenta}{HTML}{D60270}
\definecolor{Xpurple}{HTML}{9B4F96}
\definecolor{Xblue}{HTML}{0038A8}

\newcommand\sqbox[1]{{
		\setbox0=\hbox{\mbox{$\Box$}}
		\setbox1=\hbox{\mbox{\raisebox{0.35ex}{\small #1}}}
		\mbox{\raisebox{-0.2ex}{\rlap{\hbox to \wd0{\hss{\box1}\hss}}\box0}}
}}
\newcommand\ssqbox[1]{{
		\setbox0=\hbox{\mbox{$\scriptstyle\Box$}}
		\setbox1=\hbox{\mbox{\raisebox{0.35ex}{\tiny #1}}}
		\mbox{\raisebox{-0.2ex}{\rlap{\hbox to \wd0{\hss{\box1}\hss}}\box0}}
}}

\begin{document}

\hfill MIPT/TH-12/26

\hfill ITEP/TH-12/26

\hfill IITP/TH-12/26

\vskip 1.5in
\begin{center}
	
    {\bf\Large \raisebox{-0.4mm}{\color{white!40!gray}Shading}\hspace{-2.cm} Shading A-polynomials\\
    via HUGE representations of $U_q(\mathfrak{su}_N)$}

	\vskip 0.2in
	\renewcommand{\thefootnote}{\fnsymbol{footnote}}
	{Dmitry Galakhov
		\footnote[2]{e-mail: d.galakhov.pion@gmail.com, galakhov@itep.ru} and  Alexei Morozov
		\footnote[3]{e-mail: morozov@itep.ru}}\\
	
	\vskip 0.2in
	\renewcommand{\thefootnote}{\roman{footnote}}
	{\small{
			\textit{
				MIPT, 141701, Dolgoprudny, Russia
			}
			\vskip 0 cm
			\textit{
				NRC “Kurchatov Institute”, 123182, Moscow, Russia
			}
			\vskip 0 cm
			\textit{
				IITP RAS, 127051, Moscow, Russia}
			\vskip 0 cm
			\textit{
				ITEP, Moscow, Russia}
	}}
\end{center}

\vskip 0.2in
\baselineskip 16pt

\centerline{ABSTRACT}

\bigskip

{\footnotesize
	Classical A-polynomials $A(\ell,m)$ define constraints on coordinates $\ell$ and $m$ in $SL(2,\IC)$ (a complexification of $SU(2)$) character varieties associated to knot complements $S^3\setminus K$.
	Quantum A-polynomials $\hat A(\hat \ell,\hat m)$ are difference operators annihilating Jones polynomials believed to represent wave functions of 3d Chern-Simons theory with gauge group $SU(2)$ on a toroidal pipe surrounding the knot $K$ strand -- a boundary of the knot complement $S^3\setminus K$.
	In this note we suggest a construction of classical \emph{shaded} A-polynomials $A_a(\ell_b,m_c)$ associated to Lie groups $SU(N)$, assuming that color $N$ adds some extra ``depth'' and ``shade'' to the classical notion of A-polynomials.
	We exploit a formalism of \emph{Clebsch-Gordan} (CG) chords, where indices $a$, $b$, $c$ run over $1,\ldots,N-1$.
	CG chords have a \emph{natural} interpretation in terms of 2d CFTs of WZW type, or, alternatively, in terms of quantum group $U_q(\mathfrak{su}_N)$.
	In the case of $\fs\fu_2$, CG chords could be associated to Reeb chords in a knot contact homology (KCH) framework. 
	KCH suggests its own analogue of A-polynomials, known as \emph{augmentation} polynomials, which is allowed to have extra spurious roots in principle.
	Yet the CG chord formalism could be easily extended to arbitrary $\mathfrak{su}_N$ allowing us to generalize the construction of A(ugmentation)-polynomials to arbitrary $\mathfrak{su}_N$ and arbitrary representations as well.
	Here we primarily aim at \emph{classical} A-polynomials by considering a double scaling limit in which $q=e^{\hbar}$, $\hbar\to 0$ and the representations are \emph{huge}: highest weight vector components $w_i\to \infty$ while combinations $\hbar w_i\sim m_i$ remain finite.
	Still, we expect the presented techniques will be helpful in deriving quantum A-polynomials for arbitrary Lie (super)algebras $\fg$.
	Also, we discuss explicit examples of A-polynomials for knots $3_1$, $4_1$ and $5_1$ for $\fg=\fs\fu_3$.

}

\bigskip

\bigskip

\tableofcontents

\bigskip


\section{Introduction}

\begin{figure}[ht!]
	\centering
	\begin{tikzpicture}
		\node (A) at (0,0) {\scalebox{0.7}{$\begin{array}{c}
				\begin{tikzpicture}
					\begin{scope}[scale=0.6]
						\begin{scope}[shift={(0,1)}]
							\draw[fill=white!50!gray] (0,-1) circle (2.1);
							\begin{scope}[shift={(0,-1)}]
								\begin{scope}[yscale=0.2]
									\draw[thin, dashed] (0,0) circle (2.1);
								\end{scope}
							\end{scope}
							\draw[line width=1mm] (1,0) to[out=90,in=90] (-1,0.15) (-1,-0.15) to[out=270,in=135] (0,-1) to[out=315,in=90] (0.4,-1.5) to[out=270,in=45] (0.1,-1.9) (-0.1,-2.1) to[out=225,in=270] (-1.8,-1.5) to[out=90,in=180] (-1,0) -- (0.85,0) (1.15,0) to[out=0,in=90] (1.8,-1.5) to[out=270, in=315] (0,-2) to[out=135,in=270] (-0.4,-1.5) to[out=90,in=225] (-0.1,-1.1) (0.1,-0.9) to[out=45,in=270] (1,0);
							\draw[line width=0.7mm,white] (1,0) to[out=90,in=90] (-1,0.15) (-1,-0.15) to[out=270,in=135] (0,-1) to[out=315,in=90] (0.4,-1.5) to[out=270,in=45] (0.1,-1.9) (-0.1,-2.1) to[out=225,in=270] (-1.8,-1.5) to[out=90,in=180] (-1,0) -- (0.85,0) (1.15,0) to[out=0,in=90] (1.8,-1.5) to[out=270, in=315] (0,-2) to[out=135,in=270] (-0.4,-1.5) to[out=90,in=225] (-0.1,-1.1) (0.1,-0.9) to[out=45,in=270] (1,0);
						\end{scope}
					\end{scope}
				\end{tikzpicture}
			\end{array}$}};
		\node (B) at (4,0) {\scalebox{0.7}{$\begin{array}{c}
				\begin{tikzpicture}
					\begin{scope}[scale=0.4]
						\draw[thick,fill=white!50!gray] (-2.9,0.) to[out=-89.8399,in=100.017] (-2.87723,-0.26343) to[out=-79.9828,in=118.828] (-2.70107,-0.76265) to[out=-61.1725,in=134.258] (-2.40139,-1.16537) to[out=-45.742,in=146.399] (-2.02811,-1.47359) to[out=-33.6013,in=156.101] (-1.60625,-1.70413) to[out=-23.8987,in=164.195] (-1.14839,-1.86896) to[out=-15.8051,in=171.311] (-0.66453,-1.97387) to[out=-8.68872,in=184.42] (0.33791,-2.01131) to[out=4.41993,in=198.572] (1.30922,-1.81925) to[out=18.5724,in=207.161] (1.75584,-1.63275) to[out=27.1611,in=217.636] (2.16277,-1.37722) to[out=37.6355,in=230.875] (2.51464,-1.03837) to[out=50.8749,in=247.59] (2.7788,-0.60076) to[out=67.5902,in=269.84] (2.9,0.)
						(2.9,0.) to[out=90.1601,in=292.698] (2.7788,0.60076) to[out=112.698,in=309.358] (2.51464,1.03837) to[out=129.358,in=322.547] (2.16277,1.37722) to[out=142.547,in=332.985] (1.75584,1.63275) to[out=152.985,in=341.551] (1.30922,1.81925) to[out=161.551,in=355.683] (0.33791,2.01131) to[out=175.683,in=8.79496] (-0.66453,1.97387) to[out=-171.205,in=15.9226] (-1.14839,1.86896) to[out=-164.077,in=24.036] (-1.60625,1.70413) to[out=-155.964,in=33.7698] (-2.02811,1.47359) to[out=-146.23,in=45.956] (-2.40139,1.16537) to[out=-134.044,in=61.4422] (-2.70107,0.76265) to[out=-118.558,in=80.2966] (-2.87723,0.26343) to[out=-99.7034,in=89.8399] (-2.9,0.);
						\draw[thick,fill=white]
						(-1.11956,0.09908) to[out=-49.9491,in=153.423] (-0.92834,-0.03418) to[out=-26.5774,in=161.919] (-0.72701,-0.11594) to[out=-18.081,in=169.27] (-0.47349,-0.18059) to[out=-10.7297,in=175.99] (-0.18692,-0.21756) to[out=-4.01049,in=184.113] (0.18692,-0.21756) to[out=4.11276,in=190.839] (0.47349,-0.18059) to[out=10.8386,in=198.203] (0.72701,-0.11594) to[out=18.2034,in=206.723] (0.92834,-0.03418) to[out=26.7226,in=229.949] (1.11956,0.09908)
						(0.97323,0.01035) to[out=150.737,in=339.671] (0.78968,0.09405) to[out=159.671,in=347.286] (0.54899,0.16486) to[out=167.286,in=354.139] (0.2697,0.21034) to[out=174.139,in=357.416] (0.12181,0.22137) to[out=177.416,in=1.0626] (-0.04693,0.22375) to[out=-178.937,in=5.96464] (-0.2697,0.21034) to[out=-174.035,in=12.8258] (-0.54899,0.16486) to[out=-167.174,in=20.4567] (-0.78968,0.09405) to[out=-159.543,in=29.2627] (-0.97323,0.01035);
						\draw[thick,\myblue]
						(-2.,0.74437) to[out=-110.513,in=111.519] (-1.99235,0.25547) to[out=-68.4811,in=143.432] (-1.59517,-0.22917) to[out=-36.5678,in=164.608] (-0.84954,-0.57841) to[out=-15.3925,in=173.316] (-0.39091,-0.66766) to[out=-6.68404,in=182.646] (0.15119,-0.68718) to[out=2.64617,in=195.535] (0.84954,-0.57841) to[out=15.5348,in=216.766] (1.59517,-0.22917) to[out=36.7659,in=248.787] (1.99235,0.25547) to[out=68.7869,in=290.513] (2.,0.74437)
						(2.,0.74437) to[out=110.827,in=311.265] (1.85392,0.98337) to[out=131.265,in=335.001] (1.41484,1.29737) to[out=155.001,in=348.618] (0.84207,1.47882) to[out=168.618,in=357.467] (0.21479,1.55296) to[out=177.467,in=5.21746] (-0.42348,1.53888) to[out=-174.783,in=15.2103] (-1.03897,1.4325) to[out=-164.79,in=31.5016] (-1.57881,1.20973) to[out=-148.498,in=43.8516] (-1.79277,1.04724) to[out=-136.148,in=69.173] (-2.,0.74437);
						\node[\myblue, below left] at (-0.84954,-0.57841) {$\scriptstyle \fl$};
						\draw[\myblue, thick]
						(1.27781,-1.82961) to[out=18.1969,in=232.273] (1.36305,-1.76682) to[out=52.2727,in=255.341] (1.42518,-1.63128) to[out=75.3413,in=270.783] (1.4499,-1.39568) to[out=90.783,in=277.359] (1.4373,-1.22373) to[out=97.3594,in=285.102] (1.38531,-0.96416) to[out=105.102,in=295.592] (1.25201,-0.60221) to[out=115.592,in=310.13] (1.05809,-0.29197) to[out=130.13,in=326.502] (0.91736,-0.16139) to[out=146.502,in=17.7616] (0.72219,-0.11751);
						\draw[\myblue, dashed, thick]
						(0.72219,-0.11751) to[out=-161.803,in=69.0245] (0.59176,-0.26297) to[out=-110.975,in=86.938] (0.55164,-0.47351) to[out=-93.062,in=100.849] (0.58046,-0.83433) to[out=-79.151,in=109.283] (0.66132,-1.13397) to[out=-70.717,in=115.835] (0.75151,-1.35224) to[out=-64.1654,in=122.344] (0.84463,-1.5209) to[out=-57.6557,in=135.66] (0.99829,-1.71626) to[out=-44.3405,in=148.487] (1.09512,-1.79371) to[out=-31.5133,in=197.762] (1.27781,-1.82961);
						\node[\myblue, above right] at (1.4499,-1.39568) {$\scriptstyle \fm$};
					\end{scope}
				\end{tikzpicture}
			\end{array}$}};
		\path (A) edge[<->] node[above] {\scriptsize inversion} (B);
		\node(C) at (8,0) {$\begin{array}{c}
				\begin{tikzpicture}[scale=0.7]
					\draw[white,line width = 1.5mm] (-0.5,0) to[out=90,in=270] (1,1.5) to[out=90,in=0] (0,2);
					\draw[gray, line width =  1mm] (-0.5,0) to[out=90,in=270] (1,1.5) to[out=90,in=0] (0,2);
					\begin{scope}[yscale=0.5]
						\draw[thick, \myblue] ([shift={(0:0.3)}]-1.5,0) arc (0:180:0.3);
					\end{scope}
					\node[left, \myblue] at (-1.8,0) {$\fm$};
 					\draw[white,line width = 1.5mm] (0,1) to[out=180,in=90] (-1.5,0) to[out=270,in=180] (-0.5,-0.7) to[out=0,in=270] (0.5,0);
					\draw[gray, line width =  1mm] (0,1.2) to[out=180,in=90] (-1.5,0) to[out=270,in=180] (-0.5,-0.7) to[out=0,in=270] (0.5,0);
					\draw[white,line width = 1.5mm] (0,2) to[out=180,in=90] (-1,1.5) to[out=270,in=90] (0.5,0);
					\draw[gray, line width =  1mm] (0,2) to[out=180,in=90] (-1,1.5) to[out=270,in=90] (0.5,0);
					\begin{scope}[xscale=-1]
						\draw[white,line width = 1.5mm] (0,1.2) to[out=180,in=90] (-1.5,0) to[out=270,in=180] (-0.5,-0.7) to[out=0,in=270] (0.5,0);
						\draw[gray, line width =  1mm] (0,1.2) to[out=180,in=90] (-1.5,0) to[out=270,in=180] (-0.5,-0.7) to[out=0,in=270] (0.5,0);
					\end{scope}
					\begin{scope}[yscale=0.5]
						\draw[white, line width = 0.7mm] ([shift={(180:0.3)}]-1.5,0) arc (180:360:0.3);
						\draw[thick, \myblue] ([shift={(180:0.3)}]-1.5,0) arc (180:360:0.3);
					\end{scope}
				\end{tikzpicture}
			\end{array}$};
			\node(D) at (12,0) {$\begin{array}{c}
					\begin{tikzpicture}[scale=0.7]
						\draw[white,line width = 1.5mm] (-0.5,0) to[out=90,in=270] (1,1.5) to[out=90,in=0] (0,2);
						\draw[gray, line width =  1mm] (-0.5,0) to[out=90,in=270] (1,1.5) to[out=90,in=0] (0,2);
						\begin{scope}[shift={(0.2,-0.2)}]
							\draw[thick, white,line width = 0.7mm] (-0.5,0) to[out=90,in=270] (1,1.5) to[out=90,in=0] (0,2);
							\draw[thick, \myblue] (-0.5,0) to[out=90,in=270] (1,1.5) to[out=90,in=0] (0,2);
						\end{scope}
						\draw[white,line width = 1.5mm] (0,1.2) to[out=180,in=90] (-1.5,0) to[out=270,in=180] (-0.5,-0.7) to[out=0,in=270] (0.5,0);
						\draw[gray, line width =  1mm] (0,1.2) to[out=180,in=90] (-1.5,0) to[out=270,in=180] (-0.5,-0.7) to[out=0,in=270] (0.5,0);
						\begin{scope}[shift={(0.2,-0.2)}]
							\draw[thick, white,line width = 0.7mm] (0,1.2) to[out=180,in=90] (-1.5,0) to[out=270,in=180] (-0.5,-0.7) to[out=0,in=270] (0.5,0);
							\draw[thick, \myblue] (0,1.2) to[out=180,in=90] (-1.5,0) to[out=270,in=180] (-0.5,-0.7) to[out=0,in=270] (0.5,0);
						\end{scope}
						\draw[white,line width = 1.5mm] (0,2) to[out=180,in=90] (-1,1.5) to[out=270,in=90] (0.5,0);
						\draw[gray, line width =  1mm] (0,2) to[out=180,in=90] (-1,1.5) to[out=270,in=90] (0.5,0);
						\begin{scope}[shift={(0.2,-0.2)}]
							\draw[thick, white,line width = 0.7mm] (0,2) to[out=180,in=90] (-1,1.5) to[out=270,in=90] (0.5,0);
							\draw[thick, \myblue] (0,2) to[out=180,in=90] (-1,1.5) to[out=270,in=90] (0.5,0);
							\node[right, \myblue] at (1.5,0) {$\fl$};
						\end{scope}
						\begin{scope}[xscale=-1]
							\draw[white,line width = 1.5mm] (0,1.2) to[out=180,in=90] (-1.5,0) to[out=270,in=180] (-0.5,-0.7) to[out=0,in=270] (0.5,0);
							\draw[gray, line width =  1mm] (0,1.2) to[out=180,in=90] (-1.5,0) to[out=270,in=180] (-0.5,-0.7) to[out=0,in=270] (0.5,0);
						\end{scope}
						\begin{scope}[shift={(0.2,-0.2)}]
							\begin{scope}[xscale=-1]
								\draw[thick, white,line width = 0.7mm] (0,1.2) to[out=180,in=90] (-1.5,0) to[out=270,in=180] (-0.5,-0.7) to[out=0,in=270] (0.5,0);
								\draw[thick, \myblue] (0,1.2) to[out=180,in=90] (-1.5,0) to[out=270,in=180] (-0.5,-0.7) to[out=0,in=270] (0.5,0);
							\end{scope}
						\end{scope}
					\end{tikzpicture}
				\end{array}$};
	\end{tikzpicture}
	\caption{Meridian and longitude operators on the knot complement $S^3\setminus K$} \label{fig:lambda_mu}
\end{figure}
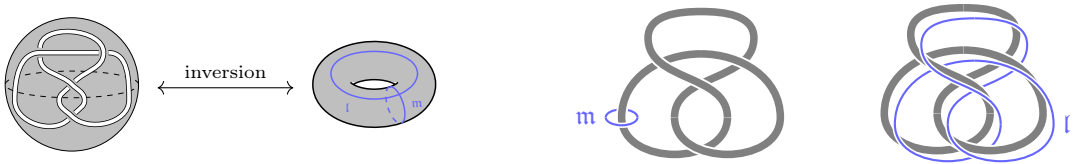

A \emph{classical} A-polynomial of a knot $K$ was defined in \cite{Cooper1994Plane} as a relation in an $SL(2,\IC)$ character variety of the fundamental group of the knot complement $\pi_1(S^3\setminus K)$.
A small tubular neighborhood of a knot $K$ embedded in $S^3$ has a torus boundary.
One considers two generators of the fundamental group as loops corresponding to a ``meridian'' and a ``longitude'' of the torus boundary (see Fig.~\ref{fig:lambda_mu}).
Images of these loops in the $SL(2,\IC)$ character variety are diagonal matrices:
\begin{equation}\label{holo}
	\left(\begin{array}{cc}
		m & 0\\
		0 & m^{-1}\\
	\end{array}\right),\quad \left(\begin{array}{cc}
	\ell & 0\\
	0 & \ell^{-1}\\
	\end{array}\right)\,.
\end{equation}
The relations in the fundamental group of a knot impose a polynomial condition $A^K(\ell,m)=0$, where the polynomial $A^K(\ell,m)$ is called an A-polynomial.

The notion of a classical A-polynomial was deformed to the notion of a \emph{quantum} A-polynomial \cite{frohman1998a,Gukov:2003na,garoufalidis2004characteristic,garoufalidis2005colored,garoufalidis2008non,fuji2012super,Aganagic:2012jb,Aganagic:2013jpa,larraguivel2020nahm,gukov2021two,ekholm2022branches,Galakhov:2024eco,Galakhov:2025ehn} as a difference operator annihilating the Jones polynomial of a knot $K$ as a function of spin $r$:
\begin{equation}
	\mathscr{A}^K(\hat\ell,\hat m|q)J_K(r|q)=0,\quad \hat\ell\, J_K(r|q)=J_K(r+1|q),\quad \hat m \, J_K(r|q)=q^r J_K(r|q)\,.
\end{equation}
The classical A-polynomial can be reconstructed from the quantum one in a ``quasi-classical'' limit, where operators are replaced by commuting variables and $q\to 1$:
\begin{equation}
	A^K(\ell,m)=\mathscr{A}^K(\ell,m|1)\,.
\end{equation}
One can also consider ``dual'' difference operators called C-polynomials \cite{garoufalidis2006c,mironov2020algebra}.

An interpretation of this construction in the broader context of HOMFLY-PT polynomials \cite{freyd1985new,przytycki1988invariants} is the following.
A HOMFLY-PT polynomial is considered to be given by a (possibly continued analytically to complexified gauge groups \cite{Witten:1989ip,Witten:2010cx}) 3d Chern-Simons path integral on the knot complement $S^3\setminus K$.
To regularize this integral it is natural to cut a tubular neighborhood of the knot out of $S^3$, then $S^3\setminus K$ becomes a manifold with a torus boundary, and the path integral corresponding to the HOMFLY-PT polynomial becomes a wave function $\Psi_K$ in the Chern-Simons Hilbert space on this torus.
Implementing the Ward identities for the Chern-Simons path integral leads to  combinations of quantum operator annihilating this wave function:
\begin{equation}
	\mathscr{A}_{a}^K(\hat \fl_R,\hat \fm_Q)\Psi_K=0\,,
\end{equation}
where $\hat \fm_Q$, $\hat \fl_R$ are quantum Wilson loop operators lying on the meridian and the longitude in the respective representations $Q$ and $R$.
These operators are quantum versions of the traces of holonomies \eqref{holo}, so in the quasi-classical limit for $\fs\fu_2$ we have:
\begin{equation}
	\fl_{\Box}=\ell+\ell^{-1},\quad \fm_{\Box}=m+m^{-1}\,.
\end{equation}

We call these Ward identities quantum \emph{shaded} A-polynomials, and here we will be interested in their quasi-classical versions for $\fs\fu_n$:
\begin{equation}
	A_{a}(\lambda_{b},\mu_{c})=0, \quad a,b,c=1,\ldots,n-1,\,
\end{equation}
where we consider a \emph{double} scaling limit, where the quantum parameter $q\to 1$ and the highest weight components of the representations become \emph{huge} $w_a\to \infty$, so that the combinations $\mu_c:=q^{w_c}$ remain finite, while the shift operators $\hat \lambda_{b}\Psi_K(w_a):=\Psi_K(w_a+\delta_{ab})$ become commutating variables $\lambda_b$.

\begin{figure}[ht!]
	\centering
	\begin{tikzpicture}
		\node[draw, rounded corners=3] (11) at (0,0) {$\begin{array}{c}
				\mbox{KZ equations}\\
				\left(\p_{x_a}-\frac{1}{\hbar}\sum\lm_{b\neq a}\frac{\eta_{ij}\;R_a(t^i)\otimes R_b(t^j)}{x_a-x_b}\right)\Psi=0
			\end{array}$};
		\node[draw, rounded corners=3] (31) at (8,0) {Skein relations};
		\path (11) edge[stealth-stealth] node[pos=0.5,above] {$\begin{array}{c}
				\mbox{\scriptsize RT formalism}\\
				\mbox{\scriptsize for $U_q(\fg)$}
			\end{array}$} (31); 
		\node[draw, rounded corners=3] (12) at (-1,-3) {$\begin{array}{c}
				\mbox{ODE}\\
				\left(\p_z-\CB(z)\right)\Psi=0
			\end{array}$};
		\node[draw, rounded corners=3] (13) at (-1,-4.5) {FG/FN coordinates};
		\draw[-stealth] (11.south west) to[out=270,in=120] node[pos=0.5,right=0.2] {$\begin{array}{l}
				\mbox{\scriptsize Limit of}\\
				\mbox{\scriptsize large spins}\\
				\scriptstyle x_0 = z,\; R_0={\rm small},\; R_a={\rm large}
			\end{array}$} (12.north west);
		\path (12) edge[-stealth] node[right] {\scriptsize Stokes lines} (13) ;
		\node[draw, rounded corners=3, fill=burgundy] (21) at (4,-2) {$\begin{array}{c}
				\mbox{\color{white}Reeb/CG}\\
				\mbox{\color{white}chords}
			\end{array}$};
		\node[draw, rounded corners=3] (32) at (9,-2) {$\begin{array}{c}
				\mbox{Relations}\\
				\mbox{for link symbols}\\
				\mbox{in the knot background}
			\end{array}$};
		\node[draw, rounded corners=3] (22) at (4,-4) {$\begin{array}{c}
				\mbox{Relations}\\
				\mbox{for Reeb/CG chords}\\
				\mbox{in the knot background}
			\end{array}$};
		\path (21) edge[-stealth] node[right] {\scriptsize braid group} (22) ;
		\draw[stealth-stealth] (12.east) to[out=0,in=180] node[pos=0.5,above=0.7,rotate=90] {\scriptsize CG=${\rm Pexp}\int\CB$} (21.west);
		\draw[-stealth] (31.south) to[out=270,in=90] (32.north);
		\draw[stealth-stealth] ([shift={(1.8,0)}]22.north) to[out=90,in=180] node[pos=0.7,above=1,rotate=90] {\scriptsize same algorithm} (32.west);
		\node[draw, rounded corners=3] (33) at (9,-5) {$\begin{array}{c}
				\mbox{Shifting spins =}\\
				\mbox{= cabling}
			\end{array}$};
		\node[draw, rounded corners=3] (14) at (-1,-6) {$\begin{array}{c}
				\mbox{Cluster mutations}\\
				\mu_{\gamma'}(X_{\gamma})\sim (X_{\gamma'}|q)_{\infty}^{-1}X_{\gamma}(X_{\gamma'}|q)_{\infty}
			\end{array}$};
		\path (13) edge[-stealth] node[right] {\scriptsize braid action} (14) ;
		\node[draw, rounded corners=3] (15) at (-1,-8) {$\begin{array}{c}
				\mbox{Relations on $X_{\gamma}$}\\
				\mbox{for various temporal slices}
			\end{array}$};
		\path (14) edge[-stealth] node[right] {\scriptsize knot} (15) ;
		\node[draw, rounded corners=3, below, fill=burgundy] (23) at (4,-9) {$\begin{array}{c}
				\mbox{\color{white}``Shaded''}\\
				\mbox{\color{white}A-polynomials}\\
				\color{white}\CA_a(m_b,\ell_c)=0\\
				\color{white}\#(a,b,c) = {\rm rk}\,\fg
			\end{array}$};
		\node[draw, rounded corners=3, below] (34) at (9,-6) {$\begin{array}{c}
				\mbox{Cable ops}=\\
				=F\left(\mbox{Link symbols}\right)
			\end{array}$};
		\node[draw, rounded corners=3, below] (35) at (6,-8) {+};
		\draw[-stealth] ([shift={(-1.8,0)}]32.south) to[out=270,in=90] (35.north);
		\path (33) edge[-stealth] (34) ;
		\draw[-stealth] (34.south) to[out=270,in=0] node[pos=0.5, below right] {$\begin{array}{l}
				\mbox{\scriptsize Ops, symbols}\\
				\mbox{\scriptsize become commutative}\\
				\mbox{\scriptsize in the quasi-classical limit}
			\end{array}$} (35.east);
		\draw[-stealth] (35.south) to[out=270,in=0] node[pos=0.8,right] {\scriptsize Exclude link symbols} (23.east);
		\draw[-stealth] (15.south) to[out=270,in=180] node[pos=0.5,below left] {\scriptsize Exclude FG coords} (23.west);
		\draw[-stealth] (22.south) to[out=270,in=90] node[pos=0.9,right=0.2, rotate=90] {\scriptsize Exclude Reeb/CG chords} (23.north);
	\end{tikzpicture}
	\caption{All roads lead to A-polynomials}\label{fig:scheme}
\end{figure}
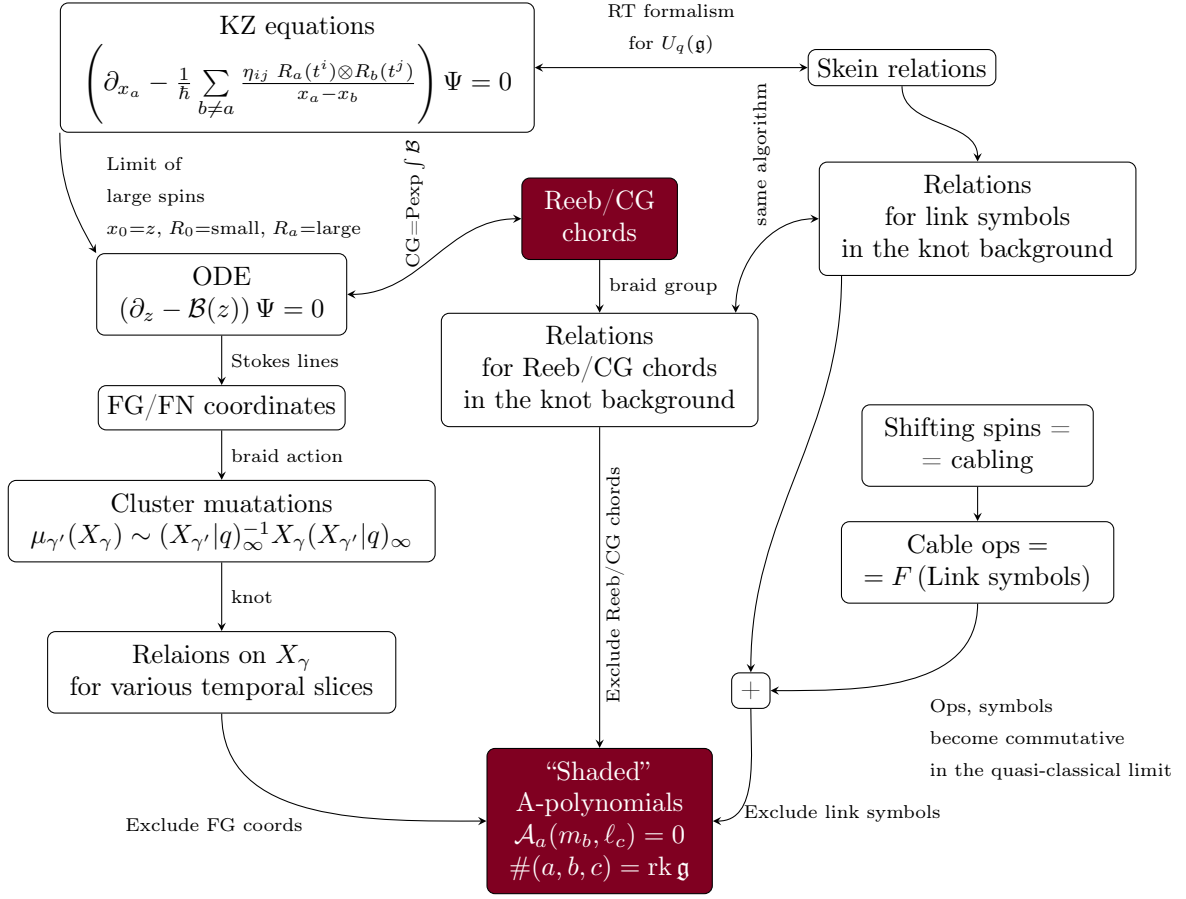

There are various ways of deriving such Ward identities (see Fig.~\ref{fig:scheme}), which we discuss separately in Sec.~\ref{sec:contexts}.
In addition, there are brute-force phenomenological methods of guessing difference operators annihilating known expressions for HOMFLY polynomials \cite{Garoufalidis:2010ek}.

In this note we extend a method of Clebsch-Gordan (CG) chords proposed in \cite{Galakhov:2024eco} to $\fs\fu_n$, $n>2$.
For $\fs\fu_2$, this method is almost identical to the extraction of \emph{augmentation} polynomials for knot contact homology and can be easily extended to generic $\fs\fu_n$.
Eventually, we would expect that this extension could be continued to arbitrary Lie (super)algebras $\fg$.
We will derive transformation rules for CG chords under the action of the braid group.
Then we will construct a non-linear eigen value problem \eqref{main_eqs} for CG chord variables in the presence of knot $K$ in its closed braid representation.
A constraint for this system to have a solution, a \emph{generalized resultant}, would take the form of shaded A-polynomials.

We will test this proposal and compute some roots of shaded A-polynomials for the trefoil $3_1$, the figure-eight $4_1$, and the cinquefoil $5_1$ knots in the case of $\fs\fu_3$ explicitly.

The paper is organized as follows.
In Sec.~\ref{sec:RTintro} we review some aspects of the quantum group representation theory and of the Reshetikhin-Turaev formalism necessary to compute HOMFLY-PT polynomials.
In Sec.~\ref{sec:CG} we introduce a notion of colored CG chords, construct the action of the braid group on them, and present a scheme of constructing shaded A-polynomials.
Also, we discuss why and in what sense A-polynomials are knot invariants.
In Sec.~\ref{sec:contexts} we discuss relations between the presented CG chord formalism and other contexts for constructing A-polynomials: knot contact homology, WZW models, 3d TQFTs and link symbols introduced in \cite{Galakhov:2025ehn}.
finally, in Sec.~\ref{sec:examples} we construct examples of shaded A-polynomials for some knots in $\fs\fu_3$.


\section{Quantum group \texorpdfstring{$U_q(\fg)$}{Uq(g)} and Reshetikhin-Turaev formalism} \label{sec:RTintro}

\subsection{Quantum group \texorpdfstring{$U_q(\fg)$}{Uq(g)}}

Let us start with a brief review of the quantum group construction.
For a simple Lie algebra $\fg$ defined by Cartan matrix $a_{ij}$, where $i,j=1,\ldots,{\rm rk}\,\fg$, one defines quantum group $U_q(\fg)$ \cite{drinfeld1986quantum,Jimbo1985,etingof1998lectures} as an enveloping algebra of canonical Chevalley generators with $q$-deformed relations:
\begin{equation}
	\begin{split}
		q^{h_j}e_iq^{-h_j}=q^{a_{ij}}e_i,\quad q^{h_j}f_iq^{-h_j}=q^{-a_{ij}}f_i,\quad \left[e_i,f_j\right]=\delta_{ij}\frac{q^{h_i}-q^{-h_i}}{q-q^{-1}}\,,\\
		\sum\lm_{n=0}^{1-a_{ij}}\frac{(-1)^n}{[n]_i![1-a_{ij}-n]_i!}e_i^ne_je_i^{1-a_{ij}-n}=0,\quad \sum\lm_{n=0}^{1-a_{ij}}\frac{(-1)^n}{[n]_i![1-a_{ij}-n]_i!}f_i^nf_jf_i^{1-a_{ij}-n}=0\,,
	\end{split}
\end{equation}
where $[n]_i=\frac{q^{nd_i}-q^{-nd_i}}{q^{d_i}-q^{-d_i}}$ for $d_i=\langle\alpha_i,\alpha_i\rangle/2$.

Here, for practical use we will operate mostly with algebras belonging to the Dynkin A-series $U_q(\fs\fu_N)$.
The respective Cartan matrix is $a_{ij}=2\delta_{i,j}-\delta_{i,j+1}-\delta_{i,j-1}$, and this leads to the following relations among the generators of $U_q(\fs\fu_N)$:
\begin{equation}\label{QG}
	\begin{aligned}
		&q^{h_j}e_iq^{-h_j}=q^{a_{ij}}e_i,\quad q^{h_j}f_iq^{-h_j}=q^{-a_{ij}}f_i,\quad \left[e_i,f_j\right]=\delta_{ij}\frac{q^{h_i}-q^{-h_i}}{q-q^{-1}}\,,\\
		&[e_i,e_j]=0,\quad [f_i,f_j]=0,\quad \mbox{if }|i-j|>1\,,\\
		&e_i^2e_j-(q+q^{-1})e_ie_je_i+e_je_i^2=0,\;f_i^2f_j-(q+q^{-1})f_if_jf_i+f_jf_i^2=0,\quad \mbox{if }|i-j|=1\,.
	\end{aligned}
\end{equation}

It is well-known (see e.g. \cite{Jantzen:1996:LQG}) that the representation theory of finite-dimensional representations of $U_q(\fg)$ is the same as that of $\fg$ if $q$ is not a root of unity.\footnote{When $q$ is a root of unity, the representation theory of $U_q(\fg)$ becomes rather sophisticated, tensor products might include reducible yet indecomposable representations etc. \cite{Bishler:2022ozd,Bishler:2022ffv}.}
Finite-dimensional representations are constructed as truncated highest weight Verma modules for the integral highest weights.

$U_q(\fg)$ carries a Hopf algebra structure, therefore representations of $U_q(\fg)$ could be organized into tensor products by applying one of two available inequivalent co-product structures:
\begin{equation}\label{coprod}
	\begin{aligned}
		&\Delta(e_i)=e_i\otimes q^{\frac{h_i}{2}}+q^{-\frac{h_i}{2}}\otimes e_i,\quad \Delta(f_i)=f_i\otimes q^{\frac{h_i}{2}}+q^{-\frac{h_i}{2}}\otimes f_i,\quad \Delta(h_i)=h_i\otimes 1+1\otimes h_i\,;\\
		&\tilde\Delta(e_i)=e_i\otimes q^{-\frac{h_i}{2}}+q^{\frac{h_i}{2}}\otimes e_i,\quad \tilde\Delta(f_i)=f_i\otimes q^{-\frac{h_i}{2}}+q^{\frac{h_i}{2}}\otimes f_i,\quad \tilde\Delta(h_i)=h_i\otimes 1+1\otimes h_i\,.
	\end{aligned}
\end{equation}
In constructing tensor products in this paper, we will, without loss of generality, always apply $\Delta$.

Similarly to the case of $\fg$, their quantized analogs $U_q(\fg)$ admit an isotypical decomposition over irreducible representations for tensor powers if $q$ is not a root of unity:
\begin{equation}
	\rho_1\otimes \rho_2=\bigoplus\lm_{\rho_3\in{\rm irr}(\fg)}D_{\rho_1 \rho_2}^{\rho_3}\otimes \rho_3\,,
\end{equation}
where $D$ are invariant multiplicity subspaces.
Their dimensions ${\rm dim}\,D$ are called Hall-Littlewood coefficients.

Let us assume that in each representation $\rho_i$ an orthonormal basis of vectors $v_k^{(i)}$, where $k=1,\ldots,{\rm dim}\,\rho_i$, is chosen.
Then we call $q$-\emph{Clebsch-Gordan} (CG) coefficients (Wigner $3j$-symbols) and inverse CG coefficients respectively the following scalar products:
\begin{equation}\label{CG}
	C_{k}^{i,j}\left[\begin{array}{c}
		\rho_1 \;\rho_2\\
		\rho_3
	\end{array}\right]=\left\langle v_i^{(1)}\otimes v_j^{(2)}\Big|v_k^{(3)}\right\rangle,\quad \bar C^{k}_{i,j}\left[\begin{array}{c}
	\rho_3\\
	\rho_1 \;\rho_2
	\end{array}\right]=\left\langle v_k^{(3)}\Big|v_i^{(1)}\otimes v_j^{(2)}\right\rangle\,,
\end{equation}
that depend explicitly on the choice of representations $\rho_1$, $\rho_2$ and $\rho_3$.

A universal R-matrix $\check R$ is an invertible element of $U_q(\fg)\otimes U_q(\fg)$ conjugating one co-product structure \eqref{coprod} into the other, in particular:
\begin{equation}\label{Rdelta}
	\check R \circ \Delta(a)=\tilde \Delta(a)\circ \check R,\quad \forall\,a\in U_q(\fg)\,.
\end{equation}
Also, we consider an R-matrix defined by the following relation:
\begin{equation}
	R=P\circ\check R\,,
\end{equation}
where $P$ is a permutation of tensor factors $v\otimes w\to w\otimes v$.
This $R$-matrix intertwines the co-product:
\begin{equation}\label{intertw}
	R\circ\Delta(a)=\Delta(a)\circ R,\quad \forall\,a\in U_q(\fg)\,.
\end{equation}
In this form $R$ satisfies the Yang-Baxter equation on a tensor cube:
\begin{equation}
	R_{12}R_{23}R_{12}=R_{23}R_{12}R_{23}\,,
\end{equation}
where $R_{ij}$ acts on the $i^{\rm th}$ and the $j^{\rm th}$ factors in the tensor power.

An existence of the universal R-matrix for quantized Lie super-algebras was determined in \cite{khoroshkin1991universal} and has the following form:\footnote{
It is easy to observe that the zeroth order of this series delivers right shifts for exponents using the following simple relations: 
\begin{equation}\nn
	\begin{aligned}
		&q^{\lambda a_{ij}^{-1}h_i\otimes h_j}(e_k\otimes 1)\,q^{-\lambda a_{ij}^{-1}h_i\otimes h_j}=e_k\otimes q^{\lambda h_k}, \quad q^{\lambda a_{ij}^{-1}h_i\otimes h_j}(1\otimes e_k)\,q^{-\lambda a_{ij}^{-1}h_i\otimes h_j}= q^{\lambda h_k} \otimes e_k\,,\\
		&q^{\lambda a_{ij}^{-1}h_i\otimes h_j}(f_k\otimes 1)\,q^{-\lambda a_{ij}^{-1}h_i\otimes h_j}=f_k\otimes q^{-\lambda h_k}, \quad q^{\lambda a_{ij}^{-1}h_i\otimes h_j}(1\otimes f_k)\,q^{-\lambda a_{ij}^{-1}h_i\otimes h_j}= q^{-\lambda h_k} \otimes f_k\,.
	\end{aligned}
\end{equation}
}

\begin{equation}\label{uni_R-mat} 
	\check R=q^{-\frac{1}{2}a_{ij}^{-1}h_i\otimes h_j}\left(1+\sum\lm_{n=1}^{\infty}\sum\lm_{\gamma,\gamma'} c_{\gamma,\gamma'} F_{\gamma}\otimes E_{\gamma'}\right) q^{-\frac{1}{2}a_{ij}^{-1}h_i\otimes h_j}\,,
\end{equation}
where $F_{\gamma}$ and $E_{\gamma'}$ are products of generators $f_k$ and $e_k$ of length $n$, respectively, corresponding to the same root.


\subsection{Reshetikhin-Turaev formalism}\label{sec:RT}

The Reshetikhin-Turaev (RT) formalism \cite{RT1,RT2,1112.2654,Mironov:2015aia,Mironov:2015qma,Anokhina:2024lbn,Galakhov:2014sha} allows one to compute HOMFLY-PT polynomials \cite{freyd1985new,przytycki1988invariants} for knot (more generally, link) diagrams by mapping diagram elements into objects and morphisms in a braided tensor category of $U_q(\fg)$.

In general, the RT formalism allows one to map an \emph{open tangle}, that is, an open braid picture with $M$ open ends, into morphism tensors ${\rm Rep}\left(U_q(\fg)\right)^{\otimes m}\to {\rm Rep}\left(U_q(\fg)\right)^{\otimes (M-m)}$ that are invariant with respect to Reidemeister moves II and III and invariant under Reidemeister move I up to a $q$-monomial scale factor (a framing anomaly factor).
Passing from tangles to knots (or, more generally, to links) reduces this map to a morphism $\varnothing\to\varnothing$, where $\varnothing$ denotes the trivial representation of $U_q(\fg)$.
This morphism is a plain number $\Psi_K(\rho_0|q)$ depending on the knot $K$ in question, a chosen representation $\rho_0$ coloring the knot strand, and the quantum parameter $q$ of the quantum group.
$\Psi_K(\rho_0|q)$ is identified with the HOMFLY-PT polynomial of $K$ in representation $\rho_0$.
Here we simply recap some basic elements of this construction.

In a plane where the knot diagram is drawn, one chooses a ``time flow'' direction (assuming that the diagram implements an embedding of the knot in a 3d space, where a 3d Chern-Simons theory is quantized in the Heisenberg picture by choosing one of three coordinates as a time coordinate) and an orientation of the knot strand.
In this note, we assume that the time flows upwards in all diagrams.

With a generic constant time slice that does not contain any vertices and intersects the knot along a sheaf of strands, one associates a tensor product of the corresponding representations.
If a strand is co-aligned with the time flow, then it contributes as a factor $\rho$ in the tensor power, otherwise it contributes as a complex conjugate representation $\bar \rho$:
\begin{equation}
	\begin{array}{c}
		\begin{tikzpicture}
			\draw[thick, -stealth] (0,0) -- (0,0.5);
			\draw[thick, stealth-] (0.8,0) -- (0.8,0.5);
			\draw[thick, -stealth] (1.6,0) -- (1.6,0.5);
			\draw[thick, stealth-] (2.4,0) -- (2.4,0.5);
			\draw[thick, -stealth] (3.2,0) -- (3.2,0.5);
			\node[left] at (0,0.25) {$\scriptstyle \rho_1$};
			\node[left] at (0.8,0.25) {$\scriptstyle \rho_2$};
			\node[left] at (1.6,0.25) {$\scriptstyle \rho_3$};
			\node[left] at (2.4,0.25) {$\scriptstyle \rho_4$};
			\node[left] at (3.2,0.25) {$\scriptstyle \rho_5$};
		\end{tikzpicture}
	\end{array}\;\mapsto\; \rho_1\otimes \bar \rho_2\otimes \rho_3\otimes\bar \rho_4\otimes \rho_5\,.
\end{equation}

In the RT formalism one introduces 3-valent vertices corresponding to 2-to-1 and 1-to-2 functors constructed with the help of $q$-Clebsch-Gordan coefficients \eqref{CG} (see also \cite{Kirillov:1989:q6j} for similar notations):
\begin{equation}
	\begin{array}{c}
		\begin{tikzpicture}[scale=0.5, yscale=-1.5]
			\draw[thick,stealth-] (-0.5,-0.5) to[out=90,in=210] (0,0);
			\draw[thick,stealth-] (0.5,-0.5) to[out=90,in=330] (0,0);
			\draw[thick,stealth-] (0,0) -- (0,0.5);
			\node[left] at (-0.5,-0.5) {$\scriptstyle \rho_1$};
			\node[right] at (0.5,-0.5) {$\scriptstyle \rho_2$};
			\node[right] at (0,0.5) {$\scriptstyle \rho_3$};
		\end{tikzpicture}
	\end{array}\;\mapsto\;\begin{array}{c}
	(\rho_3\to \rho_1\otimes \rho_2)\\
	\\
	v_k^{(3)}\mapsto C^{k}_{i,j}\left[\begin{array}{c}
		\rho_3\\
		\rho_1 \;\rho_2
	\end{array}\right] v_i^{(1)}\otimes v_j^{(2)}
	\end{array},\quad 
	\begin{array}{c}
		\begin{tikzpicture}[scale=0.5,yscale=1.5]
			\draw[thick,-stealth] (-0.5,-0.5) to[out=90,in=210] (0,0);
			\draw[thick,-stealth] (0.5,-0.5) to[out=90,in=330] (0,0);
			\draw[thick,-stealth] (0,0) -- (0,0.5);
			\node[left] at (-0.5,-0.5) {$\scriptstyle \rho_1$};
			\node[right] at (0.5,-0.5) {$\scriptstyle \rho_2$};
			\node[right] at (0,0.5) {$\scriptstyle \rho_3$};
		\end{tikzpicture}
	\end{array}\;\mapsto\;\begin{array}{c}
	(\rho_1\otimes \rho_2\to \rho_3)\\
	\\
	v_i^{(1)}\otimes v_j^{(2)}\mapsto \bar C^{k}_{i,j}\left[\begin{array}{c}
		\rho_3\\
		\rho_1 \;\rho_2
	\end{array}\right]
	\end{array}\,.
\end{equation}

To crossings we assign R-matrices:
\begin{equation}
	\begin{array}{c}
		\begin{tikzpicture}[scale=0.7]
			\draw[thick,-stealth] (0.5,-0.5) to[out=90,in=270] (-0.5,0.5);
			\draw[white, line width = 1.5mm] (-0.5,-0.5) to[out=90,in=270] (0.5,0.5);
			\draw[thick,-stealth] (-0.5,-0.5) to[out=90,in=270] (0.5,0.5);
			\node[left] at (-0.5,-0.5) {$\scriptstyle \rho_1$};
			\node[right] at (0.5,-0.5) {$\scriptstyle \rho_2$};
		\end{tikzpicture}
	\end{array}\;\mapsto\;\left(R:\,\rho_1\otimes \rho_2\to \rho_2\otimes \rho_1\right),\quad 
	\begin{array}{c}
		\begin{tikzpicture}[scale=0.7, xscale=-1]
			\draw[thick,-stealth] (0.5,-0.5) to[out=90,in=270] (-0.5,0.5);
			\draw[white, line width = 1.5mm] (-0.5,-0.5) to[out=90,in=270] (0.5,0.5);
			\draw[thick,-stealth] (-0.5,-0.5) to[out=90,in=270] (0.5,0.5);
			\node[right] at (-0.5,-0.5) {$\scriptstyle \rho_2$};
			\node[left] at (0.5,-0.5) {$\scriptstyle \rho_1$};
		\end{tikzpicture}
	\end{array}\;\mapsto\;\left(R^{-1}:\,\rho_1\otimes \rho_2\to \rho_2\otimes \rho_1\right)\,.
\end{equation}
The fact that universal R-matrices of $U_q(\fg)$ satisfy the system of Yang-Baxter equations translates diagrammatically into the invariance of $\Psi_K(\rho_0|q)$ under Reidemeister move III.

The co-product as a homomorphism of algebras $\Delta:\,U_q(\fg)\to U_q(\fg)\otimes U_q(\fg)$ satisfies additional relations with the universal R-matrix \cite{RT1}:
\begin{equation}
	(\Delta \otimes \bbone)\circ R=R_{13}R_{23},\quad \left(\bbone\otimes \Delta\right)\circ R=R_{13}R_{12}\,.
\end{equation}
These relations lead to a diagrammatic ``fork'' relation implying that a 3-valent vertex can be pulled through a crossing strand:
\begin{equation}
	\begin{array}{c}
		\begin{tikzpicture}[scale=0.5]
			\draw[thick] (0,0) -- (0,0.5) (0,0.5) to[out=30,in=270] (0.5,1.5) (0,0.5) to[out=150,in=270] (-0.5,1.5);
			\draw[white, line width = 1mm] (-1,0.75) to[out=0,in=180] (0,1) to[out=0,in=180] (1,0.75);
			\draw[thick] (-1,0.75) to[out=0,in=180] (0,1) to[out=0,in=180] (1,0.75); 
		\end{tikzpicture}
	\end{array}=\begin{array}{c}
	\begin{tikzpicture}[scale=0.5]
		\draw[thick] (0,0) -- (0,1) (0,1) to[out=30,in=270] (0.5,1.5) (0,1) to[out=150,in=270] (-0.5,1.5);
		\draw[white, line width = 1mm] (-1,0.75) to[out=0,in=180] (0,0.5) to[out=0,in=180] (1,0.75);
		\draw[thick] (-1,0.75) to[out=0,in=180] (0,0.5) to[out=0,in=180] (1,0.75);
	\end{tikzpicture}
	\end{array},\; \begin{array}{c}
	\begin{tikzpicture}[scale=0.5,yscale=-1]
		\draw[thick] (0,0) -- (0,0.5) (0,0.5) to[out=30,in=270] (0.5,1.5) (0,0.5) to[out=150,in=270] (-0.5,1.5);
		\draw[white, line width = 1mm] (-1,0.75) to[out=0,in=180] (0,1) to[out=0,in=180] (1,0.75);
		\draw[thick] (-1,0.75) to[out=0,in=180] (0,1) to[out=0,in=180] (1,0.75); 
	\end{tikzpicture}
	\end{array}=\begin{array}{c}
	\begin{tikzpicture}[scale=0.5,yscale=-1]
		\draw[thick] (0,0) -- (0,1) (0,1) to[out=30,in=270] (0.5,1.5) (0,1) to[out=150,in=270] (-0.5,1.5);
		\draw[white, line width = 1mm] (-1,0.75) to[out=0,in=180] (0,0.5) to[out=0,in=180] (1,0.75);
		\draw[thick] (-1,0.75) to[out=0,in=180] (0,0.5) to[out=0,in=180] (1,0.75);
	\end{tikzpicture}
	\end{array},\;
	\begin{array}{c}
		\begin{tikzpicture}[scale=0.5]
			\draw[thick] (-1,0.75) to[out=0,in=180] (0,1) to[out=0,in=180] (1,0.75); 
			\draw[white, line width = 1mm] (0,0) -- (0,0.5) (0,0.5) to[out=30,in=270] (0.5,1.5) (0,0.5) to[out=150,in=270] (-0.5,1.5);
			\draw[thick] (0,0) -- (0,0.5) (0,0.5) to[out=30,in=270] (0.5,1.5) (0,0.5) to[out=150,in=270] (-0.5,1.5);
		\end{tikzpicture}
	\end{array}=\begin{array}{c}
		\begin{tikzpicture}[scale=0.5]
			\draw[thick] (-1,0.75) to[out=0,in=180] (0,0.5) to[out=0,in=180] (1,0.75);
			\draw[white, line width = 1mm] (0,0) -- (0,1) (0,1) to[out=30,in=270] (0.5,1.5) (0,1) to[out=150,in=270] (-0.5,1.5);
			\draw[thick] (0,0) -- (0,1) (0,1) to[out=30,in=270] (0.5,1.5) (0,1) to[out=150,in=270] (-0.5,1.5);
		\end{tikzpicture}
	\end{array},\; \begin{array}{c}
		\begin{tikzpicture}[scale=0.5,yscale=-1]
			\draw[thick] (-1,0.75) to[out=0,in=180] (0,1) to[out=0,in=180] (1,0.75); 
			\draw[white, line width = 1mm] (0,0) -- (0,0.5) (0,0.5) to[out=30,in=270] (0.5,1.5) (0,0.5) to[out=150,in=270] (-0.5,1.5);
			\draw[thick] (0,0) -- (0,0.5) (0,0.5) to[out=30,in=270] (0.5,1.5) (0,0.5) to[out=150,in=270] (-0.5,1.5);
		\end{tikzpicture}
	\end{array}=\begin{array}{c}
		\begin{tikzpicture}[scale=0.5,yscale=-1]
			\draw[thick] (-1,0.75) to[out=0,in=180] (0,0.5) to[out=0,in=180] (1,0.75);
			\draw[white, line width = 1mm] (0,0) -- (0,1) (0,1) to[out=30,in=270] (0.5,1.5) (0,1) to[out=150,in=270] (-0.5,1.5);
			\draw[thick] (0,0) -- (0,1) (0,1) to[out=30,in=270] (0.5,1.5) (0,1) to[out=150,in=270] (-0.5,1.5);
		\end{tikzpicture}
	\end{array}\,.
\end{equation}

Traditionally, the knot diagram does not contain 3-valent vertices.
Instead, when the time flow direction is chosen there appear critical 2-valent vertices where the orientation of the knot strand is reversed: ``cups'' and ``caps''.
To cups and caps we simply assign 3-valent vertices with the third leg colored by the trivial representation:
\begin{equation}
	\begin{array}{c}
		\begin{tikzpicture}[scale=0.7]
			\draw[thick, -stealth] (-0.5,0.5) to[out=270,in=180] (0,0) to[out=0,in=270] (0.5,0.5);
			\node[right] at (0.5,0.5) {$\scriptstyle \rho$};
		\end{tikzpicture}
	\end{array}=\begin{array}{c}
	\begin{tikzpicture}[scale=0.7]
		\draw[thick, -stealth] (0,0) -- (0,0.5); 
		\draw[thick, -stealth] (0,0.5) to[out=30,in=270] (0.5,1.2);
		\draw[thick, -stealth] (0,0.5) to[out=150,in=270] (-0.5,1.2);
		\node[left] at (0,0) {$\scriptstyle \varnothing$};
		\node[left] at (-0.5,1) {$\scriptstyle \bar\rho$};
		\node[right] at (0.5,1) {$\scriptstyle \rho$};
	\end{tikzpicture}
	\end{array},\quad \begin{array}{c}
	\begin{tikzpicture}[scale=0.7,xscale=-1]
		\draw[thick, -stealth] (-0.5,0.5) to[out=270,in=180] (0,0) to[out=0,in=270] (0.5,0.5);
		\node[left] at (0.5,0.5) {$\scriptstyle \rho$};
	\end{tikzpicture}
	\end{array}=\begin{array}{c}
	\begin{tikzpicture}[scale=0.7]
		\draw[thick, -stealth] (0,0) -- (0,0.5); 
		\draw[thick, -stealth] (0,0.5) to[out=30,in=270] (0.5,1.2);
		\draw[thick, -stealth] (0,0.5) to[out=150,in=270] (-0.5,1.2);
		\node[left] at (0,0) {$\scriptstyle \varnothing$};
		\node[left] at (-0.5,1) {$\scriptstyle \rho$};
		\node[right] at (0.5,1) {$\scriptstyle \bar\rho$};
	\end{tikzpicture}
	\end{array},\quad
	\begin{array}{c}
		\begin{tikzpicture}[scale=0.7,yscale=-1]
			\draw[thick, -stealth] (-0.5,0.5) to[out=270,in=180] (0,0) to[out=0,in=270] (0.5,0.5);
			\node[right] at (0.5,0.5) {$\scriptstyle \rho$};
		\end{tikzpicture}
	\end{array}=\begin{array}{c}
		\begin{tikzpicture}[scale=0.7,yscale=-1]
			\draw[thick, stealth-] (0,0) -- (0,0.5); 
			\draw[thick, stealth-] (0,0.5) to[out=30,in=270] (0.5,1.2);
			\draw[thick, stealth-] (0,0.5) to[out=150,in=270] (-0.5,1.2);
			\node[left] at (0,0) {$\scriptstyle \varnothing$};
			\node[left] at (-0.5,1) {$\scriptstyle \rho$};
			\node[right] at (0.5,1) {$\scriptstyle \bar\rho$};
		\end{tikzpicture}
	\end{array},\quad \begin{array}{c}
		\begin{tikzpicture}[scale=-0.7]
			\draw[thick, -stealth] (-0.5,0.5) to[out=270,in=180] (0,0) to[out=0,in=270] (0.5,0.5);
			\node[left] at (0.5,0.5) {$\scriptstyle \rho$};
		\end{tikzpicture}
	\end{array}=\begin{array}{c}
		\begin{tikzpicture}[scale=0.7, yscale=-1]
			\draw[thick, stealth-] (0,0) -- (0,0.5); 
			\draw[thick, stealth-] (0,0.5) to[out=30,in=270] (0.5,1.2);
			\draw[thick, stealth-] (0,0.5) to[out=150,in=270] (-0.5,1.2);
			\node[left] at (0,0) {$\scriptstyle \varnothing$};
			\node[left] at (-0.5,1) {$\scriptstyle \bar\rho$};
			\node[right] at (0.5,1) {$\scriptstyle \rho$};
		\end{tikzpicture}
	\end{array}\,.
\end{equation}


\section{Clebsch-Gordan chords \& A-polynomials} \label{sec:CG}

\subsection{CG chords}\label{sec:chords}

Clebsch-Gordan (CG) chords proposed in \cite{Galakhov:2024eco} play the role of operators ``measuring'' the quantum theory, or of coordinates in the classical limit.
We will motivate the present construction by addressing possible connections with similar formulation in 2d CFT, knot contact homology, and 3d TQFT in Sec.~\ref{sec:contexts}.
In this section we will generalize the setting proposed in \cite{Galakhov:2024eco} for $\fs\fu_2$ to arbitrary $\fs\fu_N$.

Let us choose a specific irreducible representation $\rho$ of $U_q(\fs\fu_N)$.
We assume that $\rho$ does not produce any non-trivial multiplicities in the tensor product with any other irrep $Q$ of $U_q(\fs\fu_N)$.
For example, for $\fs\fu_n$ we could choose any anti-symmetric power $\rho=\wedge^k\Box$, $k<n$.\footnote{For generic simple Lie algebras an analysis of dominant weights delivering multiplicity-free tensor products is presented in \cite{stembridge2003multiplicity}.}
Then we will label irreducible components in the isotypical decomposition by Greek indices:
\begin{equation}\label{multi-free}
	\rho\otimes Q=\bigoplus\lm_{\alpha} Q^{\alpha}\,.
\end{equation}
In what follows we will exploit CG coefficients for this particular isotypical decomposition, therefore we want to simplify notation of the previous section a bit.
We will mark strands colored with representations $Q$ and $Q^{\alpha}$ by thick black lines, and those colored with $\rho$ by thinner blue lines.
Also we omit label $Q$ assuming that it is clear from the context what representation is used:
\begin{equation}\label{notations}
	\begin{array}{c}
		\begin{tikzpicture}[scale=0.4]
			\draw[thick] (0,0) -- (-0.8,0.8);
			\draw[thick] (0,0) -- (0.8,0.8);
			\draw[thick] (0,0) -- (0,-1);
			\node[below] at (0,-1) {$\scriptstyle Q^\alpha$};
			\node[left] at (-0.8,0.8) {$\scriptstyle Q$};
			\node[right] at (0.8,0.8) {$\scriptstyle \rho$};
		\end{tikzpicture}
	\end{array}\to
	\begin{array}{c}
		\begin{tikzpicture}[scale=0.4]
			\draw[ultra thick] (0,0) -- (-0.8,0.8);
			\draw[thick,\myblue] (0,0) -- (0.8,0.8);
			\draw[ultra thick] (0,0) -- (0,-1);
			\node[below] at (0,-1) {$\scriptstyle \alpha$};
			\node[right] at (0.8,0.8) {$\scriptstyle \rho$};
		\end{tikzpicture}
	\end{array},\; \begin{array}{c}
	\begin{tikzpicture}[scale=0.4,xscale=-1]
		\draw[thick] (0,0) -- (-0.8,0.8);
		\draw[thick] (0,0) -- (0.8,0.8);
		\draw[thick] (0,0) -- (0,-1);
		\node[below] at (0,-1) {$\scriptstyle Q^\alpha$};
		\node[right] at (-0.8,0.8) {$\scriptstyle Q$};
		\node[left] at (0.8,0.8) {$\scriptstyle \rho$};
	\end{tikzpicture}
	\end{array}\to
	\begin{array}{c}
	\begin{tikzpicture}[scale=0.4,xscale=-1]
		\draw[ultra thick] (0,0) -- (-0.8,0.8);
		\draw[thick,\myblue] (0,0) -- (0.8,0.8);
		\draw[ultra thick] (0,0) -- (0,-1);
		\node[below] at (0,-1) {$\scriptstyle \alpha$};
		\node[left] at (0.8,0.8) {$\scriptstyle \rho$};
	\end{tikzpicture}
	\end{array},\;\begin{array}{c}
	\begin{tikzpicture}[scale=0.4,yscale=-1]
		\draw[thick] (0,0) -- (-0.8,0.8);
		\draw[thick] (0,0) -- (0.8,0.8);
		\draw[thick] (0,0) -- (0,-1);
		\node[above] at (0,-1) {$\scriptstyle Q^\alpha$};
		\node[left] at (-0.8,0.8) {$\scriptstyle Q$};
		\node[right] at (0.8,0.8) {$\scriptstyle \rho$};
	\end{tikzpicture}
	\end{array}\to
	\begin{array}{c}
	\begin{tikzpicture}[scale=0.4,yscale=-1]
		\draw[ultra thick] (0,0) -- (-0.8,0.8);
		\draw[thick,\myblue] (0,0) -- (0.8,0.8);
		\draw[ultra thick] (0,0) -- (0,-1);
		\node[above] at (0,-1) {$\scriptstyle \alpha$};
		\node[right] at (0.8,0.8) {$\scriptstyle \rho$};
	\end{tikzpicture}
	\end{array},\; \begin{array}{c}
	\begin{tikzpicture}[scale=0.4,xscale=-1,yscale=-1]
		\draw[thick] (0,0) -- (-0.8,0.8);
		\draw[thick] (0,0) -- (0.8,0.8);
		\draw[thick] (0,0) -- (0,-1);
		\node[above] at (0,-1) {$\scriptstyle Q^\alpha$};
		\node[right] at (-0.8,0.8) {$\scriptstyle Q$};
		\node[left] at (0.8,0.8) {$\scriptstyle \rho$};
	\end{tikzpicture}
	\end{array}\to
	\begin{array}{c}
	\begin{tikzpicture}[scale=0.4,xscale=-1,yscale=-1]
		\draw[ultra thick] (0,0) -- (-0.8,0.8);
		\draw[thick,\myblue] (0,0) -- (0.8,0.8);
		\draw[ultra thick] (0,0) -- (0,-1);
		\node[above] at (0,-1) {$\scriptstyle \alpha$};
		\node[left] at (0.8,0.8) {$\scriptstyle \rho$};
	\end{tikzpicture}
	\end{array}\,.
\end{equation}

Next, we would like to study how these tensor morphisms acting on products of representations interact with other morphisms, in particular with R-matrices.
The fact that the R-matrix is an intertwiner \eqref{intertw} indicates that it acts individually in different isotypical channels, with certain eigenvalues we call twist factors and denote by $\varphi$:
\begin{equation}\label{R-eigen}
	\begin{aligned}
		&\begin{array}{c}
			\begin{tikzpicture}
			\draw[ultra thick] (0.5,-0.5) -- (-0.5,0.5);
			\draw[white, line width=4] (-0.5,-0.5) -- (0.5,0.5);
			\draw[thick,\myblue] (-0.5,-0.5) -- (0.5,0.5);
			\node[above] at (-0.5,-0.5) {$\scriptstyle\rho$};
			\end{tikzpicture}
		\end{array}=\sum\lm_{\gamma}\begin{array}{c}
			\begin{tikzpicture}
			\draw[thick,\myblue] (-0.5,-0.7) -- (0,-0.3) (0,0.3) -- (0.5,0.7);
			\draw[ultra thick] (0.5,-0.7) -- (0,-0.3) (0,0.3) -- (-0.5,0.7);
			\draw[ultra thick] (0,-0.3) -- (0,0.3);
			\node[right] at (0,0) {$\scriptstyle \gamma$};
			\node[above] at (-0.5,-0.7) {$\scriptstyle\rho$};
			\node[left] at (0,0) {$\left({}^\rho\varphi_\gamma\right)\times$};
			\end{tikzpicture}
		\end{array},\quad  \begin{array}{c}
			\begin{tikzpicture}[xscale=-1]
			\draw[ultra thick] (0.5,-0.5) -- (-0.5,0.5);
			\draw[white, line width=4] (-0.5,-0.5) -- (0.5,0.5);
			\draw[thick,\myblue] (-0.5,-0.5) -- (0.5,0.5);
			\node[above] at (-0.5,-0.5) {$\scriptstyle\rho$};
			\end{tikzpicture}
		\end{array}=\sum\lm_{\gamma}\begin{array}{c}
			\begin{tikzpicture}[xscale=-1]
			\draw[thick,\myblue] (-0.5,-0.7) -- (0,-0.3) (0,0.3) -- (0.5,0.7);
			\draw[ultra thick] (0.5,-0.7) -- (0,-0.3) (0,0.3) -- (-0.5,0.7);
			\draw[ultra thick] (0,-0.3) -- (0,0.3);
			\node[right] at (0,0) {$\scriptstyle \gamma$};
			\node[above] at (-0.5,-0.7) {$\scriptstyle\rho$};
			\node[left] at (0,0) {$\left({}^\rho\varphi_\gamma\right)^{-1}\times$};
			\end{tikzpicture}
		\end{array}\,,\\
		&\begin{array}{c}
			\begin{tikzpicture}
			\draw[thick,\myblue] (0.5,-0.5) -- (-0.5,0.5);
			\draw[white, line width=4] (-0.5,-0.5) -- (0.5,0.5);
			\draw[ultra thick] (-0.5,-0.5) -- (0.5,0.5);
			\node[above] at (0.5,-0.5) {$\scriptstyle\rho$};
			\end{tikzpicture}
		\end{array}=\sum\lm_{\gamma}\begin{array}{c}
			\begin{tikzpicture}[xscale=-1]
			\draw[thick,\myblue] (-0.5,-0.7) -- (0,-0.3) (0,0.3) -- (0.5,0.7);
			\draw[ultra thick] (0.5,-0.7) -- (0,-0.3) (0,0.3) -- (-0.5,0.7);
			\draw[ultra thick] (0,-0.3) -- (0,0.3);
			\node[right] at (0,0) {$\scriptstyle \gamma$};
			\node[above] at (-0.5,-0.7) {$\scriptstyle\rho$};
			\node[left] at (0,0) {$\left({}^\rho\varphi_\gamma\right)\times$};
			\end{tikzpicture}
		\end{array},\quad  \begin{array}{c}
			\begin{tikzpicture}[xscale=-1]
			\draw[thick,\myblue] (0.5,-0.5) -- (-0.5,0.5);
			\draw[white, line width=4] (-0.5,-0.5) -- (0.5,0.5);
			\draw[ultra thick] (-0.5,-0.5) -- (0.5,0.5);
			\node[above] at (0.5,-0.5) {$\scriptstyle\rho$};
			\end{tikzpicture}
		\end{array}=\sum\lm_{\gamma}\begin{array}{c}
			\begin{tikzpicture}
			\draw[thick,\myblue] (-0.5,-0.7) -- (0,-0.3) (0,0.3) -- (0.5,0.7);
			\draw[ultra thick] (0.5,-0.7) -- (0,-0.3) (0,0.3) -- (-0.5,0.7);
			\draw[ultra thick] (0,-0.3) -- (0,0.3);
			\node[right] at (0,0) {$\scriptstyle \gamma$};
			\node[above] at (-0.5,-0.7) {$\scriptstyle\rho$};
			\node[left] at (0,0) {$\left({}^\rho\varphi_\gamma\right)^{-1}\times$};
			\end{tikzpicture}
		\end{array}\,.
	\end{aligned}
\end{equation}

In terms of the same twist factors we can rewrite ``twists'' of 3-valent vertices, for example:
\begin{equation}\label{uni_twist}
	\begin{array}{c}
		\begin{tikzpicture}[scale=0.5]
		\draw[ultra thick] (0,0) to[out=30,in=270] (0.7,0.6) to[out=90,in=315] (-0.8,1.7);
		\draw[white, line width = 4] (-0.7,0.6) to[out=90,in=225] (0.8,1.7);
		\draw[thick,\myblue] (0,0) to[out=150,in=270] (-0.7,0.6) to[out=90,in=225] (0.8,1.7);
		\draw[ultra thick] (0,0) -- (0,-0.7);
		\node[below] at (0,-0.7) {$\scriptstyle \gamma$};
		\node[right] at (0.8,1.7) {$\scriptstyle \rho$};
		\end{tikzpicture}
	\end{array}={}^{\rho}\varphi_{\gamma}\hspace{-4mm}\begin{array}{c}
		\begin{tikzpicture}[scale=0.5]
		\draw[ultra thick] (0,0) -- (-0.8,0.8);
		\draw[thick,\myblue] (0,0) -- (0.8,0.8);
		\draw[ultra thick] (0,0) -- (0,-1);
		\node[below] at (0,-1) {$\scriptstyle \gamma$};
		\node[right] at (0.8,0.8) {$\scriptstyle \rho$};
		\end{tikzpicture}
	\end{array},\;
	\begin{array}{c}
		\begin{tikzpicture}[scale=0.5,xscale=-1]
		\draw[ultra thick] (0,0) to[out=30,in=270] (0.7,0.6) to[out=90,in=315] (-0.8,1.7);
		\draw[white, line width = 4] (-0.7,0.6) to[out=90,in=225] (0.8,1.7);
		\draw[thick,\myblue] (0,0) to[out=150,in=270] (-0.7,0.6) to[out=90,in=225] (0.8,1.7);
		\draw[ultra thick] (0,0) -- (0,-0.7);
		\node[below] at (0,-0.7) {$\scriptstyle \gamma$};
		\node[left] at (0.8,1.7) {$\scriptstyle \rho$};
		\end{tikzpicture}
	\end{array}=\left({}^{\rho}\varphi_{\gamma}\right)^{-1}\hspace{-7mm}\begin{array}{c}
		\begin{tikzpicture}[scale=0.5,xscale=-1]
		\draw[ultra thick] (0,0) -- (-0.8,0.8);
		\draw[thick,\myblue] (0,0) -- (0.8,0.8);
		\draw[ultra thick] (0,0) -- (0,-1);
		\node[below] at (0,-1) {$\scriptstyle \gamma$};
		\node[left] at (0.8,0.8) {$\scriptstyle \rho$};
		\end{tikzpicture}
	\end{array},\;
	\begin{array}{c}
		\begin{tikzpicture}[scale=0.5]
		\draw[thick,\myblue] (0,0) to[out=150,in=270] (-0.7,0.6) to[out=90,in=225] (0.8,1.7);
		\draw[white, line width = 4] (0,0) to[out=30,in=270] (0.7,0.6) to[out=90,in=315] (-0.8,1.7);
		\draw[ultra thick] (0,0) to[out=30,in=270] (0.7,0.6) to[out=90,in=315] (-0.8,1.7);
		\draw[ultra thick] (0,0) -- (0,-0.7);
		\node[below] at (0,-0.7) {$\scriptstyle \gamma$};
		\node[right] at (0.8,1.7) {$\scriptstyle \rho$};
		\end{tikzpicture}
	\end{array}=\left({}^{\rho}\varphi_{\gamma}\right)^{-1}\hspace{-7mm}\begin{array}{c}
		\begin{tikzpicture}[scale=0.5]
		\draw[ultra thick] (0,0) -- (0.8,0.8);
		\draw[thick,\myblue] (0,0) -- (-0.8,0.8);
		\draw[ultra thick] (0,0) -- (0,-1);
		\node[below] at (0,-1) {$\scriptstyle \gamma$};
		\node[left] at (-0.8,0.8) {$\scriptstyle \rho$};
		\end{tikzpicture}
	\end{array},\;
	\begin{array}{c}
		\begin{tikzpicture}[scale=0.5, xscale=-1]
		\draw[thick,\myblue] (0,0) to[out=150,in=270] (-0.7,0.6) to[out=90,in=225] (0.8,1.7);
		\draw[white, line width = 4] (0,0) to[out=30,in=270] (0.7,0.6) to[out=90,in=315] (-0.8,1.7);
		\draw[ultra thick] (0,0) to[out=30,in=270] (0.7,0.6) to[out=90,in=315] (-0.8,1.7);
		\draw[ultra thick] (0,0) -- (0,-0.7);
		\node[below] at (0,-0.7) {$\scriptstyle \gamma$};
		\node[left] at (0.8,1.7) {$\scriptstyle \rho$};
		\end{tikzpicture}
	\end{array}={}^{\rho}\varphi_{\gamma}\hspace{-4mm}\begin{array}{c}
		\begin{tikzpicture}[scale=0.5, xscale=-1]
		\draw[ultra thick] (0,0) -- (0.8,0.8);
		\draw[thick,\myblue] (0,0) -- (-0.8,0.8);
		\draw[ultra thick] (0,0) -- (0,-1);
		\node[below] at (0,-1) {$\scriptstyle \gamma$};
		\node[right] at (-0.8,0.8) {$\scriptstyle \rho$};
		\end{tikzpicture}
	\end{array}\,.
\end{equation}

As we can see, CG chords have many indices.
As the rank ${\rm rk}\,\fg$, the number of strands in a braid, the dimension of $\rho$ grow, the number of CG chords also grows quite rapidly.
In what follows we would like to consider relations among CG chords, and as the number of these variables grows, the relations in question become cumbersome quite rapidly.
Therefore it is natural to reduce the number of relevant CG chord variables from the very beginning.
One way to do so is to exclude one isotypical index from consideration.
Let us denote this index as 0.
We claim that if we start by considering CG chords with $\alpha,\beta\neq 0$, then new chords with $\alpha=0$ or $\beta=0$ will not appear in new relations, nor in the braid group action we will construct shortly.

Having the chosen ``uninvited'' isotypical index, which we denote by 0, we can easily derive the following universal skein relations from \eqref{R-eigen}:
\begin{equation}\label{uni_skein}
	\begin{aligned}
		&\begin{array}{c}
			\begin{tikzpicture}
			\draw[ultra thick] (0.5,-0.5) -- (-0.5,0.5);
			\draw[white, line width=4] (-0.5,-0.5) -- (0.5,0.5);
			\draw[thick,\myblue] (-0.5,-0.5) -- (0.5,0.5);
			\node[left] at (-0.5,-0.5) {$\scriptstyle\rho$};
			\node[left] at (-0.2,0) {$\left({}^\rho\varphi_0\right)^{-1}$};
			\end{tikzpicture}
		\end{array}-\begin{array}{c}
			\begin{tikzpicture}
			\draw[thick,\myblue] (-0.5,-0.5) -- (0.5,0.5);
			\draw[white, line width=4] (0.5,-0.5) -- (-0.5,0.5);
			\draw[ultra thick] (0.5,-0.5) -- (-0.5,0.5);
			\node[left] at (-0.5,-0.5) {$\scriptstyle\rho$};
			\node[left] at (-0.2,0) {$\left({}^\rho\varphi_0\right)$};
			\end{tikzpicture}
		\end{array}=\sum\lm_{\alpha\neq 0}\begin{array}{c}
			\begin{tikzpicture}
			\draw[thick,\myblue] (-0.5,-0.7) -- (0,-0.3) (0,0.3) -- (0.5,0.7);
			\draw[ultra thick] (0.5,-0.7) -- (0,-0.3) (0,0.3) -- (-0.5,0.7);
			\draw[ultra thick] (0,-0.3) -- (0,0.3);
			\node[right] at (0,0) {$\scriptstyle \alpha$};
			\node[left] at (-0.5,-0.7) {$\scriptstyle\rho$};
			\node[left] at (0,0) {$\left[\left({}^\rho\varphi_0\right)^{-1}\left({}^\rho\varphi_\alpha\right)-\left({}^\rho\varphi_0\right)\left({}^\rho\varphi_\alpha\right)^{-1}\right]$};
			\end{tikzpicture}
		\end{array}\,,\\
		&\begin{array}{c}
			\begin{tikzpicture}[xscale=-1]
			\draw[thick,\myblue] (-0.5,-0.5) -- (0.5,0.5);
			\draw[white, line width=4] (0.5,-0.5) -- (-0.5,0.5);
			\draw[ultra thick] (0.5,-0.5) -- (-0.5,0.5);
			\node[right] at (-0.5,-0.5) {$\scriptstyle\rho$};
			\node[left] at (0.2,0) {$\left({}^\rho\varphi_0\right)^{-1}$};
			\end{tikzpicture}
		\end{array}-\begin{array}{c}
			\begin{tikzpicture}[xscale=-1]
			\draw[ultra thick] (0.5,-0.5) -- (-0.5,0.5);
			\draw[white, line width=4] (-0.5,-0.5) -- (0.5,0.5);
			\draw[thick,\myblue] (-0.5,-0.5) -- (0.5,0.5);
			\node[right] at (-0.5,-0.5) {$\scriptstyle\rho$};
			\node[left] at (0.2,0) {$\left({}^\rho\varphi_0\right)$};
			\end{tikzpicture}
		\end{array}=\sum\lm_{\alpha\neq 0}\begin{array}{c}
			\begin{tikzpicture}[xscale=-1]
			\draw[thick,\myblue] (-0.5,-0.7) -- (0,-0.3) (0,0.3) -- (0.5,0.7);
			\draw[ultra thick] (0.5,-0.7) -- (0,-0.3) (0,0.3) -- (-0.5,0.7);
			\draw[ultra thick] (0,-0.3) -- (0,0.3);
			\node[right] at (0,0) {$\scriptstyle \alpha$};
			\node[right] at (-0.5,-0.7) {$\scriptstyle\rho$};
			\node[left] at (0,0) {$\left[\left({}^\rho\varphi_0\right)^{-1}\left({}^\rho\varphi_\alpha\right)-\left({}^\rho\varphi_0\right)\left({}^\rho\varphi_\alpha\right)^{-1}\right]$};
			\end{tikzpicture}
		\end{array}\,.
	\end{aligned}
\end{equation}

In these terms we introduce CG chords as tensors acting as ${\rm Rep}\left(U_q(\fs\fu_N)\right)^M \to {\rm Rep}\left(U_q(\fs\fu_N)\right)^M$ by choosing \emph{basic}\footnote{
By the term ``basic'' here we imply only that any CG chord can be expanded over basic ones not necessarily uniquely.
However, this set of CG chords is not minimal; there are mutual relations between these chords within this set.
In Sec.~\ref{sec:WZW} we will discuss the relation between these chords and parameters of conformal blocks in WZW models.
This number of parameters scales as $M$ for a sufficiently large number $M$ of strands, whereas the number of basic CG chords apparently scales as $\sim M^2$.
So mutual relations among them are inevitable.
} ones in the following diagrammatic way:
\begin{equation}\label{CGdef}
	{}^{\rho}\Xi\scalebox{0.8}{$\left[\begin{array}{cc}
			\alpha & \beta\\
			k & m
		\end{array}\right]$}:=\begin{array}{c}
		\begin{tikzpicture}[scale=0.7]
			\draw[ultra thick] (0,-1) -- (0,1) (2,-1) -- (2,1) (3,-1) -- (3,1) (5,-1) -- (5,1);
			\draw[white, line width = 4] (1,-0.5) -- (4,0.5);
			\draw[thick,\myblue] (1,-0.5) -- (4,0.5);
			\draw[ultra thick] (1,-1) -- (1,1) (4,-1) -- (4,1);
			\node[below] at (2.5,0) {$\scriptstyle \rho$};
			\node[left] at (1,-0.9) {$\scriptstyle \alpha$};
			\node[right] at (4,0.9) {$\scriptstyle \beta$};
			\node[below] at (1,-1) {$\scriptstyle k$};
			\node[below] at (4,-1) {$\scriptstyle m$};
		\end{tikzpicture}
	\end{array},\;\mbox{if }k<m;\quad {}^{\rho}\Xi\scalebox{0.8}{$\left[\begin{array}{cc}
			\alpha & \beta\\
			k & m
		\end{array}\right]$}:=\begin{array}{c}
		\begin{tikzpicture}[scale=0.7,xscale=-1]
			\draw[ultra thick] (0,-1) -- (0,1) (2,-1) -- (2,1) (3,-1) -- (3,1) (5,-1) -- (5,1);
			\draw[white, line width = 4] (1,-0.5) -- (4,0.5);
			\draw[thick,\myblue] (1,-0.5) -- (4,0.5);
			\draw[ultra thick] (1,-1) -- (1,1) (4,-1) -- (4,1);
			\node[below] at (2.5,0) {$\scriptstyle \rho$};
			\node[right] at (1,-0.9) {$\scriptstyle \alpha$};
			\node[left] at (4,0.9) {$\scriptstyle \beta$};
			\node[below] at (1,-1) {$\scriptstyle k$};
			\node[below] at (4,-1) {$\scriptstyle m$};
		\end{tikzpicture}
	\end{array},\;\mbox{if }k>m\,,
\end{equation}
where Latin indices $k,m=1,\ldots, M$ enumerate strands.
In this formulation CG chords are close to the geometric construction of Drinfeld associators \cite{Dunin-Barkowski:2012cuv}.
In fact, the entries of tensor $\Xi$ are actual functions of the representations and vectors running in all the strands of a braid, however we omit these dependencies to obtain more compact notations.
Surely, one can consider more exotic CG chords by inserting arbitrary braids of the blue $\rho$-line with other strands, yet all these chords can be decomposed over the \emph{basic} ones with the help of the skein relations \eqref{uni_skein}.
For example, let us consider a blue line passing below the black strand rather than above:
\begin{equation}
	\begin{aligned}
		& \begin{array}{c}
			\begin{tikzpicture}[scale=0.7]
				\draw[thick, \myblue] (0,-0.5) -- (2,0.5);
				\draw[white, line width=2mm] (1,-1) -- (1,1);
				\draw[ultra thick] (0,-1) -- (0,1) (1,-1) -- (1,1) (2,-1) -- (2,1);
				\node[below] at (0,-1) {$\scriptstyle 1$};
				\node[below] at (1,-1) {$\scriptstyle 2$};
				\node[below] at (2,-1) {$\scriptstyle 3$};
				\node[left] at (0,-0.75) {$\scriptstyle \alpha$};
				\node[right] at (2,0.75) {$\scriptstyle \beta$};
			\end{tikzpicture}
		\end{array}=\left({}^\rho\varphi_0^{(2)}\right)^{-2}\begin{array}{c}
		\begin{tikzpicture}[scale=0.7]
			\draw[ultra thick] (1,-1) -- (1,1);
			\draw[white, line width=2mm] (0,-0.5) -- (2,0.5);
			\draw[thick, \myblue] (0,-0.5) -- (2,0.5);
			\draw[ultra thick] (0,-1) -- (0,1) (2,-1) -- (2,1);
			\node[below] at (0,-1) {$\scriptstyle 1$};
			\node[below] at (1,-1) {$\scriptstyle 2$};
			\node[below] at (2,-1) {$\scriptstyle 3$};
			\node[left] at (0,-0.75) {$\scriptstyle \alpha$};
			\node[right] at (2,0.75) {$\scriptstyle \beta$};
		\end{tikzpicture}
		\end{array}+\sum\lm_{\gamma\neq 0}\left[\left({}^\rho\varphi_\gamma^{(2)}\right)^{-1}-\left({}^\rho\varphi_0^{(2)}\right)^{-2}\left({}^\rho\varphi_\gamma^{(2)}\right)\right]\begin{array}{c}
		\begin{tikzpicture}[scale=0.7]
			\draw[thick, \myblue] (0,-0.5) -- (1,-0.3) (1,0.3) -- (2,0.5);
			\draw[ultra thick] (0,-1) -- (0,1) (1,-1) -- (1,1) (2,-1) -- (2,1);
			\node[below] at (0,-1) {$\scriptstyle 1$};
			\node[below] at (1,-1) {$\scriptstyle 2$};
			\node[below] at (2,-1) {$\scriptstyle 3$};
			\node[left] at (0,-0.75) {$\scriptstyle \alpha$};
			\node[right] at (2,0.75) {$\scriptstyle \beta$};
			\node[right] at (1,0) {$\scriptstyle \gamma$};
		\end{tikzpicture}
		\end{array}=\\
		&=\left({}^\rho\varphi_0^{(2)}\right)^{-2}{}^{\rho}\Xi\scalebox{0.8}{$\left[\begin{array}{cc}
				\alpha & \beta\\
				1 & 3
			\end{array}\right]$}+\sum\lm_{\gamma\neq 0}\left[\left({}^\rho\varphi_\gamma^{(2)}\right)^{-1}-\left({}^\rho\varphi_0^{(2)}\right)^{-2}\left({}^\rho\varphi_\gamma^{(2)}\right)\right] {}^{\rho}\Xi\scalebox{0.8}{$\left[\begin{array}{cc}
			\gamma & \beta\\
			2 & 3
			\end{array}\right]$}{}^{\rho}\Xi\scalebox{0.8}{$\left[\begin{array}{cc}
			\alpha & \gamma\\
			1 & 2
			\end{array}\right]$} \,,
	\end{aligned}
\end{equation}
where superscript $(k)$ on the twist factor ${}^\rho\varphi_\alpha^{(k)}$ indicates that it corresponds to the $k^{\rm th}$ strand.

\subsection{Braid group action}\label{sec:braid_group}

Now we derive a braid group action on CG chords simply as an \emph{adjoint} action of the R-matrices.
This formulation is dictated by a physical analogy between $U_q(\fg)$ tensor powers and conformal blocks in 2d WZW models we discuss further in Sec.~\ref{sec:WZW}.
Our CG chords may be considered as operators acting on conformal blocks.
On the other hand, conformal blocks are wave functions, elements of the Hilbert space, in a topological theory one dimension higher -- a 3d Chern-Simons theory -- on a 2d slice represented by a horizontal line in the diagrams.
This theory has vanishing Hamiltonian, yet its Hilbert space is quite non-trivial, and braiding strands representing insertions of vertex operators in the WZW model induces a non-trivial evolution operator -- the R-matrix.
The evolution of an operator in the Heisenberg picture is described by the adjoint action of the evolution operator.
Pictorially, the adjoint action by the R-matrix is represented by adding opposite R-matrix twists to both ends of a braid.

Let us first demonstrate this action on a chord connecting strands $k$ and $m=i+1$ for $k<i$:
\begin{equation}
	R_{i,i+1}\left({}^{\rho}\Xi\scalebox{0.8}{$\left[\begin{array}{cc}
			\alpha & \beta\\
			k & m
		\end{array}\right]$}\right):=R_{i,i+1}\circ{}^{\rho}\Xi\scalebox{0.8}{$\left[\begin{array}{cc}
		\alpha & \beta\\
		k & m
		\end{array}\right]$}\circ R_{i,i+1}^{-1}=\begin{array}{c}
		\begin{tikzpicture}[yscale=0.7]
			\draw[ultra thick] (1,-0.5) -- (1,0.5);
			\draw[white, line width=4] (0,-0.5) -- (2,0.5);
			\draw[thick,\myblue] (0,-0.5) -- (2,0.5);
			\draw[ultra thick] (0,-1.5) -- (0,1.5) (1,1.5) to[out=270,in=90] (2,0.5) -- (2,-0.5) to[out=270,in=90] (1,-1.5);
			\draw[white, line width=4]  (1,0.5) to[out=90,in=270] (2,1.5) (2,-1.5) to[out=90,in=270] (1,-0.5);
			\draw[ultra thick]  (1,0.5) to[out=90,in=270] (2,1.5) (2,-1.5) to[out=90,in=270] (1,-0.5);
			\node[left] at (0,-1.5) {$\scriptstyle k$};
			\node[left] at (1,-1.5) {$\scriptstyle i$};
			\node[right] at (2,-1.5) {$\scriptstyle i+1$};
			\node[left] at (0,-0.7) {$\scriptstyle \alpha$};
			\node[right] at (2,0.7) {$\scriptstyle \beta$};
		\end{tikzpicture}
	\end{array}=\begin{array}{c}
	\begin{tikzpicture}
		\draw[ultra thick] (1,-0.8) -- (1,0.8);
		\draw[white, line width=4] (0.75,0.25) to[out=270,in=180] (1,0) -- (1.5,0);
		\draw[thick,\myblue] (1,0.5) to[out=180,in=90] (0.75,0.25) to[out=270,in=180] (1,0) -- (2,0) to[out=0,in=90] (2.25,-0.25);
		\draw[white, line width=4] (2,0.8) -- (2,-0.8);
		\draw[ultra thick] (2,0.8) -- (2,-0.8);
		\draw[white, line width=4]  (2.25,-0.25) to[out=270,in=0] (2,-0.5) -- (0,-0.5);
		\draw[thick,\myblue] (2.25,-0.25) to[out=270,in=0] (2,-0.5) -- (0,-0.5);
		\draw[ultra thick] (0,0.8) -- (0,-0.8);
		\node[right] at (0,-0.8) {$\scriptstyle k$};
		\node[left] at (1,-0.8) {$\scriptstyle i$};
		\node[right] at (2,-0.8) {$\scriptstyle i+1$};
		\node[left] at (0,-0.7) {$\scriptstyle \alpha$};
		\node[right] at (1,0.7) {$\scriptstyle \beta$};
	\end{tikzpicture}
	\end{array}\,.
\end{equation}
By applying skein relations \eqref{uni_skein}, we obtain:
\begin{equation}
	\scalebox{0.85}{$
	R_{i,i+1}\left({}^{\rho}\Xi\scalebox{0.8}{$\left[\begin{array}{cc}
			\alpha & \beta\\
			k & m
		\end{array}\right]$}\right)=\left({}^\rho\varphi_0^{(i+1)}\right)^2\begin{array}{c}
		\begin{tikzpicture}
			\draw[thick,\myblue] (0,-0.5) -- (1,0.5);
			\draw[ultra thick] (0,-0.8) -- (0,0.8) (1,-0.8) -- (1,0.8) (2,-0.8) -- (2,0.8);
			\node[right] at (0,-0.8) {$\scriptstyle k$};
			\node[left] at (1,-0.8) {$\scriptstyle i$};
			\node[right] at (2,-0.8) {$\scriptstyle i+1$};
			\node[left] at (0,-0.7) {$\scriptstyle \alpha$};
			\node[right] at (1,0.7) {$\scriptstyle \beta$};
		\end{tikzpicture}
	\end{array}+\sum\lm_{\gamma\neq 0} \left[\left({}^\rho\varphi_\alpha^{(i+1)}\right)-\left({}^\rho\varphi_0^{(i+1)}\right)^2\left({}^\rho\varphi_\alpha^{(i+1)}\right)^{-1}\right]\begin{array}{c}
	\begin{tikzpicture}
		\draw[ultra thick] (1,-1) -- (1,1);
		\draw[white, line width=4] (0.75,0.45) to[out=270,in=180] (1,0.2) -- (1.5,0.2);
		\draw[thick,\myblue] (1,0.7) to[out=180,in=90] (0.75,0.45) to[out=270,in=180] (1,0.2) -- (2,0.2) to[out=0,in=90] (2.25,-0.05) to[out=270,in=0] (2,-0.3);
		\draw[white, line width=4] (2,1) -- (2,0);
		\draw[ultra thick] (2,1) -- (2,0);
		\draw[white, line width=4]  (2,-0.7) -- (0,-0.7);
		\draw[thick,\myblue] (2,-0.7) -- (0,-0.7);
		\draw[ultra thick] (0,1) -- (0,-1) (2,0) -- (2,-1);
		\node[right] at (0,-1) {$\scriptstyle k$};
		\node[left] at (1,-1) {$\scriptstyle i$};
		\node[right] at (2,-1) {$\scriptstyle i+1$};
		\node[right] at (2,-0.5) {$\scriptstyle \gamma$};
		\node[left] at (0,-0.9) {$\scriptstyle \alpha$};
		\node[right] at (1,0.9) {$\scriptstyle \beta$};
	\end{tikzpicture}
	\end{array}\,.$}
\end{equation}
In this expression the latter term can be reduced to a product of two CG chords after applying the respective twists \eqref{uni_twist}.

We can simplify these relations slightly by instead rescaling CG chord coordinates:
\begin{equation}\label{Rmorph}
	{}^{\rho}\Theta\scalebox{0.8}{$\left[\begin{array}{cc}
			\alpha & \beta\\
			k & m
		\end{array}\right]$}:=\frac{\left({}^{\rho}\varphi_{\alpha}^{(k)}\right)^{-2}-\left({}^{\rho}\varphi_{0}^{(k)}\right)^{-2}}{\left({}^{\rho}\varphi_{\alpha}^{(k)}\right)^{-1}\left({}^{\rho}\varphi_{0}^{(k)}\right)^{-1}}{}^{\rho}\Xi\scalebox{0.8}{$\left[\begin{array}{cc}
			\alpha & \beta\\
			k & m
		\end{array}\right]$}\,.
\end{equation}

In terms of these coordinates the action of $R_{i,i+1}$ on basic CG chords reads:
\begin{subequations}
\begin{equation}\label{Rmorph_beg}
		{}^{\rho}\varphi_{\alpha}^{(k)}\mapsto {}^{\rho}\varphi_{\alpha}^{(\sigma_{i,i+1}(k))}\,,
\end{equation}
\begin{equation}
		{}^{\rho}\Theta\scalebox{0.8}{$\left[\begin{array}{cc}
				\alpha & \beta\\
				k & m
			\end{array}\right]$}\mapsto {}^{\rho}\Theta\scalebox{0.8}{$\left[\begin{array}{cc}
				\alpha & \beta\\
				k & m
			\end{array}\right]$},\;\mbox{if }k,m\neq i, i+1\,,
\end{equation}
\begin{equation}
		{}^{\rho}\Theta\scalebox{0.8}{$\left[\begin{array}{cc}
				\alpha & \beta\\
				i & i+1
			\end{array}\right]$}\mapsto {}^{\rho}\Theta\scalebox{0.8}{$\left[\begin{array}{cc}
				\alpha & \beta\\
				i+1 & i
			\end{array}\right]$}\left({}^{\rho}\varphi_{\alpha}^{(i+1)}\right)\left({}^{\rho}\varphi_{\beta}^{(i)}\right)^{-1}\,,
\end{equation}
\begin{equation}
		{}^{\rho}\Theta\scalebox{0.8}{$\left[\begin{array}{cc}
				\alpha & \beta\\
				i+1 & i
			\end{array}\right]$}\mapsto {}^{\rho}\Theta\scalebox{0.8}{$\left[\begin{array}{cc}
				\alpha & \beta\\
				i & i+1
			\end{array}\right]$}\left({}^{\rho}\varphi_{\alpha}^{(i)}\right)\left({}^{\rho}\varphi_{\beta}^{(i+1)}\right)^{-1}\,,
\end{equation}
\begin{equation}
		{}^{\rho}\Theta\scalebox{0.8}{$\left[\begin{array}{cc}
				\alpha & \beta\\
				k & i
			\end{array}\right]$}\mapsto {}^{\rho}\Theta\scalebox{0.8}{$\left[\begin{array}{cc}
				\alpha & \beta\\
				k & i+1
			\end{array}\right]$},\quad k\neq i,i+1\,,
\end{equation}
\begin{equation}
	{}^{\rho}\Theta\scalebox{0.8}{$\left[\begin{array}{cc}
				\alpha & \beta\\
				i & k
			\end{array}\right]$}\mapsto {}^{\rho}\Theta\scalebox{0.8}{$\left[\begin{array}{cc}
				\alpha & \beta\\
				i+1 & k
			\end{array}\right]$},\quad k\neq i,i+1\,,
\end{equation}
\begin{equation}
	\begin{aligned}
		&{}^{\rho}\Theta\scalebox{0.8}{$\left[\begin{array}{cc}
				\alpha & \beta\\
				k & i+1
			\end{array}\right]$}\mapsto\left({}^{\rho}\varphi_{0}^{(i+1)}\right)^2{}^{\rho}\Theta\scalebox{0.8}{$\left[\begin{array}{cc}
				\alpha & \beta\\
				k & i
			\end{array}\right]$}-\sum\lm_{\gamma\neq 0}\left({}^{\rho}\varphi_{\gamma}^{(i+1)}\right)\left({}^{\rho}\varphi_{0}^{(i+1)}\right)\times\\
		&\hspace{5cm}\times {}^{\rho}\Theta\scalebox{0.8}{$\left[\begin{array}{cc}
				\gamma & \beta\\
				i+1 & i
			\end{array}\right]$}\; {}^{\rho}\Theta\scalebox{0.8}{$\left[\begin{array}{cc}
			\alpha & \gamma\\
			k & i+1
			\end{array}\right]$}\times\left\{\begin{array}{ll}
			\left({}^{\rho}\varphi_{\beta}^{(i)}\right)^{-1},&\mbox{if }k<i;\\
			\left({}^{\rho}\varphi_{\gamma}^{(i+1)}\right)^{-1},&\mbox{if }k>i+1;
		\end{array}\right.
	\end{aligned}
\end{equation}
\begin{equation}\label{Rmorph_end}
	\begin{aligned}
		&{}^{\rho}\Theta\scalebox{0.8}{$\left[\begin{array}{cc}
				\alpha & \beta\\
				i+1 & k
			\end{array}\right]$}\mapsto\left({}^{\rho}\varphi_{0}^{(i+1)}\right)^{-2}{}^{\rho}\Theta\scalebox{0.8}{$\left[\begin{array}{cc}
				\alpha & \beta\\
				i & k
			\end{array}\right]$}+\sum\lm_{\gamma\neq 0}\left({}^{\rho}\varphi_{\gamma}^{(i+1)}\right)^{-1}\left({}^{\rho}\varphi_{0}^{(i+1)}\right)^{-1}\times\\
		&\hspace{5cm}\times {}^{\rho}\Theta\scalebox{0.8}{$\left[\begin{array}{cc}
				\gamma & \beta\\
				i+1 & k
			\end{array}\right]$}\; {}^{\rho}\Theta\scalebox{0.8}{$\left[\begin{array}{cc}
				\alpha & \gamma\\
				i & i+1
			\end{array}\right]$}\times\left\{\begin{array}{ll}
			\left({}^{\rho}\varphi_{\alpha}^{(i)}\right),&\mbox{if }k<i;\\
			\left({}^{\rho}\varphi_{\gamma}^{(i+1)}\right),&\mbox{if }k>i+1;
		\end{array}\right.
	\end{aligned}
\end{equation}
\end{subequations}
where $\sigma_{i,i+1}$ is a simple permutation.
And similarly, for $R_{i,i+1}^{-1}$ we obtain the following expressions:
\begin{subequations}
\begin{equation}\label{RRmorph_beg}
	{}^{\rho}\varphi_{\alpha}^{(k)}\mapsto {}^{\rho}\varphi_{\alpha}^{(\sigma_{i,i+1}(k))}\,,
\end{equation}
\begin{equation}
	{}^{\rho}\Theta\scalebox{0.8}{$\left[\begin{array}{cc}
				\alpha & \beta\\
				k & m
			\end{array}\right]$}\mapsto {}^{\rho}\Theta\scalebox{0.8}{$\left[\begin{array}{cc}
				\alpha & \beta\\
				k & m
			\end{array}\right]$},\;\mbox{if }k,m\neq i, i+1\,,
\end{equation}
\begin{equation}
		{}^{\rho}\Theta\scalebox{0.8}{$\left[\begin{array}{cc}
				\alpha & \beta\\
				i & i+1
			\end{array}\right]$}\mapsto {}^{\rho}\Theta\scalebox{0.8}{$\left[\begin{array}{cc}
				\alpha & \beta\\
				i+1 & i
			\end{array}\right]$}\left({}^{\rho}\varphi_{\alpha}^{(i+1)}\right)^{-1}\left({}^{\rho}\varphi_{\beta}^{(i)}\right)\,,
\end{equation}
\begin{equation}
		{}^{\rho}\Theta\scalebox{0.8}{$\left[\begin{array}{cc}
				\alpha & \beta\\
				i+1 & i
			\end{array}\right]$}\mapsto {}^{\rho}\Theta\scalebox{0.8}{$\left[\begin{array}{cc}
				\alpha & \beta\\
				i & i+1
			\end{array}\right]$}\left({}^{\rho}\varphi_{\alpha}^{(i)}\right)^{-1}\left({}^{\rho}\varphi_{\beta}^{(i+1)}\right)\,,
\end{equation}
\begin{equation}
		{}^{\rho}\Theta\scalebox{0.8}{$\left[\begin{array}{cc}
				\alpha & \beta\\
				k & i+1
			\end{array}\right]$}\mapsto {}^{\rho}\Theta\scalebox{0.8}{$\left[\begin{array}{cc}
				\alpha & \beta\\
				k & i
			\end{array}\right]$},\quad k\neq i,i+1\,,
\end{equation}
\begin{equation}
		{}^{\rho}\Theta\scalebox{0.8}{$\left[\begin{array}{cc}
				\alpha & \beta\\
				i+1 & k
			\end{array}\right]$}\mapsto {}^{\rho}\Theta\scalebox{0.8}{$\left[\begin{array}{cc}
				\alpha & \beta\\
				i & k
			\end{array}\right]$},\quad k\neq i,i+1\,,
\end{equation}
\begin{equation}
	\begin{aligned}
		&{}^{\rho}\Theta\scalebox{0.8}{$\left[\begin{array}{cc}
				\alpha & \beta\\
				k & i
			\end{array}\right]$}\mapsto\left({}^{\rho}\varphi_{0}^{(i)}\right)^{-2}{}^{\rho}\Theta\scalebox{0.8}{$\left[\begin{array}{cc}
				\alpha & \beta\\
				k & i+1
			\end{array}\right]$}+\sum\lm_{\gamma\neq 0}\left({}^{\rho}\varphi_{\gamma}^{(i)}\right)^{-1}\left({}^{\rho}\varphi_{0}^{(i)}\right)^{-1}\times\\
		&\hspace{5cm}\times {}^{\rho}\Theta\scalebox{0.8}{$\left[\begin{array}{cc}
				\gamma & \beta\\
				i & i+1
			\end{array}\right]$}\; {}^{\rho}\Theta\scalebox{0.8}{$\left[\begin{array}{cc}
				\alpha & \gamma\\
				k & i
			\end{array}\right]$}\times\left\{\begin{array}{ll}
			\left({}^{\rho}\varphi_{\beta}^{(i+1)}\right),&\mbox{if }k>i+1;\\
			\left({}^{\rho}\varphi_{\gamma}^{(i)}\right),&\mbox{if }k<i;
		\end{array}\right.\,,
	\end{aligned}
\end{equation}
\begin{equation}\label{RRmorph_end}
	\begin{aligned}
		&{}^{\rho}\Theta\scalebox{0.8}{$\left[\begin{array}{cc}
				\alpha & \beta\\
				i & k
			\end{array}\right]$}\mapsto\left({}^{\rho}\varphi_{0}^{(i)}\right)^{2}{}^{\rho}\Theta\scalebox{0.8}{$\left[\begin{array}{cc}
				\alpha & \beta\\
				i+1 & k
			\end{array}\right]$}-\sum\lm_{\gamma\neq 0}\left({}^{\rho}\varphi_{\gamma}^{(i)}\right)\left({}^{\rho}\varphi_{0}^{(i)}\right)\times\\
		&\hspace{5cm}\times {}^{\rho}\Theta\scalebox{0.8}{$\left[\begin{array}{cc}
				\gamma & \beta\\
				i & k
			\end{array}\right]$}\; {}^{\rho}\Theta\scalebox{0.8}{$\left[\begin{array}{cc}
				\alpha & \gamma\\
				i+1 & i
			\end{array}\right]$}\times\left\{\begin{array}{ll}
			\left({}^{\rho}\varphi_{\alpha}^{(i+1)}\right)^{-1},&\mbox{if }k>i+1;\\
			\left({}^{\rho}\varphi_{\gamma}^{(i)}\right)^{-1},&\mbox{if }k<i.
		\end{array}\right.
	\end{aligned}
\end{equation}
\end{subequations}

These morphisms of CG chords represent the braid group; in other words, the following relations hold:
\begin{tcolorbox}
\begin{equation}\label{R_identities}
	R_{i,i+1}\circ R_{i+1,i+2}\circ R_{i,i+1}=R_{i+1,i+2}\circ R_{i,i+1}\circ R_{i+1,i+2},\quad R_{i,i+1}^{-1}\circ R_{i,i+1}=\bbone\,.
\end{equation}
\end{tcolorbox}


\subsection{Constructing shaded A-polynomials}\label{sec:A-poly}

In this note we concentrate on classical A-polynomials.
They are obtained from quantum A-polynomials -- difference operators annihilating HOMFLY-PT polynomials -- in a double scaling limit, when $q\to1$ and one considers huge representations of $U_q(\fg)$.
In particular, let us assume that all the highest weight vector $\vec w$ has all components $w_a$ non-zero.
We consider the following parametrization $q=e^{\hbar}$, $\hbar\to 0$.
Then $w_a\to\infty$, while the combinations $\hbar w_a=:\log\,\mu_a$ remain finite.
The representation $\rho$ marking CG chords remains fixed and finite.

This is a quasi-classical limit where difference operators shifting weights $w_a\mapsto w_a+1$ become the momenta $\lambda_a$.

There is no critical necessity to consider the quasi-classical case; it is a mere measure of simplification.
We expect that in this limit two simplifications take place:
\begin{enumerate}
	\item {\bf Quantum operators become commuting coordinates.}
	\item {\bf We could close open braids to obtain knots with CG chords inserted.} 
\end{enumerate}
We will discuss physical arguments in favor of these effects in Sec.~\ref{sec:WZW}.
For now we accept these statements as assumptions.

Under these assumptions classical A-polynomials can be obtained as consistency constraints on a system of mutual relations between CG chord quasi-classical values.
An inspiration for considering these particular systems of equations goes back to Reeb chords in the knot contact homology framework.
We will discuss relations between CG chords and Reeb chords further in Sec.~\ref{sec:KCH}.

\begin{figure}[ht!]
	\centering
	\scalebox{0.8}{\begin{tikzpicture}
		\begin{scope}
			\foreach \i in {0,1,2} {
				\begin{scope}[shift={(0,\i)}]
					\draw[ultra thick] (1,0) to[out=90,in=270] (0,1);
					\draw[white, line width=2mm] (0,0) to[out=90,in=270] (1,1);
					\draw[ultra thick] (0,0) to[out=90,in=270] (1,1);
				\end{scope}
			}
			\draw[ultra thick] (1,0) to[out=270,in=0] (-0.5,-1) to[out=180,in=270] (-2,0) -- (-2,3) to[out=90,in=180] (-0.5,4) to[out=0,in=90] (1,3);
			\draw[burgundy,line width = 1.5mm] (0,0) to[out=270,in=0] (-0.5,-0.5) to[out=180,in=270] (-1,0) -- (-1,3) to[out=90,in=180] (-0.5,3.5) to[out=0,in=90] (0,3);
			\draw[white,line width = 0.5mm] (0,0) to[out=270,in=0] (-0.5,-0.5) to[out=180,in=270] (-1,0) -- (-1,3) to[out=90,in=180] (-0.5,3.5) to[out=0,in=90] (0,3);
			\draw[\myblue, thick] (0,0) -- (1,0);
			\draw[\myblue, fill=\myblue] (0,0) circle (0.1) (1,0) circle (0.1);
		\end{scope}
		\begin{scope}[shift={(5,0)}]
			\begin{scope}
				\foreach \i in {0,1,2} {
					\begin{scope}[shift={(0,\i)}]
						\draw[ultra thick] (1,0) to[out=90,in=270] (0,1);
						\draw[white, line width=2mm] (0,0) to[out=90,in=270] (1,1);
						\draw[ultra thick] (0,0) to[out=90,in=270] (1,1);
					\end{scope}
				}
				\draw[ultra thick] (1,0) to[out=270,in=0] (-0.5,-1) to[out=180,in=270] (-2,0) -- (-2,3) to[out=90,in=180] (-0.5,4) to[out=0,in=90] (1,3);
				\draw[ultra thick](0,0) to[out=270,in=0] (-0.5,-0.5) to[out=180,in=270] (-1,0) -- (-1,1.5);
				\draw[burgundy,line width = 1.5mm] (-1,1.5) -- (-1,3) to[out=90,in=180] (-0.5,3.5) to[out=0,in=90] (0,3);
				\draw[white,line width = 0.5mm] (-1,1.5) -- (-1,3) to[out=90,in=180] (-0.5,3.5) to[out=0,in=90] (0,3);
				\draw[white, line width=1mm] (-1,1.5) to[out=0,in=180] (1,0);
				\draw[\myblue, thick] (-1,1.5) to[out=0,in=180] (1,0);
				\draw[\myblue, fill=\myblue] (-1,1.5) circle (0.08) (1,0) circle (0.08);
			\end{scope}
		\end{scope}
		\begin{scope}[shift={(10,0)}]
			\begin{scope}
				\foreach \i in {0,1,2} {
					\begin{scope}[shift={(0,\i)}]
						\draw[ultra thick] (1,0) to[out=90,in=270] (0,1);
						\draw[white, line width=2mm] (0,0) to[out=90,in=270] (1,1);
						\draw[ultra thick] (0,0) to[out=90,in=270] (1,1);
					\end{scope}
				}
				\draw[ultra thick] (0,0) to[out=270,in=0] (-0.5,-0.5) to[out=180,in=270] (-1,0) -- (-1,3) to[out=90,in=180] (-0.5,3.5) to[out=0,in=90] (0,3);
				\draw[burgundy,line width = 1.5mm]  (1,0) to[out=270,in=0] (-0.5,-1) to[out=180,in=270] (-2,0) -- (-2,3) to[out=90,in=180] (-0.5,4) to[out=0,in=90] (1,3);
				\draw[white,line width = 0.5mm] (1,0) to[out=270,in=0] (-0.5,-1) to[out=180,in=270] (-2,0) -- (-2,3) to[out=90,in=180] (-0.5,4) to[out=0,in=90] (1,3);
				\draw[\myblue, thick] (0,3) -- (1,3);
				\draw[\myblue, fill=\myblue] (0,3) circle (0.1) (1,3) circle (0.1);
			\end{scope}
		\end{scope}
		\begin{scope}[shift={(15,0)}]
			\begin{scope}
				\begin{scope}[shift={(0,0)}]
					\draw[ultra thick] (1,0) to[out=90,in=270] (0,1);
					\draw[thick, \myblue] (1.2,0) to[out=90,in=270] (0.2,1);
					\draw[white, line width=2mm] (0,0) to[out=90,in=270] (1,1);
					\draw[ultra thick] (0,0) to[out=90,in=270] (1,1);
				\end{scope}
				\begin{scope}[shift={(0,1)}]
					\draw[ultra thick] (1,0) to[out=90,in=270] (0,1);
					\draw[white, line width=2mm] (0,0) to[out=90,in=270] (1,1);
					\draw[ultra thick] (0,0) to[out=90,in=270] (1,1);
					\draw[white,line width = 1mm] (0.2,0) to[out=90,in=270] (1.2,1);
					\draw[thick, \myblue] (0.2,0) to[out=90,in=270] (1.2,1);
				\end{scope}
				\begin{scope}[shift={(0,2)}]
					\draw[ultra thick] (1,0) to[out=90,in=270] (0,1);
					\draw[thick, \myblue] (1.2,0) to[out=90,in=0] (0,0.8);
					\draw[white, line width=2mm] (0,0) to[out=90,in=270] (1,1);
					\draw[ultra thick] (0,0) to[out=90,in=270] (1,1);
				\end{scope}
				\draw[ultra thick] (1,0) to[out=270,in=0] (-0.5,-1) to[out=180,in=270] (-2,0) -- (-2,3) to[out=90,in=180] (-0.5,4) to[out=0,in=90] (1,3);
				\draw[ultra thick] (0,0) to[out=270,in=0] (-0.5,-0.5) to[out=180,in=270] (-1,0) -- (-1,3) to[out=90,in=180] (-0.5,3.5) to[out=0,in=90] (0,3);
				\draw[white, line width = 1mm] (-0.5,1.5) to[out=270,in=180] (0.5,-0.5) to[out=0,in=270] (1.2,0);
				\draw[thick, \myblue] (0,3.2) to[out=180,in=90] (-0.5,1.5) to[out=270,in=180] (0.5,-0.5) to[out=0,in=270] (1.2,0);
				\draw[\myblue, fill=\myblue] (0,2.8) circle (0.1) (0,3.2) circle (0.1);
			\end{scope}
		\end{scope}
		\draw[thick, -stealth] (1.3,1.5) -- (2.7,1.5);
		\draw[thick, -stealth] (6.3,3.5) to[out=30,in=150] (12.7,3.5);
		\draw[thick, -stealth] (6.3,-0.5) to[out=330,in=210] (7.7,-0.5);
		\node at (12,1.5) {\scalebox{1.5}{$=$}};
		\node[rotate=90] at (16.5,1.5) {Braid group action};
		\node[below] at (-0.5,-1.2) {(a)};
		\node[below] at (4.5,-1.2) {(b)};
		\node[below] at (9.5,-1.2) {(c1)};
		\node[below] at (14.5,-1.2) {(c2)};
	\end{tikzpicture}}
	\caption{Deriving equations on CG chords via equivalences of moves. Here, the sequences (a)--(b)--(c1) and (a)--(b)--(c2) denote frames of two homotopy movies. Since both movies start with the same frame (a), the two final frames (c1) and (c2) must be equivalent.
	}\label{fig:eqs}
\end{figure}
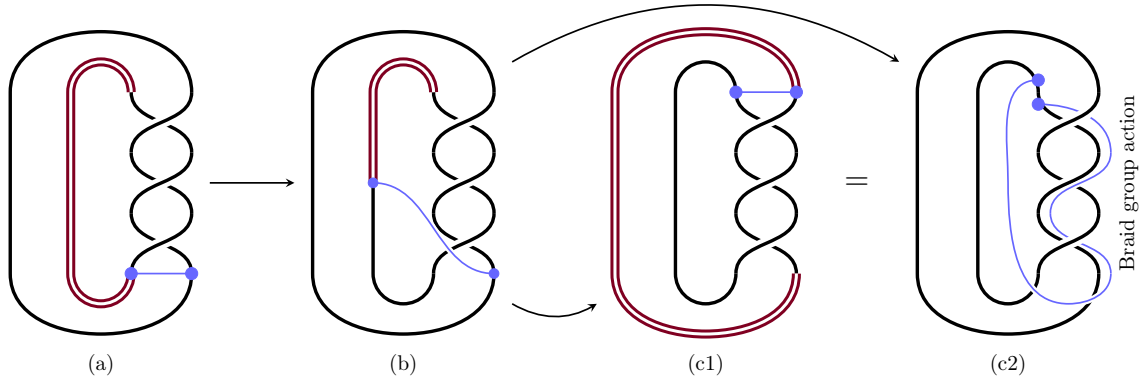

A construction of these equations follows a scheme depicted in Fig.\ref{fig:eqs}:
\begin{enumerate}
	\item We consider a knot diagram with a CG chord inserted in diagram (a).
	\item Using the fact that the knot is represented as a closed braid, we pull one of the chord ends along the closure to arrive at diagram (b).
	\item The second end of the chosen CG chord can be pulled upwards in two distinct yet topologically equivalent ways.
	\item In one way we pull it through the closure again to arrive at diagram (c1).
	However, here we have to bear in mind that after passing through a 3-valent Clebsch-Gordan vertex the representation is shifted.
	Strands whose representations are shifted relative to the original one are marked by a double {\bf \color{burgundy} red} line in the diagrams.
	\item The other way is to pull the second end directly through the braid.
	This move results in an action on the CG chord by morphisms $R^{\pm 1}$ according to the braid structure.
	In this way we arrive at diagram (c2).
	\item Finally, we arrive at an equivalence of diagrams (c1) and (c2) relating a certain action of the braid morphisms to shifts of representations in separate strands.
\end{enumerate}

We can distinguish two types of equations for all CG chords by choosing which end of a CG chord \eqref{CGdef} is pulled first in move (a)$\to$(b) in Fig.~\ref{fig:eqs}, the first one labeled by $k$ in \eqref{CGdef}, or the second one labeled by $m$ in \eqref{CGdef}.
Depending on this, we distinguish two modifications: $\chi_{I}$ in the first case, and $\chi_{II}$ in the second case.

In both cases we assume that a strand in the closure on the left carries index $\bf\color{\mygreen} 0$, which is smaller than any other index in the braid.
So, for these modifications we find:
\begin{equation}
	\begin{aligned}
		&\chi_I\left({}^{\rho}\Theta\scalebox{0.8}{$\left[\begin{array}{cc}
				\alpha & \beta\\
				k & m
			\end{array}\right]$}\right)={}^{\rho}\Theta\scalebox{0.8}{$\left[\begin{array}{cc}
			\alpha & \beta\\
			{\bf\color{\mygreen} 0} & m
			\end{array}\right]$}\times \left\{\begin{array}{ll}
			  \left({}^{\rho}\varphi_{\alpha}^{(k)}\right)\left({}^{\rho}\varphi_{\beta}^{(m)}\right),& \mbox{if }k>m\,;\\
			 1,& \mbox{otherwise}\,;
		\end{array}\right.\\
		&\chi_I^{-1}\left({}^{\rho}\Theta\scalebox{0.8}{$\left[\begin{array}{cc}
				\alpha & \beta\\
				{\bf\color{\mygreen}\color{\mygreen} 0} & m
			\end{array}\right]$}\right)={}^{\rho}\Theta\scalebox{0.8}{$\left[\begin{array}{cc}
				\alpha & \beta\\
				k & m
			\end{array}\right]$}\times \left\{\begin{array}{ll}
			\left({}^{\rho}\varphi_{\alpha}^{(k)}\right)^{-1}\left({}^{\rho}\varphi_{\beta}^{(m)}\right)^{-1},& \mbox{if }k>m\,;\\
			1,& \mbox{otherwise}\,;
		\end{array}\right.\\
		&\chi_{II}\left({}^{\rho}\Theta\scalebox{0.8}{$\left[\begin{array}{cc}
				\alpha & \beta\\
				k & m
			\end{array}\right]$}\right)={}^{\rho}\Theta\scalebox{0.8}{$\left[\begin{array}{cc}
				\alpha & \beta\\
				k & {\bf\color{\mygreen} 0}
			\end{array}\right]$}\times \left\{\begin{array}{ll}
			\left({}^{\rho}\varphi_{\alpha}^{(k)}\right)^{-1}\left({}^{\rho}\varphi_{\beta}^{(m)}\right)^{-1},& \mbox{if }k<m\,;\\
			1,& \mbox{otherwise}\,;
		\end{array}\right.\\
		&\chi_{II}^{-1}\left({}^{\rho}\Theta\scalebox{0.8}{$\left[\begin{array}{cc}
				\alpha & \beta\\
				k & {\bf\color{\mygreen} 0}
			\end{array}\right]$}\right)={}^{\rho}\Theta\scalebox{0.8}{$\left[\begin{array}{cc}
				\alpha & \beta\\
				k & m
			\end{array}\right]$}\times \left\{\begin{array}{ll}
			\left({}^{\rho}\varphi_{\alpha}^{(k)}\right)\left({}^{\rho}\varphi_{\beta}^{(m)}\right),& \mbox{if }k<m\,;\\
			1,& \mbox{otherwise}\,;
		\end{array}\right.
	\end{aligned}
\end{equation}
where we have to take into account the twists of the CG chord ends \eqref{uni_twist}.
Let us stress here again that maps $\chi$ are not simply multiplications by factors; rather, they move one of the CG chord ends to strand $\bf 0$ located in the closure on the left.

Let us denote a shift of the representation in strand $k$ parametrized by an isotypical index $\alpha$ as ${}^{\rho}\lambda_{\alpha}^{(k)}$ (this shift of a representation in a knot strand is depicted as a double red line in Fig.~\ref{fig:eqs}), then a shift of the knot representations corresponds to simultaneous shifts in all strands:
\begin{equation}
	{}^{\rho}\lambda_{\alpha}:=\prod\lm_{k=1}^{\#\;{\rm of\; strands}}{}^{\rho}\lambda_{\alpha}^{(k)}\,.
\end{equation}
Also we denote a morphism associated to a braid element $\fB$ decomposed into elementary morphisms \eqref{Rmorph_beg}-\eqref{Rmorph_end} and \eqref{RRmorph_beg}-\eqref{RRmorph_end} as ${\bf R}_{\fB}$.
Then we arrive at the following equations for CG chords in a background of knot $K$ represented as a closure of braid $\fB$:
\begin{tcolorbox}
\begin{equation}\label{main_eqs}
	\begin{aligned}
	&\Upsilon_I:={}^{\rho}\lambda_\beta^{(m)}\cdot {}^{\rho}\Theta\scalebox{0.8}{$\left[\begin{array}{cc}
			\alpha & \beta\\
			k & m
		\end{array}\right]$}-\chi_I^{-1}\circ{\bf R}_{\fB}\circ\chi_I\left({}^{\rho}\Theta\scalebox{0.8}{$\left[\begin{array}{cc}
		\alpha & \beta\\
		k & m
		\end{array}\right]$}\right)=0\,,\\
	&\Upsilon_{II}:=\left({}^{\rho}\lambda_\alpha^{(k)}\right)^{-1}\cdot {}^{\rho}\Theta\scalebox{0.8}{$\left[\begin{array}{cc}
			\alpha & \beta\\
			k & m
		\end{array}\right]$}-\chi_{II}^{-1}\circ{\bf R}_{\fB}\circ\chi_{II}\left({}^{\rho}\Theta\scalebox{0.8}{$\left[\begin{array}{cc}
			\alpha & \beta\\
			k & m
		\end{array}\right]$}\right)=0\,.
	\end{aligned}
\end{equation}
\end{tcolorbox}

We should note that even from Fig.~\ref{fig:eqs} we encounter situations when after applying $\chi^{-1}$ there might appear new chords with \emph{coincident} ends.
Moreover, there is no need to avoid such chords in constructing \eqref{main_eqs}.
Yet these are not new chord variables; rather, we can calculate those a priori:
\begin{equation}\label{Kroneker}
	\begin{aligned}
	&\mbox{Type I:}\quad {}^{\rho}\Theta\scalebox{0.8}{$\left[\begin{array}{cc}
			\alpha & \beta\\
			k & k
		\end{array}\right]$}=\begin{array}{c}
		\begin{tikzpicture}[scale=0.7]
			\draw[thick, \myblue] (0,-0.5) to[out=30,in=270] (0.5,-0.2) (-0.5,0.2) to[out=90,in=210] (0,0.5);
			\draw[ultra thick] (0,-0.8) -- (0,0.8);
			\draw[white, line width=1mm] (0.5,-0.2) to[out=90,in=270] (-0.5,0.2);
			\draw[thick, \myblue] (0.5,-0.2) to[out=90,in=270] (-0.5,0.2); 
			\node[left] at (0,-0.8) {$\scriptstyle \alpha$};
			\node[right] at (0,0.8) {$\scriptstyle \beta$};
		\end{tikzpicture}
	\end{array}=\delta_{\alpha\beta}\frac{\left({}^{\rho}\varphi_{\alpha}^{(k)}\right)^{-2}-\left({}^{\rho}\varphi_{0}^{(k)}\right)^{-2}}{\left({}^{\rho}\varphi_{\alpha}^{(k)}\right)^{-1}\left({}^{\rho}\varphi_{0}^{(k)}\right)^{-1}}\left({}^{\rho}\varphi_{\alpha}^{(k)}\right)^{-1}\,,\\
	&\mbox{Type II:}\quad {}^{\rho}\Theta\scalebox{0.8}{$\left[\begin{array}{cc}
			\alpha & \beta\\
			k & k
		\end{array}\right]$}=\begin{array}{c}
		\begin{tikzpicture}[scale=0.7, xscale=-1]
			\draw[thick, \myblue] (0,-0.5) to[out=30,in=270] (0.5,-0.2) (-0.5,0.2) to[out=90,in=210] (0,0.5);
			\draw[ultra thick] (0,-0.8) -- (0,0.8);
			\draw[white, line width=1mm] (0.5,-0.2) to[out=90,in=270] (-0.5,0.2);
			\draw[thick, \myblue] (0.5,-0.2) to[out=90,in=270] (-0.5,0.2); 
			\node[right] at (0,-0.8) {$\scriptstyle \alpha$};
			\node[left] at (0,0.8) {$\scriptstyle \beta$};
		\end{tikzpicture}
	\end{array}=\delta_{\alpha\beta}\frac{\left({}^{\rho}\varphi_{\alpha}^{(k)}\right)^{-2}-\left({}^{\rho}\varphi_{0}^{(k)}\right)^{-2}}{\left({}^{\rho}\varphi_{\alpha}^{(k)}\right)^{-1}\left({}^{\rho}\varphi_{0}^{(k)}\right)^{-1}}\left({}^{\rho}\varphi_{\alpha}^{(k)}\right)\,.
	\end{aligned}
\end{equation}

Clearly, adding chords with coinciding tips into consideration ``overloads'' a system of equations on CG chords of either type of equations, not to mention that types I and II must be consistent with each other, and with different choices of irrep $\rho$.
Eventually, the number of equations exceeds the number of unknown variables by more than ${\rm rk}\,\fg$.
So it is natural to hope that this overdetermined system has consistency constraints, we call roots of \emph{shaded} A-polynomials, for a solution to exist.
We hope that these relations can be rewritten purely in terms of ${}^{\rho}\varphi_{\alpha}$ and ${}^{\rho}\lambda_{\alpha}$ for a given knot, which can be re-expressed further in terms of $\mu_a$ and $\lambda_a$:
\begin{tcolorbox}
\begin{equation}\label{A-poly}
	\eqref{main_eqs}\;\overset{{\rm exclude}\;\Theta}{\Longrightarrow}\; A_a\left({}^{\rho}\lambda_{\alpha},{}^{\rho}\varphi_{\alpha}\right)=0\; \Longrightarrow\;A_a(\lambda_b,\mu_c)=0,\quad a,b,c=1,\ldots,{\rm rk}\,\fg\,.
\end{equation}
\end{tcolorbox}

At first glance it might seem that the procedure of deriving such A-polynomials is cumbersome.
On the contrary, this is a rather algorithmic, albeit tedious, procedure!
Although the equations in the system are non-linear, we still can solve for CG chords explicitly one by one by exploiting the fact that the system is overdetermined.

Let us order CG chord variables in some way.
Then we regard the first variable in our ordering as a variable $x$ and the remaining CG chords as parameters.
With respect to this choice the system takes the following form:
\begin{equation}
	Q_i(x)=\sum\lm_{k=0}^{p_i} a_{ik}x^k = 0\,.
\end{equation}
Let us choose an equation where a degree of a polynomial in $x$ is minimal but non-zero, suppose its index is $h$; we call its degree $p_h=\fp$ the degree of the system in $x$.
There are two options: either the coefficient $a_{hp}=0$, or we can solve for 
\begin{equation}\label{xp}
	x^\fp=:y_{\fp-1}(x)=\frac{Q_h(x)}{a_{h\fp}}-x^\fp\,,
\end{equation}
as a polynomial of smaller degree.
In the first case we add a new constraint $a_{hp}=0$ to the set of equations, eliminate this coefficient and recalculate $\fp$.
In the second case we simplify all the remaining equations by the substitution:
\begin{equation}
	x^a\to x^{a\;{\rm mod}\;\fp} y_{\fp-1}(x)^{\frac{a-(a\;{\rm mod}\;\fp)}{\fp}}\,,
\end{equation}
until the degrees of the remaining polynomials are strictly less than $\fp$.

We repeat this procedure iteratively.
For sufficiently generic polynomials it leads to a situation where a single polynomial is linear in $x$; this gives a solution to $x$ trivially, while the remaining polynomials are of degree 0 and provide constraints for this root to satisfy all the equations simultaneously.
For example, for a pair of sufficiently generic polynomials $Q_1(x)$ and $Q_2(x)$ in a single variable $x$, apparently, this procedure extracts the resultant \cite{Dolotin:2006zr} as a single constraint.

After excluding one variable, before continuing to the next one, the system must be transformed back into polynomials form, since the substitution \eqref{xp} may have made it rational instead.
So after reducing to a common denominator the degree in the next variable might increase, which might make this procedure rather expensive computationally, unfortunately.

To conclude this part, let us note that possible moves of CG chords along the knot leading to relations are not restricted to those reflected in \eqref{main_eqs}.
There are many more moves.
For example, we could pull a CG chord as a whole once through the braid and the other time through the closure.
This equivalence indicates that CG chord solutions to \eqref{main_eqs} are simultaneously solutions to a potentially nonlinear eigenvalue problem of ${\bf R}_{\fB}$:
\begin{equation}\label{eigen}
	{\bf R}_{\fB}\left({}^{\rho}\Theta\scalebox{0.8}{$\left[\begin{array}{cc}
			\alpha & \beta\\
			k & m
		\end{array}\right]$}\right)=\left({}^{\rho}\lambda_\alpha^{(k)}\right)^{-1}\left({}^{\rho}\lambda_\beta^{(m)}\right)\cdot {}^{\rho}\Theta\scalebox{0.8}{$\left[\begin{array}{cc}
			\alpha & \beta\\
			k & m
		\end{array}\right]$}\,.
\end{equation}


\subsection{Reference chords and questions of A-polynomial covariance} \label{sec:covariant}

We expect that roots of shaded A-polynomials are invariants of knot diagrams up to a simple modification due to framing.
This expectation follows naturally from the fact that our construction for A-polynomial roots is auxiliary to HOMFLY polynomials, which are true knot invariants, we simply consider a quasi-classical limit of the equations that HOMFLY polynomials ought to satisfy.
Nevertheless, it is intriguing to observe effects of A-polynomial covariance directly within the CG chord formalism.

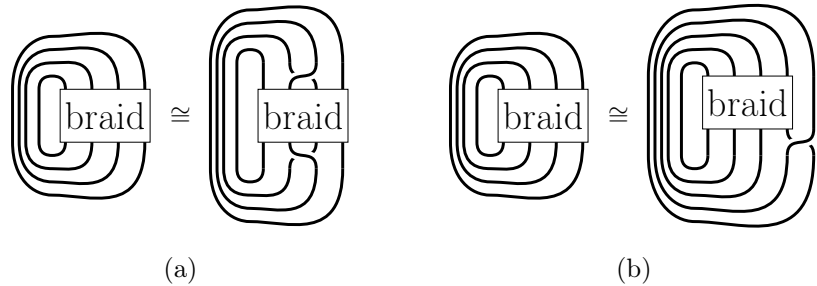
\begin{figure}[ht!]
	\centering
	\begin{tikzpicture}
		\node{\scalebox{0.7}{$\begin{array}{c}
				\begin{tikzpicture}[xscale=0.5]
					\draw[ultra thick] (0,-0.5) -- (0,0.5) to[out=90,in=0] (-0.5,0.75) to[out=180,in=90] (-1,0.5) -- (-1,-0.5) to [out=270,in=180] (-0.5,-0.75) to [out=0,in=270] (0,-0.5);
					\draw[ultra thick] (1,-0.5) -- (1,0.5) to[out=90,in=0] (-0.5,1) to[out=180,in=90] (-1.5,0.5) -- (-1.5,-0.5) to [out=270,in=180] (-0.5,-1) to [out=0,in=270] (1,-0.5);
					\draw[ultra thick] (2,-0.5) -- (2,0.5) to[out=90,in=0] (-0.5,1.25) to[out=180,in=90] (-1.75,0.5) -- (-1.75,-0.5) to [out=270,in=180] (-0.5,-1.25) to [out=0,in=270] (2,-0.5);
					\draw[ultra thick] (3,-0.5) -- (3,0.5) to[out=90,in=0] (-0.5,1.5) to[out=180,in=90] (-2,0.5) -- (-2,-0.5) to [out=270,in=180] (-0.5,-1.5) to [out=0,in=270] (3,-0.5);
					\draw[fill=white] (-0.2,-0.5) -- (3.2,-0.5) -- (3.2,0.5) -- (-0.2,0.5) -- cycle;
					\node at (1.5,0) {\huge braid};
				\end{tikzpicture}
			\end{array}$} $\cong$ \scalebox{0.7}{$\begin{array}{c}
			\begin{tikzpicture}[xscale=0.5]
				\draw[ultra thick] (0,0.5) -- (0,1) (2,0.5) to[out=90,in=270] (1,1) (3,0.5) -- (3,1);
				\draw[white,line width = 2mm] (1,0.5) to[out=90,in=270] (2,1);
				\draw[ultra thick] (1,0.5) to[out=90,in=270] (2,1);
				\begin{scope}[yscale=-1]
					\draw[ultra thick] (0,0.5) -- (0,1) (2,0.5) to[out=90,in=270] (1,1) (3,0.5) -- (3,1);
					\draw[white,line width = 2mm] (1,0.5) to[out=90,in=270] (2,1);
					\draw[ultra thick] (1,0.5) to[out=90,in=270] (2,1);
				\end{scope}
				\draw[ultra thick] (0,1) to[out=90,in=0] (-0.5,1.25) to[out=180,in=90] (-1,1) -- (-1,-1) to [out=270,in=180] (-0.5,-1.25) to [out=0,in=270] (0,-1);
				\draw[ultra thick] (1,1) to[out=90,in=0] (-0.5,1.5) to[out=180,in=90] (-1.5,1) -- (-1.5,-1) to [out=270,in=180] (-0.5,-1.5) to [out=0,in=270] (1,-1);
				\draw[ultra thick] (2,1) to[out=90,in=0] (-0.5,1.75) to[out=180,in=90] (-1.75,1) -- (-1.75,-1) to [out=270,in=180] (-0.5,-1.75) to [out=0,in=270] (2,-1);
				\draw[ultra thick] (3,1) to[out=90,in=0] (-0.5,2) to[out=180,in=90] (-2,1) -- (-2,-1) to [out=270,in=180] (-0.5,-2) to [out=0,in=270] (3,-1);
				\draw[fill=white] (-0.2,-0.5) -- (3.2,-0.5) -- (3.2,0.5) -- (-0.2,0.5) -- cycle;
				\node at (1.5,0) {\huge braid};
			\end{tikzpicture}
			\end{array}$}};
		\node at (6,0){\scalebox{0.7}{$\begin{array}{c}
					\begin{tikzpicture}[xscale=0.5]
						\draw[ultra thick] (0,-0.5) -- (0,0.5) to[out=90,in=0] (-0.5,0.75) to[out=180,in=90] (-1,0.5) -- (-1,-0.5) to [out=270,in=180] (-0.5,-0.75) to [out=0,in=270] (0,-0.5);
						\draw[ultra thick] (1,-0.5) -- (1,0.5) to[out=90,in=0] (-0.5,1) to[out=180,in=90] (-1.5,0.5) -- (-1.5,-0.5) to [out=270,in=180] (-0.5,-1) to [out=0,in=270] (1,-0.5);
						\draw[ultra thick] (2,-0.5) -- (2,0.5) to[out=90,in=0] (-0.5,1.25) to[out=180,in=90] (-1.75,0.5) -- (-1.75,-0.5) to [out=270,in=180] (-0.5,-1.25) to [out=0,in=270] (2,-0.5);
						\draw[ultra thick] (3,-0.5) -- (3,0.5) to[out=90,in=0] (-0.5,1.5) to[out=180,in=90] (-2,0.5) -- (-2,-0.5) to [out=270,in=180] (-0.5,-1.5) to [out=0,in=270] (3,-0.5);
						\draw[fill=white] (-0.2,-0.5) -- (3.2,-0.5) -- (3.2,0.5) -- (-0.2,0.5) -- cycle;
						\node at (1.5,0) {\huge braid};
					\end{tikzpicture}
				\end{array}$} $\cong$ \scalebox{0.7}{$\begin{array}{c}
					\begin{tikzpicture}[xscale=0.5]
						\draw[ultra thick] (0,0.5) to[out=90,in=0] (-0.5,0.75) to[out=180,in=90] (-1,0.5) -- (-1,-1) to [out=270,in=180] (-0.5,-1.25) to [out=0,in=270] (0,-1);
						\draw[ultra thick] (1,0.5) to[out=90,in=0] (-0.5,1) to[out=180,in=90] (-1.5,0.5) -- (-1.5,-1) to [out=270,in=180] (-0.5,-1.5) to [out=0,in=270] (1,-1);
						\draw[ultra thick] (2,0.5) to[out=90,in=0] (-0.5,1.25) to[out=180,in=90] (-1.75,0.5) -- (-1.75,-1) to [out=270,in=180] (-0.5,-1.75) to [out=0,in=270] (2,-1);
						\draw[ultra thick] (3,0.5) to[out=90,in=0] (-0.5,1.5) to[out=180,in=90] (-2,0.5) -- (-2,-1) to [out=270,in=180] (-0.5,-2) to [out=0,in=270] (3,-1);
						\draw[fill=white] (-0.2,-0.5) -- (3.2,-0.5) -- (3.2,0.5) -- (-0.2,0.5) -- cycle;
						\draw[ultra thick] (4,0.5) to[out=90,in=0] (-0.5,1.75) to[out=180,in=90] (-2.25,0.5) -- (-2.25,-1) to [out=270,in=180] (-0.5,-2.25) to [out=0,in=270] (4,-1);
						\draw[ultra thick] (0,-0.5) -- (0,-1) (1,-0.5) -- (1,-1) (2,-0.5) -- (2,-1) (3,-0.5) to[out=270,in=90] (4,-1) (4,0.5) -- (4,-0.5);
						\draw[white, line width=2mm] (3,-1) to[out=90,in=270] (4,-0.5);
						\draw[ultra thick] (3,-1) to[out=90,in=270] (4,-0.5);
						\node at (1.5,0) {\huge braid};
					\end{tikzpicture}
				\end{array}$}};
			\node at (0,-2) {(a)};
			\node at (6,-2) {(b)};
	\end{tikzpicture}
	\caption{Markov moves.}\label{fig:Markov}
\end{figure}

Since we work initially with a braid representation of knot diagrams, we should prove an invariance of the construction with respect to the Markov moves \cite{Markov:1936,Birman:1974}: in addition to the invariance with respect to Reidemeister moves II and III, which is trivial in our construction due to morphism identities \eqref{R_identities}, one has to prove equivalences also with respect to a conjugation of a braid by arbitrary elements $R_{i,i+1}^{\pm 1}$, and with respect to adding a new strand with a twist (see Fig.~\ref{fig:Markov}).
It should be stressed that the latter move modifies knot framing, so the procedure does not deliver the same A-polynomials; rather, it delivers A-polynomials with changed coordinates taking framing into account.
Here we will not present a complete proof of our construction, instead we will map out possible pathways towards this goal.

Unfortunately, it is not immediately clear why different braids connected by Markov moves would produce equivalent consistency constraints for \eqref{main_eqs} constructed from these braid representations.
It would be somewhat easier to rewrite these relations in terms of new variables we call reference chords.

Let us introduce a new \emph{reference} strand $\bf\color{\mygreen} 0$ not taking part in the initial braid formation, so it is never entangled with the other strands.
Then any \emph{basic} CG chord can be expanded as a quadratic combination of \emph{reference} chords starting from or ending on strand $\bf\color{\mygreen} 0$:
\begin{equation}\label{reference}
	\begin{aligned}
	&\scalebox{0.88}{${}^{\rho}\Theta\scalebox{0.8}{$\left[\begin{array}{cc}
			\alpha & \beta\\
			k & m
		\end{array}\right]$}=\begin{array}{c}
		\begin{tikzpicture}[scale=0.5]
			\draw[ultra thick] (2,-1) -- (2,1);
			\draw[white, line width=2mm] (1,-0.5) -- (3,0.5);
			\draw[thick,\myblue] (1,-0.5) -- (3,0.5);
			\draw[ultra thick] (0,-1) -- (0,1) (1,-1) -- (1,1) (3,-1) -- (3,1) (4,-1) -- (4,1);
			\draw[ultra thick, \mygreen] (-1,-1) -- (-1,1);
			\node[below] at (-1,-1) {$\bf\color{\mygreen}\scriptstyle 0$};
			\node[below] at (1,-1) {$\scriptstyle k$};
			\node[below] at (3,-1) {$\scriptstyle m$};
			\node[left] at (1,-0.75) {$\scriptstyle \alpha$};
			\node[right] at (3,0.75) {$\scriptstyle \beta$};
		\end{tikzpicture}
	\end{array}=\sum\lm_{\delta}{}^{\rho}\varphi_{\alpha}^{(k)}\begin{array}{c}
	\begin{tikzpicture}[scale=0.5]
		\draw[ultra thick] (0,-1) -- (0,1) (1,0) -- (1,1) (2,-1) -- (2,1) (4,-1) -- (4,1);
		\draw[white, line width=2mm] (1,-0.7) -- (-1,-0.3) (-1,0.3) -- (3,0.7);
		\draw[thick,\myblue] (1,-0.7) -- (-1,-0.3) (-1,0.3) -- (3,0.7);
		\draw[ultra thick]  (1,-1) -- (1,0) (3,-1) -- (3,1);
		\draw[ultra thick, \mygreen] (-1,-1) -- (-1,1);
		\node[below] at (-1,-1) {$\bf\color{\mygreen}\scriptstyle 0$};
		\node[below] at (1,-1) {$\scriptstyle k$};
		\node[below] at (3,-1) {$\scriptstyle m$};
		\node[right] at (1,-0.75) {$\scriptstyle \alpha$};
		\node[right] at (3,0.75) {$\scriptstyle \beta$};
		\node[left] at (-1,0) {$\scriptstyle \delta$};
	\end{tikzpicture}
	\end{array}=\sum\lm_{\delta}{}^{\rho}\varphi_{\alpha}^{(k)}\;
	{}^{\rho}\Theta\scalebox{0.8}{$\left[\begin{array}{cc}
			\delta & \beta\\
			{\bf\color{\mygreen}0} & m
		\end{array}\right]$}\;
	{}^{\rho}\Theta\scalebox{0.8}{$\left[\begin{array}{cc}
		\alpha & \delta\\
		k & {\bf\color{\mygreen}0}
	\end{array}\right]$},\; k<m$}\,,\\
	&\scalebox{0.88}{${}^{\rho}\Theta\scalebox{0.8}{$\left[\begin{array}{cc}
			\alpha & \beta\\
			k & m
		\end{array}\right]$}=\begin{array}{c}
		\begin{tikzpicture}[scale=0.5]
			\draw[ultra thick] (2,-1) -- (2,1);
			\draw[white, line width=2mm] (3,-0.5) -- (1,0.5);
			\draw[thick,\myblue] (3,-0.5) -- (1,0.5);
			\draw[ultra thick] (0,-1) -- (0,1) (1,-1) -- (1,1) (3,-1) -- (3,1) (4,-1) -- (4,1);
			\draw[ultra thick, \mygreen] (-1,-1) -- (-1,1);
			\node[below] at (-1,-1) {$\bf\color{\mygreen}\scriptstyle 0$};
			\node[below] at (1,-1) {$\scriptstyle m$};
			\node[below] at (3,-1) {$\scriptstyle k$};
			\node[right] at (3,-0.75) {$\scriptstyle \alpha$};
			\node[left] at (1,0.75) {$\scriptstyle \beta$};
		\end{tikzpicture}
	\end{array}=\sum\lm_{\delta}\left({}^{\rho}\varphi_{\beta}^{(m)}\right)^{-1}\begin{array}{c}
		\begin{tikzpicture}[scale=0.5]
			\draw[ultra thick] (0,-1) -- (0,1) (1,-1) -- (1,0) (2,-1) -- (2,1) (4,-1) -- (4,1);
			\draw[white, line width=2mm] (3,-0.7) -- (-1,-0.3) (-1,0.3) -- (1,0.7);
			\draw[thick,\myblue] (3,-0.7) -- (-1,-0.3) (-1,0.3) -- (1,0.7);
			\draw[ultra thick]  (1,0) -- (1,1) (3,-1) -- (3,1);
			\draw[ultra thick, \mygreen] (-1,-1) -- (-1,1);
			\node[below] at (-1,-1) {$\bf\color{\mygreen}\scriptstyle 0$};
			\node[below] at (1,-1) {$\scriptstyle k$};
			\node[below] at (3,-1) {$\scriptstyle m$};
			\node[right] at (3,-0.75) {$\scriptstyle \alpha$};
			\node[right] at (1,0.75) {$\scriptstyle \beta$};
			\node[left] at (-1,0) {$\scriptstyle \delta$};
		\end{tikzpicture}
	\end{array}=\sum\lm_{\delta}\left({}^{\rho}\varphi_{\beta}^{(m)}\right)^{-1}\;
	{}^{\rho}\Theta\scalebox{0.8}{$\left[\begin{array}{cc}
			\delta & \beta\\
			{\bf\color{\mygreen}0} & m
		\end{array}\right]$}\;
	{}^{\rho}\Theta\scalebox{0.8}{$\left[\begin{array}{cc}
			\alpha & \delta\\
			k & {\bf\color{\mygreen}0}
		\end{array}\right]$},\; k>m$}\,.
	\end{aligned}
\end{equation}
Note that now index $\delta$ runs over all isotypical components including the uninvited one with index 0.
Also, we should mention that the number of reference chords scales as $\sim M$ with the number of strands $M$ in the braid.
This scaling is equivalent to the scaling of actual coordinates parametrizing conformal blocks in WZW models (see Sec.~\ref{sec:WZW}), so the reference chords are more suitable for the role of independent coordinates than the basic CG chords defined earlier.
However, compared with the basic CG chords, the definition of the reference chords requires an auxiliary strand.

In terms of reference chords, the knot relations \eqref{main_eqs} can be rewritten in a form that bypasses the need to introduce maps $\chi_{I,II}$:
\begin{equation}\label{ref_main}
	\begin{aligned}
		{}^{\rho}\lambda_\beta^{(m)}\cdot {}^{\rho}\Theta\scalebox{0.8}{$\left[\begin{array}{cc}
				\delta & \beta\\
				{\bf \color{\mygreen} 0} & m
			\end{array}\right]$}={\bf R}_{\fB}\left({}^{\rho}\Theta\scalebox{0.8}{$\left[\begin{array}{cc}
				\delta & \beta\\
				{\bf\color{\mygreen} 0} & m
			\end{array}\right]$}\right),\quad 
		\left({}^{\rho}\lambda_\alpha^{(k)}\right)^{-1}\cdot {}^{\rho}\Theta\scalebox{0.8}{$\left[\begin{array}{cc}
				\alpha & \delta\\
				k & {\bf\color{\mygreen}0}
			\end{array}\right]$}={\bf R}_{\fB}\left({}^{\rho}\Theta\scalebox{0.8}{$\left[\begin{array}{cc}
				\alpha & \delta\\
				k & {\bf\color{\mygreen} 0}
			\end{array}\right]$}\right)\,.
	\end{aligned}
\end{equation} 
We expect that a solution to these equations is equivalent to a solution to \eqref{main_eqs}.

Equivalences of A-polynomials as ``resultants'' for systems of type \eqref{main_eqs}, \eqref{ref_main} under Markov moves should imply that the solutions of the original system can be obtained from those of a system obtained by a direct change of variables.
In this case resultants remain invariant.

For the first Markov move depicted in Fig.~\ref{fig:Markov}(a), system \eqref{ref_main} for the r.h.s. corresponds to a morphism ${\bf R}_{\fB}$ conjugated by R-matrices:
\begin{equation}
	\begin{aligned}
		&{}^{\rho}\lambda_\beta^{(m)}\cdot {}^{\rho}\Theta\scalebox{0.8}{$\left[\begin{array}{cc}
				\delta & \beta\\
				{\bf \color{\mygreen} 0} & m
			\end{array}\right]$}=R_{i,i+1}\circ{\bf R}_{\fB}\circ R_{i,i+1}^{-1}\left({}^{\rho}\Theta\scalebox{0.8}{$\left[\begin{array}{cc}
				\delta & \beta\\
				{\bf\color{\mygreen} 0} & m
			\end{array}\right]$}\right)\,,\\
		&\left({}^{\rho}\lambda_\alpha^{(k)}\right)^{-1}\cdot {}^{\rho}\Theta\scalebox{0.8}{$\left[\begin{array}{cc}
				\alpha & \delta\\
				k & {\bf\color{\mygreen}0}
			\end{array}\right]$}=R_{i,i+1}\circ{\bf R}_{\fB}\circ R_{i,i+1}^{-1}\left({}^{\rho}\Theta\scalebox{0.8}{$\left[\begin{array}{cc}
				\alpha & \delta\\
				k & {\bf\color{\mygreen} 0}
			\end{array}\right]$}\right)\,.
	\end{aligned}
\end{equation}
Evidently, by a simple change of variables:
\begin{equation}
	{}^{\rho}\tilde\Theta\scalebox{0.8}{$\left[\begin{array}{cc}
			\delta & \beta\\
			{\bf\color{\mygreen} 0} & m
		\end{array}\right]$}:=R_{i,i+1}^{-1}\left({}^{\rho}\Theta\scalebox{0.8}{$\left[\begin{array}{cc}
			\delta & \beta\\
			{\bf\color{\mygreen} 0} & m
		\end{array}\right]$}\right),\quad {}^{\rho}\tilde\Theta\scalebox{0.8}{$\left[\begin{array}{cc}
		\alpha & \delta\\
		k & {\bf\color{\mygreen} 0}
		\end{array}\right]$}:= R_{i,i+1}^{-1}\left({}^{\rho}\Theta\scalebox{0.8}{$\left[\begin{array}{cc}
		\alpha & \delta\\
		k & {\bf\color{\mygreen} 0}
		\end{array}\right]$}\right)\,,
\end{equation}
we would arrive at system \eqref{ref_main} for variables with tildes.
These systems will produce the same roots for shaded A-polynomials.

Now let us consider another Markov move depicted in Fig.~\ref{fig:Markov}(b).
This move switches the number of strands in the braid from $M$ to $M+1$.
Let us first note that for all reference chords with $k<M$, $m<M$, equations \eqref{ref_main} are not modified.
We consider explicitly the relations for chords with $k,m=M,M+1$:
\begin{equation}\label{Markov2}
	\begin{aligned}
		&{}^{\rho}\lambda_\beta^{(M)}\cdot {}^{\rho}\Theta\scalebox{0.8}{$\left[\begin{array}{cc}
				\delta & \beta\\
				{\bf \color{\mygreen} 0} & M
			\end{array}\right]$}={}^{\rho}\Theta\scalebox{0.8}{$\left[\begin{array}{cc}
				\delta & \beta\\
				{\bf\color{\mygreen} 0} & M+1
			\end{array}\right]$}\,,\\
		&{}^{\rho}\lambda_\beta^{(M+1)}\cdot {}^{\rho}\Theta\scalebox{0.8}{$\left[\begin{array}{cc}
				\delta & \beta\\
				{\bf \color{\mygreen} 0} & M+1
			\end{array}\right]$}={\bf R}_{\fB}\circ R_{M,M+1}\left({}^{\rho}\Theta\scalebox{0.8}{$\left[\begin{array}{cc}
				\delta & \beta\\
				{\bf\color{\mygreen} 0} & M+1
			\end{array}\right]$}\right)\,,\\
		&\left({}^{\rho}\lambda_\alpha^{(M)}\right)^{-1}\cdot {}^{\rho}\Theta\scalebox{0.8}{$\left[\begin{array}{cc}
				\alpha & \delta\\
				M & {\bf\color{\mygreen}0}
			\end{array}\right]$}={}^{\rho}\Theta\scalebox{0.8}{$\left[\begin{array}{cc}
				\alpha & \delta\\
				M+1 & {\bf\color{\mygreen} 0}
			\end{array}\right]$}\,,\\
		&\left({}^{\rho}\lambda_\alpha^{(M+1)}\right)^{-1}\cdot {}^{\rho}\Theta\scalebox{0.8}{$\left[\begin{array}{cc}
				\alpha & \delta\\
				M+1 & {\bf\color{\mygreen}0}
			\end{array}\right]$}={\bf R}_{\fB}\circ R_{M,M+1}\left({}^{\rho}\Theta\scalebox{0.8}{$\left[\begin{array}{cc}
				\alpha & \delta\\
				M+1 & {\bf\color{\mygreen} 0}
			\end{array}\right]$}\right)\,.
	\end{aligned}
\end{equation}

The action of the R-twist reads in this case \eqref{Rmorph_beg}-\eqref{Rmorph_end}:
\begin{equation}
		\begin{aligned}
		&R_{M,M+1}\left({}^{\rho}\Theta\scalebox{0.8}{$\left[\begin{array}{cc}
				\delta & \beta\\
				{\bf\color{\mygreen}0} & M+1
			\end{array}\right]$}\right)=\left({}^{\rho}\varphi_{0}^{(M+1)}\right)^2{}^{\rho}\Theta\scalebox{0.8}{$\left[\begin{array}{cc}
				\delta & \beta\\
				{\bf\color{\mygreen}0} & M
			\end{array}\right]$}-\sum\lm_{\gamma\neq 0}\left({}^{\rho}\varphi_{\gamma}^{(M+1)}\right)\left({}^{\rho}\varphi_{0}^{(M+1)}\right)\times\\
		&\hspace{5cm}\times {}^{\rho}\Theta\scalebox{0.8}{$\left[\begin{array}{cc}
				\gamma & \beta\\
				M+1 & M
			\end{array}\right]$}\; {}^{\rho}\Theta\scalebox{0.8}{$\left[\begin{array}{cc}
				\delta & \gamma\\
				{\bf\color{\mygreen}0} & M+1
			\end{array}\right]$}
			\left({}^{\rho}\varphi_{\beta}^{(M)}\right)^{-1}\,,\\
			&R_{M,M+1}\left({}^{\rho}\Theta\scalebox{0.8}{$\left[\begin{array}{cc}
					\alpha & \delta\\
					M+1 & {\bf\color{\mygreen}0}
				\end{array}\right]$}\right)=\left({}^{\rho}\varphi_{0}^{(M+1)}\right)^{-2}{}^{\rho}\Theta\scalebox{0.8}{$\left[\begin{array}{cc}
					\alpha & \delta\\
					M & {\bf\color{\mygreen}0}
				\end{array}\right]$}+\sum\lm_{\gamma\neq 0}\left({}^{\rho}\varphi_{\gamma}^{(M+1)}\right)^{-1}\left({}^{\rho}\varphi_{0}^{(M+1)}\right)^{-1}\times\\
			&\hspace{5cm}\times {}^{\rho}\Theta\scalebox{0.8}{$\left[\begin{array}{cc}
					\gamma & \delta\\
					M+1 & {\bf\color{\mygreen}0}
				\end{array}\right]$}\; {}^{\rho}\Theta\scalebox{0.8}{$\left[\begin{array}{cc}
					\alpha & \gamma\\
					M & M+1
				\end{array}\right]$}\left({}^{\rho}\varphi_{\alpha}^{(M)}\right)\,.
	\end{aligned}
\end{equation}
The first and the third relations in system \eqref{Markov2} impose on the chords connecting strands $M$ and $M+1$ the constraints that they are diagonal in color indices, for example (where we use the fact $\varphi_{\beta}^{(M)}=\varphi_{\beta}^{(M+1)}$):
\begin{equation}
	\begin{aligned}
	&{}^{\rho}\Theta\scalebox{0.8}{$\left[\begin{array}{cc}
			\gamma & \beta\\
			M+1 & M
		\end{array}\right]$}\overset{\eqref{reference}}{=}\sum\lm_{\delta}\left({}^{\rho}\varphi_{\beta}^{(M)}\right)^{-1}{}^{\rho}\Theta\scalebox{0.8}{$\left[\begin{array}{cc}
		\delta & \beta\\
		{\bf\color{\mygreen}0} & M
		\end{array}\right]$}{}^{\rho}\Theta\scalebox{0.8}{$\left[\begin{array}{cc}
		\gamma & \delta\\
		M+1 & {\bf\color{\mygreen}0}
		\end{array}\right]$}\overset{\eqref{Markov2}}{=}\\
	&=\sum\lm_{\delta}\left({}^{\rho}\varphi_{\beta}^{(M)}\right)^{-1}\left({}^{\rho}\lambda_{\beta}^{(M)}\right)^{-1}{}^{\rho}\Theta\scalebox{0.8}{$\left[\begin{array}{cc}
		\delta & \beta\\
		{\bf\color{\mygreen}0} & M+1
		\end{array}\right]$}{}^{\rho}\Theta\scalebox{0.8}{$\left[\begin{array}{cc}
		\gamma & \delta\\
		M+1 & {\bf\color{\mygreen}0}
		\end{array}\right]$}=\left({}^{\rho}\lambda_{\beta}^{(M)}\right)^{-1}{}^{\rho}\Theta\scalebox{0.8}{$\left[\begin{array}{cc}
		\gamma & \beta\\
		M+1 & M+1
		\end{array}\right]$}\overset{\substack{\eqref{Kroneker}\\ {\rm Type}\;{\rm II}}}{=}\\
	&=\delta_{\gamma\beta}\frac{\left({}^{\rho}\varphi_{\gamma}^{(M+1)}\right)^{-2}-\left({}^{\rho}\varphi_{0}^{(M+1)}\right)^{-2}}{\left({}^{\rho}\varphi_{\beta}^{(M+1)}\right)^{-1}\left({}^{\rho}\varphi_{0}^{(M+1)}\right)^{-1}}\left({}^{\rho}\varphi_{\beta}^{(M+1)}\right)\,.
	\end{aligned}
\end{equation}

After excluding reference chords for strand $M+1$, in this parametrization equations \eqref{Markov2} reduce to two equations:
\begin{equation}
		{}^{\rho}\tilde\lambda_\beta^{(M)}\cdot {}^{\rho}\Theta\scalebox{0.8}{$\left[\begin{array}{cc}
				\delta & \beta\\
				{\bf \color{\mygreen} 0} & M
			\end{array}\right]$}={\bf R}_{\fB}\left({}^{\rho}\Theta\scalebox{0.8}{$\left[\begin{array}{cc}
				\delta & \beta\\
				{\bf\color{\mygreen} 0} & M
			\end{array}\right]$}\right),\quad \left({}^{\rho}\tilde\lambda_\alpha^{(M)}\right)^{-1}\cdot {}^{\rho}\Theta\scalebox{0.8}{$\left[\begin{array}{cc}
				\alpha & \delta\\
				M & {\bf\color{\mygreen}0}
			\end{array}\right]$}={\bf R}_{\fB}\left({}^{\rho}\Theta\scalebox{0.8}{$\left[\begin{array}{cc}
				\alpha & \delta\\
				M & {\bf\color{\mygreen} 0}
			\end{array}\right]$}\right)\,,
\end{equation}
where an \emph{effective} shift reads:
\begin{equation}
	{}^{\rho}\tilde\lambda_\alpha^{(M)}={}^{\rho}\lambda_\alpha^{(M)}{}^{\rho}\lambda_\alpha^{(M+1)}\left({}^{\rho}\varphi_{\alpha}^{(M+1)}\right)^{-2}\,.
\end{equation}
Therefore, the system of equations \eqref{ref_main} for a braid shifted by this Markov move is equivalent to the initial system with rescaled parameters.
Thus, we expect that if the initial system produces A-polynomials $A_{\gamma}(\lambda_{\alpha},\mu_{\beta})$, then the new system shifted by the Markov move produces A-polynomials $A_{\gamma}(\lambda_{\alpha}f_{\alpha},\mu_{\beta})$, where the functions $f_{\alpha}$ are monomials in $\mu_{\beta}$ that take framing into account.
In the case of the Markov move depicted in Fig.~\ref{fig:Markov}(b) $f_{\alpha}=\left({}^{\rho}\varphi_{\alpha}^{(M+1)}\right)^{-2}$.


\section{CG chords in different contexts}\label{sec:contexts}

\subsection{Knot contact homology}\label{sec:KCH}

The theory of knot contact homology \cite{ekholm_knot_2013,ng_framed_2008,ng_knot_2005a,ng_knot_2005b} has a natural setting in terms of topological string theory \cite{Aganagic:2012jb,Aganagic:2013jpa,Ekholm:2024ceb,Chauhan:2026pdv} when an intersection of stacks of Lagrangian D-branes is described by an embedding of a knot $K$ (or, more generically, a link) into $S^3$.
In this application topological strings naturally probe a unit cotangent bundle $U^*S^3$, which is contact, and the knot geometry inside it.

This geometry has the form of the so-called jet manifold $J^1(M)\cong T^*M\times \IR_z$ with contact form $\lambda=dz-\theta$.
The Reeb vector field $\xi_R$ is defined as $\lambda(\xi_R)=1$, $d\lambda(\xi_R,\cdot)=0$.
BPS strings in this setting resemble rigid non-vibrating chords and correspond to Reeb chords -- integral curves of $\xi_R$ starting and ending on some Legendrian D-brane leaves.
To a knot $K$ one associates a Legendrian immersion:
\begin{equation}
	\Lambda_K=\left\{(x,v)|x\in K,\;\langle v,T_xK\rangle=0\right\}\cap U^*S^3\,.
\end{equation}

In the original construction (see \cite{ng_topological_2014} for a comprehensive review) one considers a knot as a closed braid on the surface of a torus, which is embedded canonically into $S^3$.
This embedding allows one to construct both the Reeb chord elements for such a Legendrian immersion and the corresponding differential ($Q_{\rm BRST}$ for topological strings) as a sum over disk instantons \cite{Ekholm:2005:ContactHomology,Ekholm:2007:MorseFlowTrees}.
Suppose the braid is on $M$ strands.
In this case one distinguishes $M(M-1)$ Reeb chord elements $\theta_{km}$ of degree 0 (chords $a_{ij}$ in \cite{ng_topological_2014}) for $1\leq k,m \leq M$, $k\neq m$, as well as $M^2$ pairs of Reeb chord elements $\kappa_{km}^{I}$, $\kappa_{km}^{II}$ (chords $c_{ij}$ and $d_{ij}$ in \cite{ng_topological_2014}) of degree 1 for $1\leq k,m\leq M$, and other elements that we will not consider here.
Remarkably, the differential acting on $\kappa_{km}^{I,II}$ is given by polynomials in $\theta_{km}$ :
\begin{equation}\label{differential}
	d \kappa_{km}^I=\upsilon_I(\theta_{km}),\quad d\kappa_{km}^{II}=\upsilon_{II}(\theta_{km})\,.
\end{equation}

Let us return for a moment to our construction of CG chords for $\fg=\fs\fu_2$.
In this case there are no choices for $\rho$ other than $\rho=\Box$.
Also, in this case, the tensor product of any representation $V(j)$ of spin $j>0$ contains only two isotypical components, one of which we enumerate by index 1 and the other by index 0:
\begin{equation}
	\Box\otimes V(j)=\underbrace{V\left(j+\frac{1}{2}\right)}_1\;\oplus\;\underbrace{ V\left(j-\frac{1}{2}\right)}_0\,.
\end{equation}
Therefore, the only possible choice of color indices of CG chords is $\alpha,\beta=1$.
It is natural to identify CG chord variables with Reeb chords of degree 0:
\begin{equation}
	{}^{\Box}\Theta\scalebox{0.8}{$\left[\begin{array}{cc}
			1 & 1\\
			k & m
		\end{array}\right]$}=\theta_{km}\,.
\end{equation}
Moreover, the functions $\upsilon_{I,II}$ on the r.h.s. of differential \eqref{differential} coincide with our functions $\Upsilon_{I,II}$ in \eqref{main_eqs}:
\begin{tcolorbox}
	\begin{equation}
		\upsilon_I=\Upsilon_I,\quad \upsilon_{II}=\Upsilon_{II}\,.
	\end{equation}
\end{tcolorbox}

We believe that this identification of CG chords with Reeb chords in knot contact homology, together with the possibility of extending the notion of CG chords to arbitrary Lie (super)algebras $\fg$, will allow one to introduce ``coloring'' indices for Reeb chords as well (see also \cite{Ekholm:2024ceb,Chauhan:2026pdv} for some recent developments in this direction).
Presumably these indices would play the role of some novel Chan-Paton factors in topological string theory.

Furthermore, under this identification of CG and Reeb chords, nonlinear eigenvalue equation \eqref{eigen} becomes the r.h.s. of $d\theta_{km}$ under Legendrian differential $d$.
The analogue of the A-polynomial in knot contact homology is called an \emph{augmentation} polynomial in $\lambda$ and $\mu$, whose vanishing set is exactly an augmentation variety where constraints $\upsilon_I=\upsilon_{II}=0$ are satisfied.


\subsection{WZW conformal blocks} \label{sec:WZW} 

States in the Hilbert space of a 3d Chern-Simons theory with simple Lie gauge group $G$ on temporal slices cutting our knot diagrams horizontally correspond to conformal blocks in a 2d WZW model associated with Lie algebra $\fg$ \cite{Witten:1989,moore_seiberg_1990}.
Knot strands intersect the 2d temporal slice in a collection of 2d points with complex coordinates $x_a$.
We can associate with the elements of knot diagrams in Sec.~\ref{sec:RT} the corresponding elements of the WZW  model: OPE with 3-valent vertices, and conformal block monodromies with R-matrices.
In this sense 2d WZW theory is compatible with braided tensor categories of $U_q(\fg)$ \cite{etingof1998lectures}.
One is able to choose specific bases of conformal blocks where elements coincide literally with those of $U_q(\fg)$, once a specific basis is chosen.

Conformal blocks $\Psi$ in the WZW model with $M$ marked points are tensors in ${\rm Rep}(\fg)^{\otimes M}$ and satisfy Knizhnik-Zamolodchikov (KZ) equations \cite{Knizhnik:1984qf,DiFrancesco:1997,etingof1998lectures}:
\begin{equation}\label{KZ}
	\left(\p_{x_a}-\hbar\sum\lm_{b\neq a}\frac{\eta_{ij}\;\rho_a(t^i)\otimes \rho_b(t^j)}{x_a-x_b}\right)\Psi=0\,,
\end{equation}
where $t^i$ are generators of $\fg$, $\rho_a$ are respective representations, and $\eta_{ij}$ is the inverse Killing metric for $\fg$.
Quantum parameter $\hbar$ depends on the theory, and for the $\fs\fu_n$ WZW model at level $k$ it reads $\hbar=\left(k+n\right)^{-1}$.

CG chords could be defined in this setting in the following way.
Fix one distinguished puncture $x_0=z$ with associated primary field in representation $\rho$.
The KZ equations allow one to reconstruct OPE coefficients for the primary fields.
So, to define CG chords in this setting we simply use OPE coefficients for 3-valent vertices and parallel transport puncture $z$ along a trajectory prescribed by the braid with the help of the same KZ equations.

In this setting a discussion of the quasi-classical limit is more natural, as we can first consider the limit of large representations.
For the KZ equations this limit seems intuitively simpler as representations entering them are of $\fg$ rather than $U_q(\fg)$.
In the large representation limit, the matrix generators become simply commuting numbers, up to higher corrections, for Perelomov states \cite{simon1980classical,perelomov1972coherent,Perelomov1986}.
Here we review this effect explicitly solely for $SO(3)$ in App.~\ref{app:packet}.
If the highest weights of representations $\rho_a$ scale as $\vec w_a\sim\vec \omega_a/\hbar$, then up to $\hbar^{-\frac{1}{2}}$-corrections $\rho_a(t^i)\sim T^i_a/\hbar$ where $T^i_a$ are quasi-classical values of operators.
Then, for puncture $z$ the KZ equation becomes a linear system of ODEs in the limit $\hbar \to 0$:
\begin{equation}\label{flat}
	\p_z\Psi= \left(\sum\lm_a\frac{\eta_{ij}T^j_{a}}{z-x_a}\right)\rho(t^i)\;\Psi\,.
\end{equation}

In this setting CG chords can be treated as ``open'' spectral coordinates \cite{Grassi:2021wpw,Hollands:2019wbr,Hollands:2026rhz} reinterpreted in various contexts as Fock-Goncharov/Fenchel-Nielsen coordinates or their higher dimensional analogs for moduli spaces of flat $SL(n,\IC)$ connections like \eqref{flat}.
Surely, one could apply the spectral network machinery \cite{Gaiotto:2009hg,Gaiotto:2010be,Gaiotto:2011tf,Gaiotto:2012rg,Gaiotto:2012rg2} to construct exact WKB expressions for them.
In this framework our CG chords correspond to ``laminations'' -- objects connecting singularities with prescribed asymptotics near singularities.
Yet let us stress here again that this is solely a choice of coordinates; there are different choices of coordinates that are no less effective for evaluating various types of knot invariants and generic computations for $U_q(\fs\fu_n)$ \cite{Galakhov:2014aha,Galakhov:2014sha,Galakhov:2015fna,Galakhov:2015gza,Alekseev:2019klg,Alekseev:2019nzw}.

Another way to treat this limit is in the framework of known solutions to the KZ equations \eqref{KZ} in terms of the free field formalism \cite{doi:10.1142/S0217751X9000115X,etingof1998lectures,Mironov:2010zs,Dotsenko:1984:CAMP}.
In this setting actual weights of the primary fields are generated from the highest weight $w_a$ by screening operators as $w_a-H_a\omega_a$, where $H_a$ is the number of screening operators of type $a$, and $\omega_a$ are fundamental weights.
The quasi-classical limit in this setting is equivalent to the quasi-classical limit in matrix models \cite{Morozov:2010cq,Mironov:2011jn,Mironov:2010pi} at large $N$.
In this limit we can estimate the asymptotic behavior of the conformal block in the following way \cite{Zamolodchikov:1989}:
\begin{equation}
	\Psi\sim\exp\left(F(h_a)/\hbar+O(\hbar^0)\right)\,,
\end{equation}
where parameters $h_a=\hbar H_a$ remain finite in the limit $\hbar\to 0$, and function $F$ plays the role of a prepotential function in dual class-S theories \cite{Alday:2009aq}.
Weight shift operators ${}^{\rho}\lambda_{\alpha}^{(k)}$ modify parameters $h_a\to h_a + {}^{\rho,\alpha} p_a^{(k)}\hbar$.
Therefore, up to higher corrections in $\hbar$, the conformal blocks $\Psi$ are eigenfunctions of the operators ${}^{\rho}\lambda_{\alpha}^{(k)}$:
\begin{equation}
	{}^{\rho}\lambda_{\alpha}^{(k)}\times\Psi=\exp\left({}^{\rho,\alpha} p_a^{(k)}\frac{\p F}{\p h_a}+O(\hbar)\right)\times\Psi\,.
\end{equation}
So we conclude that CG chords become commuting quantities in the quasi-classical limit $\hbar\to 0$.


\subsection{3d TQFT}

\begin{figure}[ht!]
	\centering
	\begin{tikzpicture}
		\draw[draw=none, fill=white!80!gray] (0,0) -- (1,1) -- (4,1) to[out=200,in=20] cycle;
		\draw[draw=none, fill=white!80!blue] (0,0) -- (3,0) -- (4,1) to[out=200,in=20] cycle;
		\draw[thick, burgundy] (1,0.5) -- (1,1) (1,0.5) -- (0,0) (3,0.5) -- (3,0) (3,0.5) -- (4,1) (1,0.5) to[out=0,in=190] (4,1) (3,0.5) to[out=180,in=10] (0,0);
		\draw[thick] (0,0) to[out=20,in=200] (4,1);
		\draw[thick] (0,0) -- (3,0) -- (4,1) -- (1,1) -- cycle;
		\begin{scope}[shift={(1,0.5)}]
			\draw[ultra thick, \mygreen] (-0.1,-0.05) -- (0.1,0.05) (-0.1,0.05) -- (0.1,-0.05);
		\end{scope}
		\begin{scope}[shift={(3,0.5)}]
			\draw[ultra thick, \mygreen] (-0.1,-0.05) -- (0.1,0.05) (-0.1,0.05) -- (0.1,-0.05);
		\end{scope}
		\begin{scope}
			\begin{scope}[rotate=210]
				\draw (0,0) -- (0.5,0);
				\begin{scope}[rotate=15]
					\draw (0,0) -- (0.5,0);
				\end{scope}
				\begin{scope}[rotate=-15]
					\draw (0,0) -- (0.5,0);
				\end{scope}
			\end{scope}
		\end{scope}
		\begin{scope}[shift={(1,1)}]
			\begin{scope}[rotate=120]
				\draw (0,0) -- (0.5,0);
				\begin{scope}[rotate=15]
					\draw (0,0) -- (0.5,0);
				\end{scope}
				\begin{scope}[rotate=-15]
					\draw (0,0) -- (0.5,0);
				\end{scope}
			\end{scope}
		\end{scope}
		\begin{scope}[shift={(4,1)}]
			\begin{scope}[rotate=30]
				\draw (0,0) -- (0.5,0);
				\begin{scope}[rotate=15]
					\draw (0,0) -- (0.5,0);
				\end{scope}
				\begin{scope}[rotate=-15]
					\draw (0,0) -- (0.5,0);
				\end{scope}
			\end{scope}
		\end{scope}
		\begin{scope}[shift={(3,0)}]
			\begin{scope}[rotate=300]
				\draw (0,0) -- (0.5,0);
				\begin{scope}[rotate=15]
					\draw (0,0) -- (0.5,0);
				\end{scope}
				\begin{scope}[rotate=-15]
					\draw (0,0) -- (0.5,0);
				\end{scope}
			\end{scope}
		\end{scope}
		\begin{scope}[shift={(0,1.5)}]
			\draw[draw=none, fill=white!80!red] (3,0) -- (0,0) -- (1,1) to[out=300,in=120] cycle;
			\draw[draw=none, fill=white!80!green] (3,0) -- (4,1) -- (1,1) to[out=300,in=120] cycle;
			\draw[thick, burgundy] (1,0.5) -- (1,1) (1,0.5) -- (0,0) (3,0.5) -- (3,0) (3,0.5) -- (4,1) (1,0.5) to[out=0,in=150] (3,0) (3,0.5) to[out=180,in=330] (1,1);
			\draw[thick] (3,0) to[out=120,in=300] (1,1);
			\draw[thick] (0,0) -- (3,0) -- (4,1) -- (1,1) -- cycle;
			\begin{scope}[shift={(1,0.5)}]
				\draw[ultra thick, \mygreen] (-0.1,-0.05) -- (0.1,0.05) (-0.1,0.05) -- (0.1,-0.05);
			\end{scope}
			\begin{scope}[shift={(3,0.5)}]
				\draw[ultra thick, \mygreen] (-0.1,-0.05) -- (0.1,0.05) (-0.1,0.05) -- (0.1,-0.05);
			\end{scope}
			\begin{scope}
				\begin{scope}[rotate=210]
					\draw (0,0) -- (0.5,0);
					\begin{scope}[rotate=15]
						\draw (0,0) -- (0.5,0);
					\end{scope}
					\begin{scope}[rotate=-15]
						\draw (0,0) -- (0.5,0);
					\end{scope}
				\end{scope}
			\end{scope}
			\begin{scope}[shift={(1,1)}]
				\begin{scope}[rotate=120]
					\draw (0,0) -- (0.5,0);
					\begin{scope}[rotate=15]
						\draw (0,0) -- (0.5,0);
					\end{scope}
					\begin{scope}[rotate=-15]
						\draw (0,0) -- (0.5,0);
					\end{scope}
				\end{scope}
			\end{scope}
			\begin{scope}[shift={(4,1)}]
				\begin{scope}[rotate=30]
					\draw (0,0) -- (0.5,0);
					\begin{scope}[rotate=15]
						\draw (0,0) -- (0.5,0);
					\end{scope}
					\begin{scope}[rotate=-15]
						\draw (0,0) -- (0.5,0);
					\end{scope}
				\end{scope}
			\end{scope}
			\begin{scope}[shift={(3,0)}]
				\begin{scope}[rotate=300]
					\draw (0,0) -- (0.5,0);
					\begin{scope}[rotate=15]
						\draw (0,0) -- (0.5,0);
					\end{scope}
					\begin{scope}[rotate=-15]
						\draw (0,0) -- (0.5,0);
					\end{scope}
				\end{scope}
			\end{scope}
		\end{scope}
		\begin{scope}[shift={(6,0.75)}]
			\draw[dashed] (0,0) to[out=10,in=210] (4,1);
			\draw[thick] (0,0) -- (3,0) -- (4,1) -- (1,1) -- cycle;
			\draw[thick] (1,1) to[out=350,in=120] (3,0);
		\end{scope}
		\draw (4.5,0.5) to[out=0,in=180]  (5,1.25) (4.5,2) to[out=0,in=180]  (5,1.25);
		\draw[-stealth] (5,1.25) -- (5.5,1.25);
	\end{tikzpicture}
	\caption{Surface triangulations are formed by triplets of Stokes lines (red lines) starting from branching points (green crosses) of the spectral cover and running towards singularities.
	Gluing two such triangulations differing by a flip form a ``dumpling'' operator.
	}\label{fig:dumpling}
\end{figure}
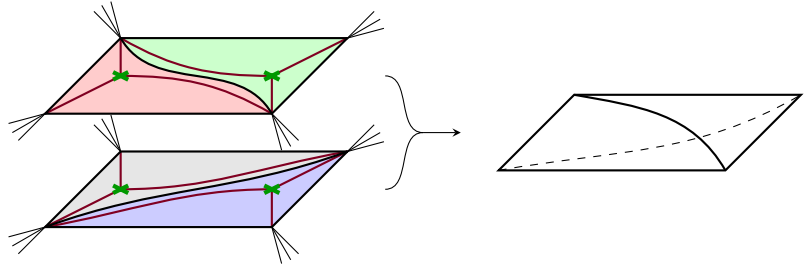

A transition from the 2d CFT to a 3d TQFT \cite{Dimofte:2009yn,Gaiotto:2011nm,Terashima:2013fg} in the quasi-classical limit is natural in terms of Fock-Goncharov coordinates associated with ideal triangulations \cite{Galakhov:2014aha}.
The exact WKB analysis of \eqref{flat} for $\fs\fu_2$ delivers a triangulation of the temporal slice by Stokes lines corresponding to triplets of triangle ``medians'' intersecting in a single point -- a branching point for the respective spectral curve.
Braiding the punctures of a conformal block results in a sequence of triangulation flips.
A single triangulation flip is interpreted as an evolution operator acting on spectral coordinates parametrizing a conformal block -- a state in the 3d Chern-Simons TQFT.
By ``raviolo''/``dumpling'' gluing\footnote{``Raviolo'' operators appear in \cite{Bullimore:2016hdc} in the context of Hecke modifications. } the initial and the flipped triangulations, one arrives at an association of this operator with a Chern-Simons path integral inside a tetrahedron (see Fig.~\ref{fig:dumpling}).
This is a reasonable conclusion for the $\fs\fu_2$ case since an expression for the flip evolution operator is given by the quantum dilogarithm, which behaves nicely $\sim e^{\frac{V}{\hbar}}$ in the limit $\hbar \to 0$; here $V$ is the hyperbolic volume of an ideal tetrahedron; this is precisely a saddle point approximation for the 3d $SL(2,\IC)$ Chern-Simons theory \cite{Gukov:2003na}.

This reasoning leads to a quasi-classical calculation of Jones polynomials for representations of large spin in the form of Hikami invariants \cite{Hikami:2001:HyperbolicStructure,Hikami:2007:GeneralizedVolume,Hikami:2014:ClusterAlgebra,Hikami:2014:BraidingOperator,Hikami:2015:Braids} where the knot complement $S^3\setminus K$ is glued from tetrahedra.
In this context CG chords are re-expressed in terms of Fock-Goncharov coordinates and evolve according to \eqref{Rmorph_beg}-\eqref{Rmorph_end} and \eqref{RRmorph_beg}-\eqref{RRmorph_end} as the knot braid evolves.

The situation seems much more complicated for the case $\fs\fu_{n>2}$, $SL(n,\IC)$ respectively, since Stokes lines accumulating in spectral networks can have quite a complicated, messy topology \cite{Galakhov:2013oja}.
However, \cite{Dimofte:2013iv} developed a canonical approach to constructing the corresponding A-polynomials based on considering canonical spectral networks in the form of so called ``snakes'' \cite{Gaiotto:2012rg2}.
We believe that our approach via CG chords and that of \cite{Dimofte:2013iv} are compatible, as we expect that CG chords \eqref{CGdef} could be expanded in terms of spectral coordinates in this case as well.
Yet the approach via CG chords might be slightly more advantageous as there is no need to rely on the existence of an auxiliary canonical spectral network construction.


\subsection{Link symbols}

Here we would like to discuss a relation between CG chord bases introduced in the present text and link symbol bases introduced in \cite{Galakhov:2024eco,Galakhov:2025ehn}.
As we have seen in Sec.~\ref{sec:braid_group}, one could try to untangle a strand colored by $\rho$ from a knot via skein relations \eqref{uni_skein} that would produce exactly CG chords.

To do this effectively, it is sufficient to introduce another ``hump'' relation (see e.g. \cite[eq. (4.9)]{Kirillov:1989:q6j} for a $\fs\fu_2$ example):
\begin{equation}
	\begin{array}{c}
		\begin{tikzpicture}
			\draw[thick, \myblue, -stealth] (0,0) to[out=330, in=180] (0.4,-0.3) to[out=0,in=270] (0.7,0.5);
			\draw[ultra thick] (-0.5,0.5) -- (0,0) -- (0,-0.5); 
			\node[left] at (-0.5,0.5) {$\scriptstyle Q^{\alpha}$};
			\node[below] at (0,-0.5) {$\scriptstyle Q$};
			\node[right] at (0.7,0.5) {$\scriptstyle \rho$};
			\node[right] at (0.1,-0.1) {$\scriptstyle \bar\rho$};
		\end{tikzpicture}
	\end{array}={}^{\rho}\eta_{\alpha}\hspace{-0.6cm}\begin{array}{c}
		\begin{tikzpicture}
			\draw[thick, -stealth, \myblue] (0,0) -- (0.5,0.5);
			\draw[ultra thick] (-0.5,0.5) -- (0,0) -- (0,-0.5);
			\node[left] at (-0.5,0.5) {$\scriptstyle Q^{\alpha}$};
			\node[below] at (0,-0.5) {$\scriptstyle Q$};
			\node[right] at (0.5,0.5) {$\scriptstyle \rho$};
		\end{tikzpicture}
	\end{array}=:{}^{\rho}\eta_{\alpha}\hspace{-0.3cm}\begin{array}{c}
	\begin{tikzpicture}
		\draw[thick, -stealth, \myblue] (0,0) -- (0.5,0.5);
		\draw[ultra thick] (-0.5,0.5) -- (0,0) -- (0,-0.5);
		\node[right] at (0,-0.5) {$\scriptstyle -\alpha$};
		\node[right] at (0.5,0.5) {$\scriptstyle \rho$};
	\end{tikzpicture}
	\end{array}\,,
\end{equation}
where we have introduced a new scalar function ${}^{\rho}\eta_{\alpha}$ and a new notation for a 3-valent vertex, so that it complies better with \eqref{notations}.
There is another apparent ``hump'' relation when representation $\rho$ is on the left:
\begin{equation}
	\begin{array}{c}
		\begin{tikzpicture}[xscale=-1]
			\draw[thick, \myblue, -stealth] (0,0) to[out=330, in=180] (0.4,-0.3) to[out=0,in=270] (0.7,0.5);
			\draw[ultra thick] (-0.5,0.5) -- (0,0) -- (0,-0.5); 
			\node[right] at (-0.5,0.5) {$\scriptstyle Q^{\alpha}$};
			\node[below] at (0,-0.5) {$\scriptstyle Q$};
			\node[left] at (0.7,0.5) {$\scriptstyle \rho$};
			\node[left] at (0.1,-0.1) {$\scriptstyle \bar\rho$};
		\end{tikzpicture}
	\end{array}={}^{\rho}\psi_{\alpha}\hspace{-0.6cm}\begin{array}{c}
		\begin{tikzpicture}[xscale=-1]
			\draw[thick, -stealth, \myblue] (0,0) -- (0.5,0.5);
			\draw[ultra thick] (-0.5,0.5) -- (0,0) -- (0,-0.5);
			\node[right] at (-0.5,0.5) {$\scriptstyle Q^{\alpha}$};
			\node[below] at (0,-0.5) {$\scriptstyle Q$};
			\node[left] at (0.5,0.5) {$\scriptstyle \rho$};
		\end{tikzpicture}
	\end{array}=:{}^{\rho}\psi_{\alpha}\hspace{-0.3cm}\begin{array}{c}
		\begin{tikzpicture}[xscale=-1]
			\draw[thick, -stealth, \myblue] (0,0) -- (0.5,0.5);
			\draw[ultra thick] (-0.5,0.5) -- (0,0) -- (0,-0.5);
			\node[right] at (0,-0.5) {$\scriptstyle -\alpha$};
			\node[left] at (0.5,0.5) {$\scriptstyle \rho$};
		\end{tikzpicture}
	\end{array}\,,
\end{equation}
however phases ${}^{\rho}\psi_{\alpha}$ and ${}^{\rho}\eta_{\alpha}$ are not independent; rather, other twist moves \eqref{uni_twist} impose an explicit relation on them.

Applying these relations to basic CG chords produces an apparent relation for chords in representations $\rho$ and $\bar\rho$:
\begin{equation}
	{}^{\rho}\Xi\scalebox{0.8}{$\left[\begin{array}{cc}
			\alpha & \beta\\
			k & m
		\end{array}\right]$}={}^{\bar\rho}\Xi\scalebox{0.8}{$\left[\begin{array}{cc}
		-\beta& -\alpha \\
		m & k
		\end{array}\right]$}\times\left\{\begin{array}{ll}
		\left({}^{\rho}\eta_{-\alpha}^{(k)}\right)^{-1}\left({}^{\rho}\psi_{\beta}^{(m)}\right),& \mbox{if }k<m\,,\\
		\left({}^{\rho}\psi_{-\alpha}^{(k)}\right)^{-1}\left({}^{\rho}\eta_{\beta}^{(m)}\right),& \mbox{if }k>m\,.
	\end{array}\right.
\end{equation}

First, we re-express the Hopf link in terms of twist phases:
\begin{equation}
	\scalebox{0.9}{$\begin{array}{c}
		\begin{tikzpicture}
			\draw[ultra thick] (0,-0.7) -- (0,0);
			\begin{scope}[yscale=0.5]
				\draw[thick, \myblue] ([shift={(0:0.5)}]0,0) arc (0:180:0.5);
				\draw[white, line width=2mm] ([shift={(180:0.5)}]0,0) arc (180:360:0.5);
				\draw[thick, \myblue, postaction={decorate},decoration={markings, mark= at position 0.8 with {\arrow{stealth}}}] ([shift={(180:0.5)}]0,0) arc (180:360:0.5);
			\end{scope}
			\draw[white, line width=2mm] (0,0) -- (0,0.7);
			\draw[ultra thick] (0,0) -- (0,0.7);
		\end{tikzpicture}
	\end{array}=\sum\lm_{\gamma}\left({}^\rho\varphi_\gamma\right)\begin{array}{c}
	\begin{tikzpicture}
		\draw[thick,\myblue, postaction={decorate},decoration={markings, mark= at position 0.5 with {\arrow{stealth}}}] (0,0) to[out=30,in=270] (0.5,0.3);
		\draw[thick, \myblue] (-0.5,0.3) to[out=270,in=180] (-0.25,-0.6) to[out=0,in=210] (0,-0.4);
		\draw[ultra thick] (0,-0.7) -- (0,0);
		\begin{scope}[shift={(0,0.3)}]
		\begin{scope}[yscale=0.5]
			\draw[thick, \myblue] ([shift={(0:0.5)}]0,0) arc (0:180:0.5);
		\end{scope}
		\end{scope}
		\draw[white, line width=2mm] (0,0.2) -- (0,0.7);
		\draw[ultra thick] (0,0) -- (0,0.7);
		\node[right] at (0,-0.2) {$\scriptstyle \gamma$};
	\end{tikzpicture}
	\end{array}=\sum\lm_{\gamma}\left({}^\rho\varphi_\gamma\right)^2\begin{array}{c}
	\begin{tikzpicture}
		\draw[thick, \myblue, postaction={decorate},decoration={markings, mark= at position 0.5 with {\arrow{stealth}}}] (0,0.4) to[out=150,in=0] (-0.25,0.6) to[out=180,in=90] (-0.5,0) to[out=270,in=180] (-0.25,-0.6) to[out=0,in=210] (0,-0.4);
		\draw[ultra thick] (0,-0.7) -- (0,0.7);
		\node[right] at (0,0) {$\scriptstyle \gamma$};
	\end{tikzpicture}
	\end{array}=\sum\lm_{\gamma}\left({}^\rho\varphi_\gamma\right)^2\left({}^{\rho}\psi_{\gamma}\right)^{-1}\begin{array}{c}
	\begin{tikzpicture}
		\draw[thick, \myblue, postaction={decorate},decoration={markings, mark= at position 0.5 with {\arrow{stealth}}}] (0,0.4) to[out=210,in=90] (-0.5,0) to[out=270,in=180] (-0.25,-0.6) to[out=0,in=210] (0,-0.4);
		\draw[ultra thick] (0,-0.7) -- (0,0.7);
		\node[right] at (0,0) {$\scriptstyle \gamma$};
	\end{tikzpicture}
	\end{array}=\sum\lm_{\gamma}\left({}^\rho\varphi_\gamma\right)^2\begin{array}{c}
	\begin{tikzpicture}
		\draw[thick, \myblue, postaction={decorate},decoration={markings, mark= at position 0.5 with {\arrow{stealth}}}] (0,0.4) to[out=210,in=90] (-0.3,0) to[out=270,in=150] (0,-0.4);
		\draw[ultra thick] (0,-0.7) -- (0,0.7);
		\node[right] at (0,0) {$\scriptstyle \gamma$};
	\end{tikzpicture}
	\end{array}=\sum\lm_{\gamma}\left({}^\rho\varphi_\gamma\right)^2\begin{array}{c}
	\begin{tikzpicture}
		\draw[ultra thick] (0,-0.7) -- (0,0.7);
	\end{tikzpicture}
	\end{array}$}\,.
\end{equation}

Acting in the same fashion for two other examples of basic link symbols \cite{Galakhov:2025ehn},  one derives:
\begin{equation}
	\begin{aligned}
		&\begin{array}{c}
			\begin{tikzpicture}
				\begin{scope}[yscale=0.5]
					\draw[thick, \myblue] ([shift={(0:0.7)}]0.5,0) arc (0:180:0.7);
				\end{scope}
				\draw[white, line width = 1.5mm] (0,-0.5) -- (0,0.5) (1,-0.5) -- (1,0.5);
				\draw[ultra thick] (0,-0.5) -- (0,0.5) (1,-0.5) -- (1,0.5);
				\begin{scope}[yscale=0.5]
					\draw[white, line width = 1.5mm] ([shift={(180:0.7)}]0.5,0) arc (180:360:0.7);
					\draw[thick, \myblue, postaction={decorate},decoration={markings, mark= at position 0.5 with {\arrow{stealth}}}] ([shift={(180:0.7)}]0.5,0) arc (180:360:0.7);
				\end{scope}
			\end{tikzpicture}
		\end{array} =\left({}^{\bar \rho}\varphi_0^{(1)}\right)^{-2}\sum\lm_{\gamma\neq 0}\left({}^{ \rho}\varphi_\gamma^{(2)}\right)^{2}+\sum\lm_{\gamma\neq 0}\left({}^{\bar \rho}\varphi_\gamma^{(1)}\right)^{-2}\left({}^{ \rho}\varphi_0^{(2)}\right)^{2}+\left({}^{\bar \rho}\varphi_0^{(1)}\right)^{-2}\left({}^{ \rho}\varphi_0^{(2)}\right)^{2}+\\
		&+\sum\lm_{\gamma\neq 0}\sum\lm_{\gamma'\neq 0}\left(\left({}^{\bar \rho}\varphi_\gamma^{(1)}\right)^{-2}-\left({}^{\bar \rho}\varphi_0^{(1)}\right)^{-2}\right)\left(\left({}^{ \rho}\varphi_\gamma^{(2)}\right)^{2}-\left({}^{ \rho}\varphi_0^{(2)}\right)^{2}\right)\left({}^{\rho}\eta_{\gamma}^{(1)}\right)\left({}^{\bar\rho}\eta_{\gamma}^{(1)}\right)^{-1}{}^{\rho}\Xi\scalebox{0.8}{$\left[\begin{array}{cc}
				\gamma & -\gamma'\\
				2 & 1
			\end{array}\right]$}{}^{\rho}\Xi\scalebox{0.8}{$\left[\begin{array}{cc}
			\gamma' & -\gamma\\
			1& 2
			\end{array}\right]$}\,.
	\end{aligned}
\end{equation}

\begin{equation}
	\begin{aligned}
		&\begin{array}{c}
			\begin{tikzpicture}
				\draw[ultra thick] (0,0) -- (0,-0.5) (1,0) -- (1,-0.5);
				\draw[white, line width = 1.5mm] (-0.2,0) to[out=270,in=180] (0.1,-0.25) to[out=0,in=180] (0.9,0.25) to[out=0,in=90] (1.2,0);
				\draw[thick, \myblue, postaction={decorate},decoration={markings, mark= at position 0.3 with {\arrow{stealth}}}] (-0.2,0) to[out=270,in=180] (0.1,-0.25) to[out=0,in=180] (0.9,0.25) to[out=0,in=90] (1.2,0);
				\begin{scope}[shift={(1,0)}]
					\begin{scope}[xscale=-1]
						\draw[white, line width = 1.5mm] (-0.2,0) to[out=270,in=180] (0.1,-0.25) to[out=0,in=180] (0.9,0.25) to[out=0,in=90] (1.2,0);
						\draw[thick, \myblue, postaction={decorate},decoration={markings, mark= at position 0.3 with {\arrow{stealth}}}] (-0.2,0) to[out=270,in=180] (0.1,-0.25) to[out=0,in=180] (0.9,0.25) to[out=0,in=90] (1.2,0);
					\end{scope}
				\end{scope}
				\draw[white, line width = 1.5mm] (0,0) -- (0,0.5) (1,0) -- (1,0.5);
				\draw[ultra thick] (0,0) -- (0,0.5) (1,0) -- (1,0.5);
			\end{tikzpicture} 
		\end{array}= \begin{array}{c}
		\begin{tikzpicture}
			\draw[ultra thick]  (1,0) -- (1,0.5);
			\begin{scope}[yscale=0.5]
				\draw[white, line width = 1.5mm] ([shift={(0:0.7)}]0.5,0) arc (0:180:0.7);
				\draw[thick, \myblue] ([shift={(0:0.7)}]0.5,0) arc (0:180:0.7);
			\end{scope}
			\draw[white, line width = 1.5mm] (0,-0.5) -- (0,0.5);
			\draw[ultra thick] (0,-0.5) -- (0,0.5);
			\begin{scope}[yscale=0.5]
				\draw[white, line width = 1.5mm] ([shift={(180:0.7)}]0.5,0) arc (180:360:0.7);
				\draw[thick, \myblue, postaction={decorate},decoration={markings, mark= at position 0.5 with {\arrow{stealth}}}] ([shift={(180:0.7)}]0.5,0) arc (180:360:0.7);
			\end{scope}
			\draw[white, line width = 1.5mm] (1,0) -- (1,-0.5);
			\draw[ultra thick] (1,0) -- (1,-0.5);
		\end{tikzpicture}
		\end{array} =\left({}^{\bar \rho}\varphi_0^{(1)}\right)^{-2}\sum\lm_{\gamma\neq 0}\left({}^{ \rho}\varphi_\gamma^{(2)}\right)^{-2}+\sum\lm_{\gamma\neq 0}\left({}^{\bar \rho}\varphi_\gamma^{(1)}\right)^{-2}\left({}^{ \rho}\varphi_0^{(2)}\right)^{-2}+\left({}^{\bar \rho}\varphi_0^{(1)}\right)^{-2}\left({}^{ \rho}\varphi_0^{(2)}\right)^{-2}+\\
		&+\sum\lm_{\gamma\neq 0}\sum\lm_{\gamma'\neq 0}\left(\left({}^{\bar \rho}\varphi_\gamma^{(1)}\right)^{-2}-\left({}^{\bar \rho}\varphi_0^{(1)}\right)^{-2}\right)\left(\left({}^{ \rho}\varphi_\gamma^{(2)}\right)^{-2}-\left({}^{ \rho}\varphi_0^{(2)}\right)^{-2}\right)\left({}^{\rho}\eta_{\gamma}^{(1)}\right)\left({}^{\bar\rho}\eta_{\gamma}^{(1)}\right)^{-1}{}^{\rho}\Xi\scalebox{0.8}{$\left[\begin{array}{cc}
				\gamma & -\gamma'\\
				2 & 1
			\end{array}\right]$}{}^{\rho}\Xi\scalebox{0.8}{$\left[\begin{array}{cc}
				\gamma' & -\gamma\\
				1& 2
			\end{array}\right]$}\,.
	\end{aligned}
\end{equation}
In the case $\fg=\fs\fu_2$, sums over $\gamma\neq 0$ and $\gamma'\neq 0$ reduce to a single term, so these link symbols are not independent; rather, they are related via the Kauffman bracket.

So we see clearly that link symbols are re-expressed directly in terms of basic CG chords.
Each of these approaches has its own advantages and disadvantages.
Link symbols are manifestly gauge invariant; they can be used with arbitrary knots and links, and there is no need to downgrade to the quasi-classical level, so one can work with actual quantum relations.
On the other hand, CG chords are simpler; the action of the braid group maps basic chords into polynomials of basic chords again.
Yet, to lift this consideration to the quantum level, one is forced to consider only ``colorless'' combinations of chords, that modify representations in the whole knot simultaneously, rather than in distinct strands of the braid.


\section{Examples for \texorpdfstring{$U_q(\fs\fu_3)$}{Uq(su3)}}\label{sec:examples}

\subsection{Some details on \texorpdfstring{$U_q(\fs\fu_3)$}{Uq(su3)}}
Quantum group $U_q(\fs\fu_3)$ is an example of quantum group $U_q(\fg)$ \eqref{QG} specified by the following Cartan matrix:
\begin{equation}
	a_{ij}=\left(\begin{array}{cc}
		2 & -1\\
		-1 & 2\\
	\end{array}\right)
\end{equation}

To perform computations with it, first we need to describe its parametrization, in particular, indices $\alpha$ in the isotypical decomposition and the twist functions \eqref{uni_twist}.
As representation $\rho$ coloring the CG diagrams we select the fundamental representation $\Box$ and the first antisymmetric representation, which is isomorphic to the anti-fundamental representation $\bar\Box$.

Fundamental representation $\Box$ reads:
\begin{equation}\label{box}
	\begin{aligned}
		&h_1=\left(\begin{array}{ccc}
			1 & 0 & 0\\
			0 & -1 & 0\\
			0 & 0 & 0\\
		\end{array}\right),\quad e_1=\left(\begin{array}{ccc}
			0 & 1 & 0\\
			0 & 0 & 0\\
			0 & 0 & 0\\
		\end{array}\right), \quad f_1=\left(\begin{array}{ccc}
			0 & 0 & 0\\
			1 & 0 & 0\\
			0 & 0 & 0\\
		\end{array}\right)\,,\\
		&h_2=\left(\begin{array}{ccc}
			0 & 0 & 0\\
			0 & 1 & 0\\
			0 & 0 & -1\\
		\end{array}\right),\quad e_2=\left(\begin{array}{ccc}
			0 & 0 & 0\\
			0 & 0 & 1\\
			0 & 0 & 0\\
		\end{array}\right), \quad f_2=\left(\begin{array}{ccc}
			0 & 0 & 0\\
			0 & 0 & 0\\
			0 & 1 & 0\\
		\end{array}\right)\,.
	\end{aligned}
\end{equation}

Anti-fundamental representation $\bar\Box$ reads:
\begin{equation}\label{anti-box}
	\begin{aligned}
		&h_1=\left(\begin{array}{ccc}
			0 & 0 & 0\\
			0 & 1 & 0\\
			0 & 0 & -1\\
		\end{array}\right),\quad e_1=\left(\begin{array}{ccc}
			0 & 0 & 0\\
			0 & 0 & 1\\
			0 & 0 & 0\\
		\end{array}\right), \quad f_1=\left(\begin{array}{ccc}
			0 & 0 & 0\\
			0 & 0 & 0\\
			0 & 1 & 0\\
		\end{array}\right)\,,\\
		&h_2=\left(\begin{array}{ccc}
			1 & 0 & 0\\
			0 & -1 & 0\\
			0 & 0 & 0\\
		\end{array}\right),\quad e_2=\left(\begin{array}{ccc}
			0 & 1 & 0\\
			0 & 0 & 0\\
			0 & 0 & 0\\
		\end{array}\right), \quad f_2=\left(\begin{array}{ccc}
			0 & 0 & 0\\
			1 & 0 & 0\\
			0 & 0 & 0\\
		\end{array}\right)\,.
	\end{aligned}
\end{equation}

Since $\Box$ and $\bar\Box$ occur only in a single channel, only monomials $e_a$, $f_a$, $e_ae_b$, $f_af_b$ for $a\neq b$ survive;
therefore the universal R-matrix series \eqref{uni_R-mat} truncate:
\begin{equation}
	\begin{aligned}
		&\check R=e^{-\frac{1}{2}a_{ij}^{-1}h_i\otimes h_j}\Big(1-(q-q^{-1})f_1\otimes e_1-(q-q^{-1})f_2\otimes e_2+\\
		&+(q-q^{-1})\left(q(f_1f_2)\otimes(e_1e_2)-(f_2f_1)\otimes(e_1e_2)-(f_1f_2)\otimes(e_2e_1)+q(f_2f_1)\otimes(e_2e_1)\right)\Big)e^{-\frac{1}{2}a_{ij}^{-1}h_i\otimes h_j}\,,\\
		&\check R^{-1}=e^{\frac{1}{2}a_{ij}^{-1}h_i\otimes h_j}\Big(1+(q-q^{-1})f_1\otimes e_1+(q-q^{-1})f_2\otimes e_2-\\
		&-(q-q^{-1})\left(q^{-1}(f_1f_2)\otimes(e_1e_2)-(f_2f_1)\otimes(e_1e_2)-(f_1f_2)\otimes(e_2e_1)+q^{-1}(f_2f_1)\otimes(e_2e_1)\right)\Big)e^{\frac{1}{2}a_{ij}^{-1}h_i\otimes h_j}\,.
	\end{aligned}
\end{equation}

We label representations of $\fs\fu_3$ by the highest weights $(w_1,w_2)$.
The fundamental and the anti-fundamental ones are labeled accordingly $\Box=(1,0)$, $\bar\Box=(0,1)$.
If $w_i\neq 0$ we have the following isotypical decompositions:
\begin{equation}
	\begin{aligned}
		& (w_1,w_2)\otimes \Box=\underbrace{(w_1+1,w_2)}_1\;\oplus\;\underbrace{(w_1-1,w_2+1)}_0\;\oplus\;\underbrace{(w_1,w_2-1)}_2\,,\\
		& (w_1,w_2)\otimes \bar\Box=\underbrace{(w_1,w_2+1)}_2\;\oplus\;\underbrace{(w_1+1,w_2-1)}_0\;\oplus\;\underbrace{(w_1-1,w_2)}_1\,,
	\end{aligned}
\end{equation}
where we have marked a desired choice of indices $\alpha$ for isotypical components (see Sec.~\ref{sec:chords} for details).
As an ``uninvited'' component in both cases we choose the intermediate channel, which implies that index 0 would never appear as a label of the ends of CG chords: $\alpha,\beta\neq 0$ in \eqref{CGdef}.
With these notations, the shifts of representations entering \eqref{main_eqs} can be rewritten in terms of weight shifts as follows:
\begin{equation}
	{}^{\Box}\lambda_1=\lambda_1,\quad {}^{\Box}\lambda_2=\lambda_2^{-1},\quad {}^{\bar \Box}\lambda_1=\lambda_1^{-1},\quad {}^{\bar\Box}\lambda_2=\lambda_2\,.
\end{equation}

Let us denote vectors in the bases \eqref{box} and \eqref{anti-box} as $u_i$ and $\bar u_i$, $i=1,2,3$ respectively.
We denote the highest weight vector of the irreducible representation $(w_1,w_2)$ by $|w_1,w_2\rangle$.

Then we construct the following useful vectors, assuming that $\vec w$ is in general position:
\begin{equation}\label{v_space}
	\begin{aligned}
		& v_0=|w_1,w_2\rangle,\quad (h_1,h_2)v_0=(w_1,w_2)v_0\,,\\
		& v_1=f_1|w_1,w_2\rangle,\quad (h_1,h_2)v_1=(w_1-2,w_2+1)v_1\,,\\
		& v_2=f_2|w_1,w_2\rangle,\quad (h_1,h_2)v_2=(w_1+1,w_2-2)v_2\,,\\
		& v_{12}=f_1f_2|w_1,w_2\rangle,\quad (h_1,h_2)v_{12}=(w_1-1,w_2-1)v_{12}\,,\\
		& v_{21}=f_2f_1|w_1,w_2\rangle,\quad (h_1,h_2)v_{21}=(w_1-1,w_2-1)v_{21}\,.
	\end{aligned}
\end{equation}

The respective Gram matrix reads:
\begin{equation}
	\Gamma=\left(\begin{array}{ccccc}
		1 & 0 & 0 & 0 & 0 \\
		0 & [w_1]_q & 0 & 0 & 0\\
		0 & 0 & [w_2]_q & 0 & 0\\
		0 & 0 & 0 & [w_1+1]_q[w_2]_q & [w_1]_q[w_2]_q\\
		0 & 0 & 0 & [w_1]_q[w_2]_q & [w_1]_q[w_2+1]_q
	\end{array}\right)\,.
\end{equation}

Thus we can construct the following effective representation operators, truncated to the subspace spanned by \eqref{v_space}:
\begin{equation}
	\begin{aligned}
		& h_1=\left(\begin{array}{ccccc}
			w_1 & 0 & 0 & 0 & 0 \\
			0 & w_1-2 & 0 & 0 & 0\\
			0 & 0 & w_1+1& 0 & 0\\
			0 & 0 & 0 & w_1-1 & 0\\
			0 & 0 & 0 & 0 & w_1-1
		\end{array}\right),\quad h_2=\left(\begin{array}{ccccc}
			w_2 & 0 & 0 & 0 & 0 \\
			0 & w_2+1 & 0 & 0 & 0\\
			0 & 0 & w_2-2& 0 & 0\\
			0 & 0 & 0 & w_2-1 & 0\\
			0 & 0 & 0 & 0 & w_2-1
		\end{array}\right)\,,\\
		& e_1=\left(\begin{array}{ccccc}
			0 & [w_1]_q & 0 & 0 & 0 \\
			0 & 0 & 0 & 0 & 0\\
			0 & 0 & 0 & [w_1+1]_q & [w_1]_q\\
			0 & 0 & 0 & 0 & 0\\
			0 & 0 & 0 & 0 & 0
		\end{array}\right),\quad e_2=\left(\begin{array}{ccccc}
			0 & 0 & [w_2]_q & 0 & 0 \\
			0 & 0 & 0 & [w_2]_q & [w_2+1]_q\\
			0 & 0 & 0 & 0 & 0\\
			0 & 0 & 0 & 0 & 0\\
			0 & 0 & 0 & 0 & 0
		\end{array}\right)\,,\\
		& f_1=\left(\begin{array}{ccccc}
			0 & 0 & 0 & 0 & 0 \\
			1 & 0 & 0 & 0 & 0\\
			0 & 0 & 0 & 0 & 0\\
			0 & 0 & 1 & 0 & 0\\
			0 & 0 & 0 & 0 & 0
		\end{array}\right),\quad f_2=\left(\begin{array}{ccccc}
			0 & 0 & 0 & 0 & 0 \\
			0 & 0 & 0 & 0 & 0\\
			1 & 0 & 0 & 0 & 0\\
			0 & 0 & 0 & 0 & 0\\
			0 & 1 & 0 & 0 & 0
		\end{array}\right)\,.
	\end{aligned}
\end{equation}

We construct the highest weight vectors in various isotypical components of $\Box\otimes\rho$ by imposing eigenvalue constraints for the Cartan operators and annihilation under $\Delta(e_{1,2})$:
\begin{equation}
	\begin{aligned}
		&|w_1+1,w_2\rangle_{\Box\otimes\rho}=u_1\otimes v_0\,,\\
		&|w_1-1,w_2+1\rangle_{\Box\otimes\rho}=u_1\otimes v_1-q^{-\frac{1+w_1}{2}}[w_1]_qu_2\otimes v_0\,,\\
		&|w_1,w_2-1\rangle_{\Box\otimes\rho}=q^{-1-\frac{w_1}{2}} \left([w_1]_q [w_2]_q-[w_1+1]_q[w_2+1]_q\right)\left(q^{-\frac{1+w_2}{2}} [w_2]_qu_3\otimes v_0-u_2\otimes v_2\right)-\\
		&\hspace{5cm}-[w_2+1]_qu_1\otimes v_{12}+[w_2]_qu_1\otimes v_{21}\,.
	\end{aligned}
\end{equation}

For the opposite order $\rho\otimes \Box$ we have:
\begin{equation}
	\begin{aligned}
		&|w_1+1,w_2\rangle_{\rho\otimes \Box}=v_0\otimes u_1\,,\\
		&|w_1-1,w_2+1\rangle_{\rho\otimes \Box}=v_1\otimes u_1-q^{\frac{1+w_1}{2}}[w_1]_q v_0\otimes u_2\,,\\
		&|w_1,w_2-1\rangle_{\rho\otimes \Box}=q^{1+\frac{w_1}{2}} \left([w_1]_q [w_2]_q-[w_1+1]_q[w_2+1]_q\right)\left(q^{\frac{1+w_2}{2}} [w_2]_q v_0\otimes u_3- v_2\otimes u_2\right)-\\
		&\hspace{5cm}-[w_2+1]_q v_{12}\otimes u_1+[w_2]_q v_{21}\otimes u_1\,.
	\end{aligned}
\end{equation}

Using these relations we derive the following twists by applying the R-matrix to $\Box\otimes \rho$:
\begin{equation}
	{}^{\Box}\varphi_{1}=q^{-\frac{2 w_1}{3}-\frac{w_2}{3}},\quad {}^{\Box}\varphi_{0}=q^{1+\frac{w_1}{3}-\frac{w_2}{3}},\quad  {}^{\Box}\varphi_{2}=q^{2+\frac{w_1}{3}+\frac{2 w_2}{3}}\,.
\end{equation}

Now we apply the same machinery to the anti-fundamental representation $\bar\Box$ for the highest weight vectors in the product $\bar\Box\otimes\rho$:
\begin{equation}
	\begin{aligned}
		&|w_1,w_2+1\rangle_{\bar\Box\otimes\rho}=\bar u_1\otimes v_0\,,\\
		&|w_1+1,w_2-1\rangle_{\bar\Box\otimes\rho}=\bar u_1\otimes v_2-q^{-\frac{1+w_2}{2}}[w_2]_q\bar u_2\otimes v_0\,,\\
		&|w_1-1,w_2\rangle_{\bar\Box\otimes\rho}=q^{-1-\frac{w_2}{2}} \left([w_1]_q [w_2]_q-[w_1+1]_q[w_2+1]_q\right)\left(q^{-\frac{1+w_1}{2}} [w_1]_q\bar u_3\otimes v_0-\bar u_2\otimes v_1\right)-\\
		&\hspace{5cm}-[w_1+1]_q\bar u_1\otimes v_{21}+[w_1]_q\bar u_1\otimes v_{12}\,.
	\end{aligned}
\end{equation}

For the opposite order $\rho\otimes \bar \Box$ we have:
\begin{equation}
	\begin{aligned}
		&|w_1,w_2+1\rangle_{\rho\otimes \bar\Box}=v_0\otimes \bar u_1\,,\\
		&|w_1+1,w_2-1\rangle_{\rho\otimes \bar\Box}=v_2\otimes \bar u_1-q^{\frac{1+w_2}{2}}[w_2]_q v_0\otimes \bar u_2\,,\\
		&|w_1-1,w_2\rangle_{\rho\otimes\bar \Box}=q^{1+\frac{w_2}{2}} \left([w_1]_q [w_2]_q-[w_1+1]_q[w_2+1]_q\right)\left(q^{\frac{1+w_1}{2}} [w_1]_q v_0\otimes\bar u_3- v_1\otimes\bar u_2\right)-\\
		&\hspace{5cm}-[w_1+1]_q v_{21}\otimes\bar u_1+[w_1]_q v_{12}\otimes\bar u_1\,.
	\end{aligned}
\end{equation}

Eventually, we arrive at the following anti-fundamental twist factors:
\begin{equation}
	{}^{\bar\Box}\varphi_{2}=q^{-\frac{w_1}{3}-\frac{2 w_2}{3}},\quad {}^{\bar\Box}\varphi_{0}=q^{1-\frac{w_1}{3}+\frac{w_2}{3}},\quad  {}^{\bar\Box}\varphi_{1}=q^{2+\frac{2 w_1}{3}+\frac{w_2}{3}}\,.
\end{equation}

In the quasi-classical limit, we introduce finite parameters as follows:
\begin{equation}
	\mu_1:=q^{\frac{1}{3}w_1},\quad \mu_2:=q^{\frac{1}{3}w_2}\,,
\end{equation}
so that
\begin{equation}
	{}^{\Box}\varphi_{1}=\mu_1^{-2}\mu_2^{-1},\quad {}^{\Box}\varphi_{0}=\mu_1\mu_2^{-1},\quad  {}^{\Box}\varphi_{2}=\mu_1\mu_2^2\,.
\end{equation}


\subsection{Trefoil knot}

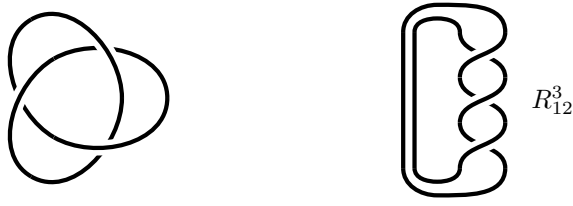
\begin{figure}[ht!]
	\centering
	\begin{tikzpicture}
		\node (A) at (0,0) {$\begin{array}{c}
				\begin{tikzpicture}[scale=0.6]
					\foreach \i in {0, ..., 2} {
						\begin{scope}[rotate=120*\i]
							\draw[ultra thick] (2,0) to[out=90,in=30] (-0.5, 0.866025);
							\draw[white, line width = 2mm] (1,0) to[out=90,in=30] (-1., 1.73205);
							\draw[ultra thick] (1,0) to[out=90,in=30] (-1., 1.73205);
						\end{scope}
					}
				\end{tikzpicture}
			\end{array}$};
		\node(B) at (5,0) {$\begin{tikzpicture}[scale=0.6]
				\foreach \i in {0, ..., 2} {
					\begin{scope}[shift={(0,\i)}]
						\draw[ultra thick] (1,0) to[out=90,in=270] (0,1);
						\draw[white, line width = 2mm] (0,0) to[out=90,in=270] (1,1);
						\draw[ultra thick] (0,0) to[out=90,in=270] (1,1);
					\end{scope}
				}
				\draw[ultra thick] (0,0) to[out=270,in=0] (-0.5,-0.3) to[out=180,in=270] (-1,0) (1,0) to[out=270,in=0] (-0.5,-0.6) to[out=180,in=270] (-1.25,0);
				\begin{scope}[shift={(0,3)}]
					\begin{scope}[yscale=-1]
						\draw[ultra thick] (0,0) to[out=270,in=0] (-0.5,-0.3) to[out=180,in=270] (-1,0) (1,0) to[out=270,in=0] (-0.5,-0.6) to[out=180,in=270] (-1.25,0);
					\end{scope}
				\end{scope}
				\draw[ultra thick] (-1,0) -- (-1,3) (-1.25,0) -- (-1.25,3);
			\end{tikzpicture}$};
		\node[right] at (B.east) {$R_{12}^3$};
	\end{tikzpicture}
	\caption{A trefoil knot $3_1$ diagram and its braid representation.}\label{fig:3_1}
\end{figure}

We start with the trefoil knot $3_1$ depicted in Fig.~\ref{fig:3_1}.
This knot has a braid representation as $R_{12}^3$.

To abbreviate the notation a bit, we substitute the CG chords with variables $X_i$ according to the following rule:
\begin{equation}
	\begin{aligned}
	&{}^{\Box}\Theta\scalebox{0.8}{$\left[\begin{array}{cc}
			1 & 1\\
			1 & 2
		\end{array}\right]$}=:X_1,\; 
	{}^{\Box}\Theta\scalebox{0.8}{$\left[\begin{array}{cc}
			1 & 2\\
			1 & 2
		\end{array}\right]$}=:X_2,\;
	{}^{\Box}\Theta\scalebox{0.8}{$\left[\begin{array}{cc}
			2 & 1\\
			1 & 2
		\end{array}\right]$}=:X_3,\;
	{}^{\Box}\Theta\scalebox{0.8}{$\left[\begin{array}{cc}
			2 & 2\\
			1 & 2
		\end{array}\right]$}=:X_4\,,\\
	&{}^{\Box}\Theta\scalebox{0.8}{$\left[\begin{array}{cc}
			1 & 1\\
			2 & 1
		\end{array}\right]$}=:X_5,\;
	{}^{\Box}\Theta\scalebox{0.8}{$\left[\begin{array}{cc}
			1 & 2\\
			2 & 1
		\end{array}\right]$}=:X_6,\;
	{}^{\Box}\Theta\scalebox{0.8}{$\left[\begin{array}{cc}
			2 & 1\\
			2 & 1
		\end{array}\right]$}=:X_7,\;
	{}^{\Box}\Theta\scalebox{0.8}{$\left[\begin{array}{cc}
			2 & 2\\
			2 & 1
		\end{array}\right]$}=:X_8\,.
	\end{aligned}
\end{equation}
The system of equations \eqref{main_eqs} for these variables reads:
\begin{subequations}
\begin{equation}
	-X_1 \mu _1^9+2 X_1 \mu _1^3+X_1^2 X_5 \mu _1^3+X_2 X_3X_8\mu _1^9\mu _2^6+X_1 X_3 X_6 \mu _1^6 \mu _2^3+X_1 X_2 X_7 \mu _1^6 \mu _2^3+\left(\mu _2^3 -\mu _1^6 \mu _2^3\right) \lambda _1^{(1)}=0\,,
\end{equation}
\begin{equation}
	\mu _2^6 \mu _1^6 X_2+\mu _2^6 \mu _1^6 X_2 X_4 X_8-\mu _1^6 X_2+\mu _2^3 \mu _1^3 X_1 X_4 X_6+\mu _2^3 \mu _1^3 X_2^2 X_7+X_2+X_1 X_2 X_5=0\,,
\end{equation}
\begin{equation}
	X_3 \mu _2^6 \mu _1^8+X_3 X_4 X_8 \mu _2^6 \mu _1^8-X_3 \mu _1^8+X_3^2 X_6 \mu _2^3 \mu _1^5+X_1 X_4 X_7 \mu _2^3 \mu _1^5+X_3 \mu _1^2+X_1 X_3 X_5 \mu _1^2=0\,,
\end{equation}
\begin{equation}
	2 \mu _2^6 \mu _1^6 X_4+\mu _2^6 \mu _1^6 X_4^2 X_8-\mu _1^6 X_4+\mu _2^3 \mu _1^3 X_3 X_4 X_6+\mu _2^3 \mu _1^3 X_2 X_4 X_7+X_2 X_3 X_5+\left(\mu _1^3 \mu _2^9-\mu _1^3 \mu _2^3\right)\left(\lambda _2^{(1)}\right)^{-1}=0\,,
\end{equation}
\begin{equation}
	\begin{aligned}
		&\mu _1^9-\mu _1^3-\lambda _1^{(2)} \mu _2^3 X_1-\mu _2^9 \mu _1^{12} X_2 X_7-\mu _2^9 \mu _1^{12} X_2 X_4 X_7 X_8+\mu _2^3 \mu _1^{12} X_2 X_7-\mu _2^6 \mu _1^9 X_2^2 X_7^2-\\
		&-\mu _2^6 \mu _1^9 X_1 X_4 X_6 X_7-\mu _2^6 \mu _1^9 X_2 X_3 X_5 X_8+\mu _1^9 X_1 X_5-\mu _2^3 \mu _1^6 X_1 X_3 X_5 X_6-2 \mu _2^3 \mu _1^6 X_2 X_7-\\
		&-2 \mu _2^3 \mu _1^6 X_1 X_2 X_5 X_7-\mu _1^3 X_1^2 X_5^2-3 \mu _1^3 X_1 X_5=0\,,
	\end{aligned}
\end{equation}
\begin{equation}
	\begin{aligned}
		&-\mu _2^9 \mu _1^9 X_2 X_4 X_8^2-2 \mu _2^9 \mu _1^9 X_2 X_8+\mu _2^3 \mu _1^9 X_2 X_8-\mu _2^6 \mu _1^6 X_1 X_6-\mu _2^6 \mu _1^6 X_2 X_3 X_6 X_8-\mu _2^6 \mu _1^6 X_1 X_4 X_6 X_8-\\
		&-\mu _2^6 \mu _1^6 X_2^2 X_7 X_8+\mu _1^6 X_1 X_6-\mu _2^3 \mu _1^3 X_1 X_3 X_6^2-\mu _2^3 \mu _1^3 X_1 X_2 X_6 X_7-\mu _2^3 \mu _1^3 X_2 X_8-\mu _2^3 \mu _1^3 X_1 X_2 X_5 X_8-\\
		&-2 X_1 X_6-X_1^2 X_5 X_6 -\mu _1^6 \mu _2^{12} X_2\left(\lambda _2^{(2)}\right)^{-1}=0\,,
	\end{aligned}
\end{equation}
\begin{equation}
	\begin{aligned}
		&-\lambda _1^{(2)} \mu _2^3 X_3-2 \mu _2^9 \mu _1^{12} X_4 X_7-\mu _2^9 \mu _1^{12} X_4^2 X_7 X_8+\mu _2^3 \mu _1^{12} X_4 X_7-\mu _2^6 \mu _1^9 X_2 X_4 X_7^2-\mu _2^6 \mu _1^9 X_3 X_5-\\
		&-\mu _2^6 \mu _1^9 X_3 X_4 X_6 X_7-\mu _2^6 \mu _1^9 X_3 X_4 X_5 X_8+\mu _1^9 X_3 X_5-\mu _2^3 \mu _1^6 X_3^2 X_5 X_6-\mu _2^3 \mu _1^6 X_4 X_7-\mu _2^3 \mu _1^6 X_2 X_3 X_5 X_7-\\
		&-\mu _2^3 \mu _1^6 X_1 X_4 X_5 X_7-\mu _1^3 X_1 X_3 X_5^2-2 \mu _1^3 X_3 X_5=0\,,
	\end{aligned}
\end{equation}
\begin{equation}
	\begin{aligned}
		&-\mu _2^9 \mu _1^9+\mu _2^3 \mu _1^9-\mu _2^9 \mu _1^9 X_4^2 X_8^2-3 \mu _2^9 \mu _1^9 X_4 X_8+\mu _2^3 \mu _1^9 X_4 X_8-2 \mu _2^6 \mu _1^6 X_3 X_6-2 \mu _2^6 \mu _1^6 X_3 X_4 X_6 X_8-\\
		&-\mu _2^6 \mu _1^6 X_2 X_4 X_7 X_8+\mu _1^6 X_3 X_6-\mu _2^3 \mu _1^3 X_3^2 X_6^2-\mu _2^3 \mu _1^3 X_1 X_4 X_6 X_7-\mu _2^3 \mu _1^3 X_2 X_3 X_5 X_8-X_3 X_6-\\
		&-X_1 X_3 X_5 X_6-\mu _1^6 \mu _2^{12} X_4\left(\lambda _2^{(2)}\right)^{-1}=0\,,
	\end{aligned}
\end{equation}
\begin{equation}\label{3_1_eq_9}
	\mu _1^6-X_3 X_6 \mu _2^3 \mu _1^3-X_5 \mu _2^3 \lambda _1^{(1)} \mu _1^3-X_1 X_5-1=0
\end{equation}
\begin{equation}
	-\lambda _2^{(1)} \mu _1^3 \mu _2^3 X_4 X_6-\lambda _2^{(1)} X_2 X_5-\mu _1^6 \mu _2^6 X_6=0\,,
\end{equation}
\begin{equation}
	-X_3 X_8 \mu _1^6 \mu _2^6-X_1 X_7 \mu _1^3 \mu _2^3-X_7 \lambda _1^{(1)}=0
\end{equation}
\begin{equation}\label{3_1_eq_12}
	-\mu _1^3 \mu _2^6+\mu _1^3-\mu _1^3 \mu _2^6 X_4 X_8-\mu _2^3 X_2 X_7-\mu _2^3 X_8 \left(\lambda _2^{(1)}\right)^{-1}=0\,,
\end{equation}
\begin{equation}
	\mu _2^6 \mu _1^9 X_4 X_6 X_7-\mu _1^9 X_5+\mu _2^3 \mu _1^6 X_3 X_5 X_6+\mu _2^3 \mu _1^6 X_2 X_5 X_7+\mu _1^3 X_1 X_5^2+2 \mu _1^3 X_5+\left(\mu _2^3-\mu _1^6 \mu _2^3\right)\lambda _1^{(2)}=0\,,
\end{equation}
\begin{equation}
	\mu _2^6 \mu _1^6 X_6+\mu _2^6 \mu _1^6 X_4 X_6 X_8-\mu _1^6 X_6+\mu _2^3 \mu _1^3 X_3 X_6^2+\mu _2^3 \mu _1^3 X_2 X_5 X_8+X_1 X_5 X_6+X_6=0\,,
\end{equation}
\begin{equation}
	\mu _2^8 \mu _1^{10} X_7+\mu _2^8 \mu _1^{10} X_4 X_7 X_8-\mu _2^2 \mu _1^{10} X_7+\mu _2^5 \mu _1^7 X_2 X_7^2+\mu _2^5 \mu _1^7 X_3 X_5 X_8+\mu _2^2 \mu _1^4 X_1 X_5 X_7+\mu _2^2 \mu _1^4 X_7=0\,,
\end{equation}
\begin{equation}
	\mu _2^6 \mu _1^6 X_4 X_8^2+2 \mu _2^6 \mu _1^6 X_8-\mu _1^6 X_8+\mu _2^3 \mu _1^3 X_3 X_6 X_8+\mu _2^3 \mu _1^3 X_2 X_7 X_8+X_1 X_6 X_7+\left(\mu _1^3 \mu _2^9-\mu _1^3 \mu _2^3\right)\left(\lambda _2^{(2)}\right)^{-1}=0\,.
\end{equation}
\end{subequations}
It is relatively simple to solve for such a small knot.
At the first step we solve \eqref{3_1_eq_9}-\eqref{3_1_eq_12} for $X_5$, $X_6$, $X_7$, $X_8$ exploiting their simplicity and linearity in these variables.
Then the remaining equation factorizes partially, delivering one root with $X_2=0$, $X_3=0$ and another root with both $X_2$ and $X_3$ nonzero.

The first option leads to four roots.

The first root reads:
\begin{equation}
	\begin{aligned}
	X_1=-\frac{\lambda _1^{(1)} \mu _2^3}{\mu _1^3},\; X_2=0,\; X_3=0,\; X_4=-\frac{\mu _2^3}{\lambda _2^{(1)} \mu _1^3},\; X_5=\frac{\mu _1^3}{\lambda _1^{(1)} \mu _2^3},\; X_6=0,\; X_7=0,\; X_8=\frac{\lambda _2^{(1)} \mu _1^3}{\mu _2^3}\,.
	\end{aligned}
\end{equation}
The respective first constraint reads:
\begin{equation}\label{3_1_root_1}
	\lambda_1=-\frac{\mu _1^6}{\mu _2^6},\quad \lambda_2=-\frac{\mu _2^6}{\mu _1^6}\,.
\end{equation}

Let us note that even though the CG chord variables themselves may depend on gauge-non-invariant operators $\lambda_a^{(k)}$ that shift representations only in strand $k$, the constraints contain only gauge invariant operators $\lambda_a$ shifting representations of the whole knot.

The second root reads:
\begin{equation}
X_1=-\frac{\lambda _1^{(1)} \mu _2^3}{\mu _1^3},\; X_2=0,\; X_3=0,\; X_4=\frac{1-\mu_2^6}{\lambda _2^{(1)} \mu _1^3 \mu _2^9}, \; X_5=\frac{\mu _1^3}{\lambda _1^{(1)} \mu _2^3},\; X_6=0,\; X_7=0,\; X_8=\lambda _2^{(1)} \mu _1^3  \mu _2^3\left(1-\mu_2^6\right)\,.
\end{equation}
The respective second constraint reads:
\begin{equation}\label{3_1_root_2}
	\lambda_1=-\frac{\mu _1^6}{\mu _2^6},\quad \lambda_2=\frac{1}{\mu _1^6 \mu _2^{12}}\,.
\end{equation}

The third root reads:
\begin{equation}
X_1=\lambda _1^{(1)}  \mu _1^3  \mu _2^3\left(\mu_1^6-1\right),\; X_2=0,\; X_3=0,\; X_4=-\frac{\mu _2^3}{\lambda _2^{(1)} \mu _1^3},\; X_5=\frac{\mu_1^6-1}{\lambda _1^{(1)} \mu _1^9 \mu _2^3},\; X_6=0,\; X_7=0,\; X_8=\frac{\lambda _2^{(1)} \mu _1^3}{\mu _2^3}\,.
\end{equation}
The respective third constraint reads:
\begin{equation}
	\lambda_1=\frac{1}{\mu _2^6 \mu _1^{12}},\quad \lambda_2=-\frac{\mu _2^6}{\mu _1^6}\,.
\end{equation}

The fourth root reads:
\begin{equation}
	\scalebox{0.9}{$\displaystyle
		X_1=\lambda _1^{(1)}  \mu _1^3  \mu _2^3\left(\mu_1^6-1\right),\; X_2=0,\; X_3=0,\; X_4=\frac{1-\mu_2^6}{\lambda _2^{(1)} \mu _1^3 \mu _2^9},\; X_5=\frac{\mu_1^6-1}{\lambda _1^{(1)} \mu _1^9 \mu _2^3},\; X_6=0,\; X_7=0,\; X_8=\lambda _2^{(1)} \mu _1^3  \mu _2^3\left(1-\mu_2^6\right)$}\,.
\end{equation}
The respective fourth constraint reads:
\begin{equation}\label{3_1_root_3}
	\lambda_1=\frac{1}{\mu _1^{12}\mu _2^6},\quad \lambda_2=\frac{1}{\mu _1^{6}\mu _2^{12}}\,.
\end{equation}

For another branch when both $X_2\neq 0$, $X_3\neq 0$ we would obtain two roots.
The fifth root reads:
\begin{equation}
	\begin{aligned}
		&\scalebox{0.9}{$\displaystyle X_1=-\frac{\lambda _1^{(1)} \left(1-\mu _1^6\right)}{\mu _1^3 \mu _2^3 \left(1-\mu _1^6 \mu _2^6\right)},\; X_2={\bf\color{burgundy} x},\; X_3=\frac{\left(1-\mu _1^6-\mu _2^6+\mu _2^6 \mu _1^6\right)\left(1-\mu _1^6 \mu _2^6+\mu _1^{12} \mu _2^{12}\right)\lambda^{(1)}}{\left(1-2 \mu _1^6 \mu _2^6+\mu _1^{12} \mu _2^{12}\right)\mu _2^6\lambda_2^{(1)}{\bf\color{burgundy}x}},\; X_4=-\frac{\mu _1^9\mu _2^3\left(1-\mu _2^6\right)}{\lambda _2^{(1)}\left(1-\mu _1^6 \mu _2^6\right) }$}\,,\\
		&\scalebox{0.9}{$\displaystyle X_5=\frac{ \mu _1^3  \mu _2^9 \left(1-\mu _1^6\right)}{\lambda _1^{(1)}\left(1-\mu _1^6 \mu _2^6\right)},\;
		X_6=-\frac{\lambda _2^{(1)} \mu _2^3 {\bf\color{burgundy} x}}{\lambda _1^{(1)} \mu _1^3},\;
		X_7=-\frac{\left(1-\mu _1^6-\mu _2^6+\mu _2^6 \mu _1^6\right) \left(1-\mu _1^6 \mu _2^6+\mu _1^{12} \mu _2^{12}\right)}{\mu _1^3 \mu _2^3  \left(1-2 \mu _1^6 \mu _2^6+\mu _1^{12} \mu _2^{12}\right) {\bf\color{burgundy} x}},\;
		X_8=\frac{\lambda _2^{(1)} \left(1-\mu _2^6\right)}{\mu _1^3 \mu _2^3 \left(1-\mu _1^6 \mu _2^6\right)}$}\,.
	\end{aligned}
\end{equation}
Clearly these solution has a scaling module we denoted as $\bf\color{burgundy} x$ appearing in CG chords with ends marked by different shift indices.
Nevertheless it imposes the fifth constraint:
\begin{equation}\label{3_1_root_4}
	\lambda_1=-\mu _1^6 \mu _2^{12},\quad \lambda_2=-\mu _1^{12} \mu _2^6\,.
\end{equation}

Finally, the sixth root also has a modulus:
\begin{equation}
	\begin{aligned}
		&X_1=-\frac{\lambda _1^{(1)} \left(1+\mu _2^6\right)}{\mu _1^3 \mu _2^3 \left(1-\mu _1^6 \mu _2^6\right)},\; 
			X_2={\bf\color{burgundy} x},\; 
			X_3=-\frac{\lambda _1^{(1)} \left(1+\mu _1^6 \mu _2^{12}+\mu _1^{12} \mu _2^6+\mu _1^{18} \mu _2^{18} \right)}{\lambda _2^{(1)} \mu _1^6 \left(1-2 \mu _1^6 \mu _2^6+\mu _1^{12} \mu _2^{12}\right){\bf\color{burgundy}x}},\; 
			X_4=\frac{\mu _1^3 \mu _2^9 \left(1+\mu _1^6\right) }{\lambda _2^{(1)} \left(1-\mu _1^6 \mu _2^6\right)}\,,\\
		&X_5=-\frac{\mu _1^9 \mu _2^3 \left(1+\mu _2^6\right)}{\lambda _1^{(1)} \left(1-\mu _1^6 \mu _2^6\right)},\;
			X_6=\frac{\lambda _2^{(1)} \mu _1^3 {\bf\color{burgundy} x}}{\lambda _1^{(1)} \mu _2^3},\;
			X_7=-\frac{\left(1+\mu _1^6 \mu _2^{12}+\mu _1^{12} \mu _2^6+\mu _1^{18} \mu _2^{18}\right)}{\mu _1^3 \mu _2^3 \left(1-2 \mu _1^6 \mu _2^6+\mu _1^{12} \mu _2^{12}\right) {\bf\color{burgundy} x}},\;
			X_8=\frac{\lambda _2^{(1)} \left(1+\mu _1^6\right)}{\mu _1^3 \mu _2^3 \left(1-\mu _1^6 \mu _2^6\right)}\,.
	\end{aligned}
\end{equation}
The respective sixth constraint reads:
\begin{equation}
	\lambda_1=\mu _1^{12} \mu _2^6,\quad \lambda_2=\mu _1^6 \mu _2^{12}\,.
\end{equation}

Another natural set of coordinates to describe A-polynomials is the eigenvalues of the $SL(3,\IC)$ connection holonomies on the  boundary torus of the fattened knot, for the A- and B-cycles (see Fig.~\ref{fig:lambda_mu}):
\begin{equation}
	\fm={\rm Hol}_A=\left(\begin{array}{ccc}
		m_1 & 0 & 0 \\
		0 & m_2 & 0\\
		0 & 0 & \frac{1}{m_1m_2}
	\end{array}\right),\quad \fl={\rm Hol}_B=\left(\begin{array}{ccc}
	\ell_1 & 0 & 0 \\
	0 & \ell_2 & 0\\
	0 & 0 & \frac{1}{\ell_1\ell_2}
	\end{array}\right)\,.
\end{equation}
Comparing these holonomies with their quantum traces \cite[eq.(6.1)]{Galakhov:2025ehn}, we find that $\mu_1=\left(m_1^2m_2\right)^{\frac{1}{6}}$, $\mu_2=\left(m_1m_2^2\right)^{-\frac{1}{6}}$.
Similarly, we can identify B-cycle holonomies with $\lambda_i$; however, there is freedom in modifying these variables by choosing different knot framings, or, equivalently, by applying Dehn twists to the fattened knot boundary surface.
We choose the following identification $\lambda_1=\ell_1/m_1^3$, $\lambda_2=m_2^3/\ell_2$.
Under this change of coordinates, five of the six roots we found for the shaded A-polynomials coincide with the roots constructed in \cite[eq.(7.13)]{Dimofte:2013iv}, while one root is spurious:
\begin{equation}
	\begin{aligned}
		&\left\{\ell_1+m_2^3 m_1^6=0, \;\ell_2+m_1^3 m_2^6=0\right\}\;{\rm (spurious)}\,,\\
		&\left\{\ell_1+m_2^3 m_1^6=0, \;\ell_2-1=0\right\}\,,\\
		&\left\{\ell_2+m_1^3 m_2^6=0,\;\ell_1-1=0\right\}\,,\\
		&\left\{\ell_1-1=0,\;\ell_2-1=0\right\}\,,\\
		&\left\{\ell_1 m_2^3+m_1^3=0,\;\ell_2 m_1^3+m_2^3=0\right\}\,,\\
		&\left\{\ell_1-m_1^6=0,\;\ell_2-m_2^6=0\right\}\,.
	\end{aligned}
\end{equation}
This observation provides support for our assumptions about th behavior of huge representations of $U_q(\fg)$ leading to the quasi-classical equations \eqref{main_eqs}.
Also, we can regard CG chords as a reliable source for \emph{shaded} A-polynomial roots beyond $\fg=\fs\fu_2$.

To conclude this subsection, let us also indicate that relation \eqref{eigen} is satisfied for all the roots determined here.


\subsection{Cinquefoil knot}

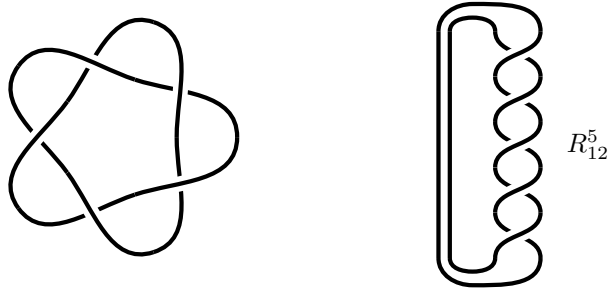
\begin{figure}[ht!]
	\centering
	\begin{tikzpicture}
		\node (A) at (0,0) {$\begin{array}{c}
				\begin{tikzpicture}[scale=0.8]
				\foreach \i in {0, ..., 4} {
					\begin{scope}[rotate=72*\i]
						\draw[ultra thick] (2,0) to[out=90,in=342] (0.309017, 0.951057);
						\draw[white, line width = 2mm] (1,0) to[out=90,in=342] (0.618034, 1.90211);
						\draw[ultra thick] (1,0) to[out=90,in=342] (0.618034, 1.90211);
					\end{scope}
				}
				\end{tikzpicture}
			\end{array}$};
		\node(B) at (5,0) {$\begin{tikzpicture}[scale=0.6]
				\foreach \i in {0, ..., 4} {
					\begin{scope}[shift={(0,\i)}]
						\draw[ultra thick] (1,0) to[out=90,in=270] (0,1);
						\draw[white, line width = 2mm] (0,0) to[out=90,in=270] (1,1);
						\draw[ultra thick] (0,0) to[out=90,in=270] (1,1);
					\end{scope}
				}
				\draw[ultra thick] (0,0) to[out=270,in=0] (-0.5,-0.3) to[out=180,in=270] (-1,0) (1,0) to[out=270,in=0] (-0.5,-0.6) to[out=180,in=270] (-1.25,0);
				\begin{scope}[shift={(0,5)}]
					\begin{scope}[yscale=-1]
						\draw[ultra thick] (0,0) to[out=270,in=0] (-0.5,-0.3) to[out=180,in=270] (-1,0) (1,0) to[out=270,in=0] (-0.5,-0.6) to[out=180,in=270] (-1.25,0);
					\end{scope}
				\end{scope}
				\draw[ultra thick] (-1,0) -- (-1,5) (-1.25,0) -- (-1.25,5);
			\end{tikzpicture}$};
		\node[right] at (B.east) {$R_{12}^5$};
	\end{tikzpicture}
	\caption{A cinquefoil knot $5_1$ diagram and its braid representation.}\label{fig:5_1}
\end{figure}

The cinquefoil knot $5_1$ depicted in Fig.~\ref{fig:5_1} in its minimal representation has five crossings, which1 is greater than the number for the figure-eight knot $4_1$.
Nevertheless $5_1$ belongs to a family of highly symmetric torus knots, therefore we can regard it as the next  simplest non-trivial knot after $3_1$.
However equations \eqref{main_eqs} written for this knot are rather involved and do not admit obvious simplifications like those for $3_1$.
We will not present these equations here due to their bulkiness.

Yet a simple exercisethat we can perform  with these equations is to check if there are ``colorless'' solutions when CG chords for $\alpha\neq\beta$ in \eqref{CGdef} vanish, similarly to roots \eqref{3_1_root_1}-\eqref{3_1_root_4} for $3_1$.
In other words, we use the following ansatz:
\begin{equation}
	\begin{aligned}
		&{}^{\Box}\Theta\scalebox{0.8}{$\left[\begin{array}{cc}
				1 & 1\\
				1 & 2
			\end{array}\right]$}=:X_1,\; 
		{}^{\Box}\Theta\scalebox{0.8}{$\left[\begin{array}{cc}
				1 & 2\\
				1 & 2
			\end{array}\right]$}=0,\;
		{}^{\Box}\Theta\scalebox{0.8}{$\left[\begin{array}{cc}
				2 & 1\\
				1 & 2
			\end{array}\right]$}=0,\;
		{}^{\Box}\Theta\scalebox{0.8}{$\left[\begin{array}{cc}
				2 & 2\\
				1 & 2
			\end{array}\right]$}=:X_4\,,\\
		&{}^{\Box}\Theta\scalebox{0.8}{$\left[\begin{array}{cc}
				1 & 1\\
				2 & 1
			\end{array}\right]$}=:X_5,\;
		{}^{\Box}\Theta\scalebox{0.8}{$\left[\begin{array}{cc}
				1 & 2\\
				2 & 1
			\end{array}\right]$}=0,\;
		{}^{\Box}\Theta\scalebox{0.8}{$\left[\begin{array}{cc}
				2 & 1\\
				2 & 1
			\end{array}\right]$}=0,\;
		{}^{\Box}\Theta\scalebox{0.8}{$\left[\begin{array}{cc}
				2 & 2\\
				2 & 1
			\end{array}\right]$}=:X_8\,.
	\end{aligned}
\end{equation}

We arrive at the following collection of equations:
\begin{subequations}
\begin{equation}\label{5_1_eq_1}
	\mu _1^6-\lambda _1^{(1)} \mu _2^5 \mu _1 X_5+\mu _1^6 X_1 X_5-X_1^2 X_5^2-3 X_1 X_5-1=0\,,
\end{equation}
\begin{equation}
	-\lambda _1^{(1)} \mu _2^5 \mu _1^6+\lambda _1^{(1)} \mu _2^5-2 \mu _1^{11} X_1-\mu _1^{11} X_1^2 X_5+\mu _1^5 X_1^3 X_5^2+3 \mu _1^5 X_1+4 \mu _1^5 X_1^2 X_5=0\,,
\end{equation}
\begin{equation}
	\mu _1^{11}-\mu _1^5-\lambda _1^{(2)} \mu _2^5 X_1+\mu _1^{11} X_1^2 X_5^2+3 \mu _1^{11} X_1 X_5-\mu _1^5 X_1^3 X_5^3-5 \mu _1^5 X_1^2 X_5^2-6 \mu _1^5 X_1 X_5=0
\end{equation}
\begin{equation}\label{5_1_eq_4}
	-\lambda _1^{(2)} \mu _2^5 \mu _1^6+\lambda _1^{(2)} \mu _2^5-\mu _1^{11} X_1 X_5^2-2 \mu _1^{11} X_5+\mu _1^5 X_1^2 X_5^3+4 \mu _1^5 X_1 X_5^2+3 \mu _1^5 X_5=0\,,
\end{equation}
\begin{equation}\label{5_1_eq_5}
	\mu _1^5 \mu _2^6-\mu _1^5+\mu _1^5 \mu _2^6 X_4^2 X_8^2+3 \mu _1^5 \mu _2^6 X_4 X_8-\mu _1^5 X_4 X_8+\mu _2^5 X_8\left(\lambda _2^{(1)}\right)^{-1}=0\,,
\end{equation}
\begin{equation}
	\mu _1^5 \mu _2^6 X_4^3 X_8^2+3 \mu _1^5 \mu _2^6 X_4+4 \mu _1^5 \mu _2^6 X_4^2 X_8-2 \mu _1^5 X_4-\mu _1^5 X_4^2 X_8+\left(\mu _2^{11}-\mu _2^5\right)\left(\lambda _2^{(1)}\right)^{-1}=0\,,
\end{equation}
\begin{equation}
	\mu _1^5 \mu _2^6 X_4^2 X_8^3+4 \mu _1^5 \mu _2^6 X_4 X_8^2+3 \mu _1^5 \mu _2^6 X_8-\mu _1^5 X_4 X_8^2-2 \mu _1^5 X_8+\left(\mu _2^{11}-\mu _2^5\right)\left(\lambda _2^{(2)}\right)^{-1}=0
\end{equation}
\begin{equation}\label{5_1_eq_8}
	\mu _1^5 \mu _2^6-\mu _1^5+\mu _1^5 \mu _2^6 X_4^3 X_8^3+5 \mu _1^5 \mu _2^6 X_4^2 X_8^2+6 \mu _1^5 \mu _2^6 X_4 X_8-\mu _1^5 X_4^2 X_8^2-3 \mu _1^5 X_4 X_8+\mu _2^{11} X_4\left(\lambda _2^{(2)}\right)^{-1}=0\,.
\end{equation}
\end{subequations}

We should note that variables are separated: the first four equations \eqref{5_1_eq_1}-\eqref{5_1_eq_4} depend only on $X_1$ and $X_5$, whereas  \eqref{5_1_eq_5}-\eqref{5_1_eq_8} depend only on $X_4$ and $X_8$.
In the first quadruplet, we eliminate $X_5$ first by finding a solution in terms of $X_1$:
\begin{equation}
	X_5=\frac{\lambda _1^{(1)} \mu _2^5 \mu _1^6-\lambda _1^{(1)} \mu _2^5+\mu _1^{11} X_1-2 \mu _1^5 X_1}{\mu _1^5 X_1 \left(X_1-\lambda _1^{(1)} \mu _1 \mu _2^5\right)}\,.
\end{equation}
Substituting this solution back produces equations solely in $X_1$, which are easy to solve explicitly, yielding three roots:
\begin{equation}
	X_1=\left(\mu _1^{13}-\mu _1^7\right)\lambda _1^{(1)} \mu _2^5,\quad X_1=\frac{\left(\pm\sqrt{5}+1\right) \lambda _1^{(1)} \mu _2^5}{2 \mu _1^5}\,.
\end{equation}
These solutions produce the first shaded A-polynomial component:
\begin{equation}\label{5_1_A_1}
	A_1=\left(\lambda _1 \mu _1^{20} \mu _2^{10}-1\right)\left(\lambda _1 \mu _2^{10}+\mu _1^{10}\right)=0\,.
\end{equation}

Applying a similar calculation to \eqref{5_1_eq_5}-\eqref{5_1_eq_8}, we arrive at the second component:
\begin{equation}\label{5_1_A_2}
	A_2=\left(\lambda _2 \mu _1^{10} \mu _2^{20}-1\right)\left(\lambda _2 \mu _1^{10}+\mu _2^{10}\right)=0
\end{equation}

Let us stress that these equations do not describe a complete set of shaded A-polynomial roots for $5_1$.
These are only the roots we found under the assumption that the ``colorful'' CG chords vanish.

As we have said at the beginning of this subsection, the knots $3_1$ and $5_1$ belong to a family of torus knots $T_{2,2k+1}$, $k=1,2,3,\ldots$.
Looking at a list of A-polynomials for $\fs\fu_2$ for this family of knots (see e.g. \cite[Tab. 2]{Gukov:2003na}), it is easy to anticipate a simple pattern for A-polynomials in this family of knots for generic $k$:
\begin{equation}
	A^{\fs\fu_2}(T_{2,2k+1})=\ell+m^{2(2k+1)}\,.
\end{equation}
If we expect that a simple pattern of A-polynomials is inherited in this knot family, then it is simple to extend our shaded A-polynomials for $\fs\fu_3$ to generic $k$ as well by comparing \eqref{5_1_A_1} and \eqref{5_1_A_2} with \eqref{3_1_root_1}-\eqref{3_1_root_4}:
\begin{equation}\label{torus_A}
	\vec A^{\fs\fu_3}(T_{2,2k+1})=\left(\begin{array}{c}
		\left(\lambda _1 \mu _1^{4(2k+1)} \mu _2^{2(2k+1)}-1\right)\left(\lambda _1 \mu _2^{2(2k+1)}+\mu _1^{2(2k+1)}\right)\\
		\left(\lambda _2 \mu _1^{2(2k+1)} \mu _2^{4(2k+1)}-1\right)\left(\lambda _2 \mu _1^{2(2k+1)}+\mu _2^{2(2k+1)}\right)
	\end{array}\right)\,.
\end{equation}
It should be stressed here again that \eqref{torus_A} covers only a part of the actual moduli space and may contain spurious roots.
The unknot, that seems to enter this family for $k=0$, however has only a \emph{single} root, which in this framing reads:
\begin{equation}
	\vec A^{\fs\fu_3}({\rm unknot})=\left(\begin{array}{c}
		\lambda _1 \mu _1^{4} \mu _2^{2}-1\\
		\lambda _2 \mu _1^{2} \mu _2^{4}-1
	\end{array}\right)\,.
\end{equation}


\subsection{Figure-eight knot}

\begin{figure}[ht!]
	\centering
	\begin{tikzpicture}
		\node(A) at (0,0) {$\begin{array}{c}
				\begin{tikzpicture}[scale=1]
					\draw[white,line width = 1.5mm] (-0.5,0) to[out=90,in=270] (1,1.5) to[out=90,in=0] (0,2);
					\draw[ultra thick] (-0.5,0) to[out=90,in=270] (1,1.5) to[out=90,in=0] (0,2);
					\draw[white,line width = 1.5mm] (0,1) to[out=180,in=90] (-1.5,0) to[out=270,in=180] (-0.5,-0.7) to[out=0,in=270] (0.5,0);
					\draw[ultra thick] (0,1.2) to[out=180,in=90] (-1.5,0) to[out=270,in=180] (-0.5,-0.7) to[out=0,in=270] (0.5,0);
					\draw[white,line width = 1.5mm] (0,2) to[out=180,in=90] (-1,1.5) to[out=270,in=90] (0.5,0);
					\draw[ultra thick] (0,2) to[out=180,in=90] (-1,1.5) to[out=270,in=90] (0.5,0);
					\begin{scope}[xscale=-1]
						\draw[white,line width = 1.5mm] (0,1.2) to[out=180,in=90] (-1.5,0) to[out=270,in=180] (-0.5,-0.7) to[out=0,in=270] (0.5,0);
						\draw[ultra thick] (0,1.2) to[out=180,in=90] (-1.5,0) to[out=270,in=180] (-0.5,-0.7) to[out=0,in=270] (0.5,0);
					\end{scope}
				\end{tikzpicture}
			\end{array}$};
		\node(B) at (5,0) {$\begin{tikzpicture}[scale=0.6]
				\begin{scope}
					\draw[ultra thick] (1,0) to[out=90,in=270] (0,1) (2,0) -- (2,1);
					\draw[white, line width = 2mm] (0,0) to[out=90,in=270] (1,1);
					\draw[ultra thick] (0,0) to[out=90,in=270] (1,1);
				\end{scope}
				\begin{scope}[shift={(0,1)}]
					\draw[ultra thick] (1,0) to[out=90,in=270] (2,1) (0,0) -- (0,1);
					\draw[white, line width = 2mm] (2,0) to[out=90,in=270] (1,1);
					\draw[ultra thick] (2,0) to[out=90,in=270] (1,1);
				\end{scope}
				\begin{scope}[shift={(0,2)}]
					\draw[ultra thick] (1,0) to[out=90,in=270] (0,1) (2,0) -- (2,1);
					\draw[white, line width = 2mm] (0,0) to[out=90,in=270] (1,1);
					\draw[ultra thick] (0,0) to[out=90,in=270] (1,1);
				\end{scope}
				\begin{scope}[shift={(0,3)}]
					\draw[ultra thick] (1,0) to[out=90,in=270] (2,1) (0,0) -- (0,1);
					\draw[white, line width = 2mm] (2,0) to[out=90,in=270] (1,1);
					\draw[ultra thick] (2,0) to[out=90,in=270] (1,1);
				\end{scope}
				\draw[ultra thick] (0,0) to[out=270,in=0] (-0.5,-0.3) to[out=180,in=270] (-1,0) (1,0) to[out=270,in=0] (-0.5,-0.6) to[out=180,in=270] (-1.25,0) (2,0) to[out=270,in=0] (-0.5,-0.9) to[out=180,in=270] (-1.5,0);
				\begin{scope}[shift={(0,4)}]
					\begin{scope}[yscale=-1]
						\draw[ultra thick] (0,0) to[out=270,in=0] (-0.5,-0.3) to[out=180,in=270] (-1,0) (1,0) to[out=270,in=0] (-0.5,-0.6) to[out=180,in=270] (-1.25,0) (2,0) to[out=270,in=0] (-0.5,-0.9) to[out=180,in=270] (-1.5,0);
					\end{scope}
				\end{scope}
				\draw[ultra thick] (-1,0) -- (-1,4) (-1.25,0) -- (-1.25,4) (-1.5,0) -- (-1.5,4);
			\end{tikzpicture}$};
		\node[right] at (B.east) {$R_{23}^{-1}R_{12}R_{23}^{-1}R_{12}$};
	\end{tikzpicture}
	\caption{A figure-eight knot $4_1$ diagram and its braid representation.}\label{fig:4_1}
\end{figure}
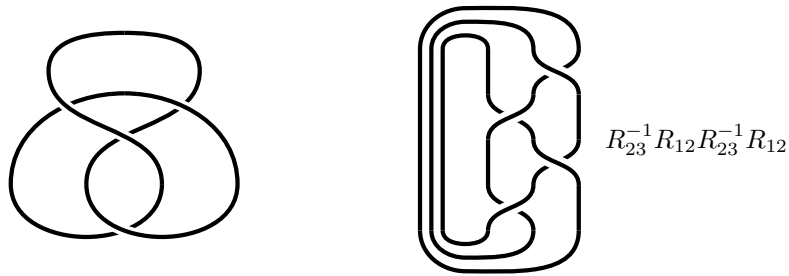

The task of computing the shaded A-polynomials for the figure-eight knot depicted in Fig.~\ref{fig:4_1} is rather involved.
Its minimal braid representation has three strands, which implies there are 24 CG chord variables; the  two types of relations \eqref{main_eqs} then produce 36 equations.
We will attempt to search for all shaded A-polynomial roots elsewhere.
Here we present one root derived under the ansatz used in the case $5_1$, when all the ``colorful'' CG chords with distinct indices $\alpha$ and $\beta$ in \eqref{CGdef} are set to zero.
This simplification leads to the following non-trivial root:
\begin{equation}
	\begin{aligned}
		&A_1=\mu _1^{12}+\left(-1+\mu _1^6 +2 \mu _1^{12}+\mu _1^{18}-\mu _1^{24}\right)\lambda_1+\mu _1^{12}\lambda_1^2\,,\\
		&A_2=\mu _2^{12}+\left(-1+\mu _2^6 +2 \mu _2^{12}+\mu _2^{18}-\mu _2^{24}\right)\lambda_2+\mu _2^{12}\lambda_2^2\,.
	\end{aligned}
\end{equation}
This is a specific root satisfying:
\begin{equation}
	A_1=A_0(\lambda_1,\mu_1),\quad A_2=A_0(\lambda_2,\mu_2)\,,
\end{equation}
where $A_0(\lambda,\mu^3)$ is the standard $4_1$ A-polynomial for $\fs\fu_2$ (see e.g. \cite[Tab. 2]{Gukov:2003na}).

One might think of this root as one where sectors of the Chern-Simons theory for two simple roots of algebra $\fs\fu_3$ are decoupled from each other.
We would like to argue that, in general, it is not unnatural for a system to have such a decoupled root.
If we substitute
\begin{equation}
	{}^{\rho}\Theta\scalebox{0.8}{$\left[\begin{array}{cc}
			\alpha & \beta\\
			k & m
		\end{array}\right]$}=0,\quad\mbox{if }\alpha\neq \beta\,,
\end{equation}
into \eqref{Rmorph_beg}-\eqref{Rmorph_end} and \eqref{RRmorph_beg}-\eqref{RRmorph_end}, then in the sum over $\gamma\neq 0$ only a single term survives.
Moreover, CG chords for different color indices $\alpha$ are never mixed in \eqref{main_eqs}.
Therefore, these equations decompose into independent components for each $\alpha$, and each component is reminiscent of this system for $\fg=\fs\fu_2$.


\section{Conclusion}

In this note, we have considered an extension $\fs\fu_n$, $n>2$, of the CG chord formalism for constructing knot A-polynomials, following a proposition in \cite{Galakhov:2024eco}.
We have defined new CG chords carrying color indices associated with roots of $\fs\fu_n$.
For these chords, the action of the braid group is derived explicitly in \eqref{Rmorph_beg}-\eqref{Rmorph_end} and \eqref{RRmorph_beg}-\eqref{RRmorph_end}.
Using this representation, we associated with a knot $K$ with its diagram in the braid representation a nonlinear eigenvalue problem \eqref{main_eqs}.
The consistency constraints for this system to have solutions \eqref{A-poly}, i.e. generalized resultants, deliver the roots of classical shaded A-polynomials for $\fs\fu_n$ that are quasi-classical limits of difference operators annihilating HOMFLY-PT polynomials.
Also, we have discussed some aspects of A-polynomial covariance with respect to Markov moves of the knot braid diagram and computed some examples of shaded A-polynomials for a few knots.

To conclude this note let us list some open problems that might seem interesting for further research:
\begin{enumerate}
	\item Due to its geometric nature, A-polynomials ought to transform covariantly with respect to Reidemeister moves of the knot diagram in the plat representation and to Markov moves in the braid representation.
	Yet this covariance is not transparent for the chord technique we presented here.
	In Sec.~\ref{sec:covariant} we have mapped out some pathways in this direction.
	However, it would be interesting to prove  a complete covariance theorem for shaded A-polynomials solely within the chord framework.
	\item It would be interesting to extend the CG chord formalism to generic simple Lie (super)algebra $\fg$.
	Especially to $\fe_8$, $\ff_4$, $\fg_2$, where there are neither minuscule nor cominuscule weights, so there are no universal multiplicity-free tensor multipliers like in \eqref{multi-free}.
	Supposedly, rather than exploiting skein relations \eqref{uni_skein}, one should apply eigenvalue decompositions for R-matrices directly to construct an action of the braid group.
	\item It is well-known  from an analysis of the knot complement fundamental group \cite{Cooper1994Plane} that there is a relation between A-polynomials and Alexander polynomials, the latter being equivalent to HOMFLY polynomials for the supergroup $GL(1|1)$ \cite{Kauffman1991FreeFA}.
	We hope that an extension of the CG chord formalism to supergroups might deliver a simpler explanation of this relation bypassing a direct reference to the knot fundamental group.
	\item It would be interesting to extend Perelomov's quasi-classical description for huge representations from classical groups to quantum ones.
	Our results suggest that such a formalism exists and is non-trivial.
	Potentially, it would allow one to construct simplified versions of universal R-matrix expressions in the double scaling limit.
	\item The action of the braid group on the CG chords is reminiscent of cluster mutations for cluster coordinates \cite{Kontsevich:2008fj}.
	The universality of the braid group action \eqref{Rmorph_beg}-\eqref{Rmorph_end} and \eqref{RRmorph_beg}-\eqref{RRmorph_end} and of the tetrahedron construction for snake spectral networks \cite{Dimofte:2013iv} suggests there might be a complete Hikami-like 3d TQFT construction for knot/link invariants associated with $SL(n,\IC)$.
	Also, one might hope to obtain universal double-scaling asymptotics for 6j-symbols/Racah coefficients of $U_q(\fs\fu_n)$ hidden behind this construction for finite representations \cite{Mironov:2016pyz,Bai:2018twf,Dhara:2018wqe,Alekseev:2019klg}.
\end{enumerate}


\section*{Acknowledgments}

We would like to thank Emil Akhmedov, Semeon Arthamonov, Lotte Hollands, Victor Mishnyakov, Mikhail Olshanetsky, Alexei Sleptsov, Andrei Zotov for illuminating discussions at various stages of the project.
The work was supported by the state assignment of the Institute for Information Transmission Problems of RAS.


\appendix

\section{Quasi-classical wave packet for huge spins of \texorpdfstring{$SO(3)$}{SO(3)}}\label{app:packet}

In spherical coordinates, with
\begin{equation}\label{raduis-vector}
	x=\cos\theta\,\sin\varphi,\quad y=\cos\theta\,\cos\varphi,\quad z=\sin\theta\,.
\end{equation}
one constructs the following angular momentum operators:
\begin{equation}
	\begin{aligned}
		&s_x=-\I(y\p_z-z\p_y)=-\I \left( \cos \varphi\, \frac{\p}{\p\theta}+ \tan \theta \, \sin \varphi \,\frac{\p}{\p\varphi}\right)\,,\\
		&s_y=-\I(z\p_x-x\p_z)=\I\left ( \sin \varphi \frac{\p}{\p\theta}- \tan \theta \, \cos \varphi \, \frac{\p}{\p\varphi}\right)\,,\\
		&s_z=-\I(x\p_y-yp_x)=\I \frac{\p}{\p\varphi}\,.
	\end{aligned}
\end{equation}
The volume form in these coordinates reads $\cos\theta\,d\theta\,d\varphi$.

It is useful to consider the raising/lowering operators:
\begin{equation}
	s_{\pm}:=s_x\pm \I s_y=-\I e^{\mp\I\varphi}\left(\frac{\p}{\p\theta}\pm \tan\theta\,\frac{\p}{\p\varphi}\right)\,.
\end{equation} 

Then, for the Laplacian, we find:
\begin{equation}
	\Delta=s_x^2+s_y^2+s_z^2=-\frac{\p^2}{\p\theta^2}+\tan\theta\,\frac{\p}{\p\theta}-\frac{1}{\cos\theta^2}\frac{\p^2}{\p\varphi^2}\,.
\end{equation}

Solutions to $\Delta\psi=\ell(\ell+1)\psi$ are well known in terms of spherical harmonics.
Yet we would like to find \emph{coherent} states approximating quasi-classical wave functions.
Let us start with a state $|\ell,\ell\rangle$.
It is annihilated by $s_+$ and has eigenvalue $\ell$ for $s_z$:
\begin{equation}\label{jj}
	|\ell,\ell\rangle=\frac{1}{2^\ell \ell!}\sqrt{\frac{(2\ell+1)!}{4\pi}}\,\cos\theta^\ell \,e^{-\I \ell\varphi}\sim e^{-\ell\frac{\theta^2}{2}+\I \ell \varphi}\,.
\end{equation}

Now we would like to construct a state that quasi-classically represents a particle spinning with a \emph{large} angular momentum $\ell$ around an axis aligned with a unit vector $\vec n$.
So it should behave as the $SO(3)$-rotated asymptotic form of \eqref{jj}:
\begin{equation}\label{quasi-psi}
	\psi(\vec n,\ell)\sim\exp\left(-\frac{\ell}{2}\left(\vec n\cdot\vec r\right)^2-\I\ell\arccos\frac{\vec\nu\cdot\vec\rho}{|\vec\nu||\vec\rho|} \right)\,,
\end{equation}
where $\vec r$ is a vector with coordinates \eqref{raduis-vector}, and
\begin{equation}
	\vec\nu=(n_y,-n_x,0),\quad \vec\rho=\vec r-(\vec n\cdot \vec r)\vec n\,.
\end{equation}
As $\ell\to\infty$, the real part of the argument of the exponential  in \eqref{quasi-psi} cuts out the solution as a delta function:
\begin{equation}
	\sin\theta=-\frac{1}{n_z}\left(n_y \cos\theta \, \cos \varphi+n_x \cos\theta\, \sin \varphi \right)\,.
\end{equation}
Using this principle, we derive:
\begin{equation}
	\vec s\cdot\psi(\vec n,\ell)=\ell\vec n\;\psi(\vec n,\ell)+O(1)\,.
\end{equation} 


\bibliographystyle{utphys}
\bibliography{biblio}

\end{document}